\documentclass[prd,nofootinbib,preprint,superscriptaddress]{revtex4}

\usepackage{amsmath, amssymb, amsthm, graphicx, epsfig, fancyhdr,epsfig, slashed}
\usepackage{tikzsymbols}
\usepackage{natbib}
\usepackage{tabularx}
\usepackage{float}
\usepackage{xcolor}
\usepackage{makecell}
\usepackage[utf8]{inputenc}
\usepackage{booktabs}
\usepackage{tabularx}
\usepackage{multirow}
\usepackage{enumitem}
\usepackage{amsmath}
\usepackage{tikz,xcolor,hyperref}
\usepackage{slashed}
\usepackage{subcaption}
\usepackage{braket}
\usepackage{hhline}
\usepackage{graphicx}   
\usepackage{multirow}   
\usepackage{makecell}   
\usepackage{float}  
\usepackage{multirow}
\definecolor{lime}{HTML}{A6CE39}
\DeclareRobustCommand{\orcidicon}{
	\begin{tikzpicture}
	\draw[lime, fill=lime] (0,0) 
	circle [radius=0.2] 
	node[white] {{\fontfamily{qag}\selectfont \tiny ID}};
	\draw[white, fill=white] (-0.0625,0.095) 
	circle [radius=0.007];
	\end{tikzpicture}
	\hspace{-2mm}
}
\foreach \x in {A, ..., Z}{\expandafter\xdef\csname orcid\x\endcsname{\noexpand\href{https://orcid.org/\csname orcidauthor\x\endcsname}
			{\noexpand\orcidicon}}
}

\newcommand{\be}{\begin{equation}}
\newcommand{\ee}{\end{equation}}
\newcommand{\bea}{\begin{eqnarray}}
\newcommand{\eea}{\end{eqnarray}}

\newcommand{\ba}{\begin{eqnarray}}
\newcommand{\ea}{\end{eqnarray}}
\newcommand{\bi}{\begin{itemize}}
\newcommand{\ei}{\end{itemize}}

\newcommand{\x}{\star}

\begin{document}
\title{Grand Unification Higgs-$\mathcal{R}^2$ Inflation: \\ \it{Complementarity between Proton Decay and CMB Observables}}
\author{Nilay Bostan\orcidA}
\email{nilay.bostan@marmara.edu.tr}
\affiliation{Department of Physics, Faculty of Science, Marmara University, 34722 Istanbul, Türkiye}
\author{Rafid H. Dejrah\orcidB{}}
\email{rafid.dejrah@gmail.com}
\thanks{he/him.}
\affiliation{Department of Physics, Faculty of Science, Ankara University,  06100 Ankara, Türkiye}
\author{Anish Ghoshal\orcidC{}}
\email{anish.ghoshal@fuw.edu.pl}
\affiliation{Institute of Theoretical Physics, Faculty of Physics,\\ University of Warsaw,
ul. Pasteura 5, 02-093 Warsaw, Poland}
\begin{abstract}
{
\renewcommand{\thefootnote}{\fnsymbol{footnote}}%
\footnotemark[0]%
}
We propose a predictive $SO(10)$ Grand Unified Theory (GUT) framework for cosmic inflation within the Palatini $\mathcal{R}^2$ formulation of gravity. In this model, a GUT Higgs field simultaneously drives inflation and induces intermediate-scale symmetry breaking, thereby linking primordial cosmology, gauge unification, and formation of topological defects. A subsequent partial inflationary phase of
$N_I \sim 10\text{--}17$ $e$-folds occurring after monopole formation can dilute magnetic monopoles to abundances
$Y_M \sim 10^{-35}\text{--}10^{-27}$.
The model predicts Cosmic Microwave Background (CMB) observables in the range
$0.955 \lesssim n_s \lesssim 0.974,$
accommodating the mild tension between the \emph{Planck}--BICEP ($n_s \simeq 0.965$) and \emph{Planck}+ACT ($n_s \simeq 0.971$) measurements via the $\phi < M$ and $\phi > M$ branches, respectively, where $M$ denotes the renormalization scale entering the Coleman--Weinberg potential. The predicted tensor-to-scalar ratio,
$r \lesssim 8 \times 10^{-4},$
is consistent with current observational bounds and may be accessible to forthcoming experiments, including the Simons Observatory and LiteBIRD. Furthermore, correlations among the unification scale $M_U$, the inflationary observables $(n_s, r)$, and proton-decay lifetimes reveal a strong complementarity between CMB measurements and proton-decay searches, with regions of parameter space testable by upcoming experiments, such as Hyper-Kamiokande and DUNE.
\end{abstract}
{
\renewcommand{\thefootnote}{\fnsymbol{footnote}}%
\footnotetext[0]{The authors contributed equally to this work.}%
\renewcommand{\thefootnote}{\arabic{footnote}}%
}
\maketitle
\tableofcontents
\flushbottom
\section{Introduction}\label{sec:intro}
Among the various candidates proposed for the scalar field responsible for driving cosmic inflation~\cite{Starobinsky:1980te,Guth:1980zm,Linde:1981mu,Albrecht:1982wi} in the early Universe (see, e.g.,~\cite{Sato:2015dga} for a review), the Higgs field of the Standard Model (SM) stands out as one of the most well-motivated. Its discovery at the Large Hadron Collider (LHC)~\cite{Aad:2012tfa,Chatrchyan:2012xdj} establishes it as a fundamental scalar field, making it a natural candidate for the inflaton.

Since the proposal of Higgs inflation, numerous extensions and variants have been investigated~\cite{Kamada:2012se}. The original Higgs inflation model, however is based on a non-minimal coupling to gravity of the form $\xi |{\cal H}|^2 \mathcal{R}$, where $\mathcal{R}$ is the Ricci scalar curvature, with $\xi = {\cal O}(10^4)$~\cite{CervantesCota:1995tz,Bezrukov:2007ep,Barvinsky:2008ia} with embeddings in scale-invariant extension of the SM within General Relativity (GR) being quite successful with CMB predictions~\cite{Akrami:2018odb}.

Remarkably, the inflationary predictions of Higgs inflation closely resemble those of the well-known Starobinsky model~\cite{Starobinsky:1980te}, despite their markedly different theoretical origins. Following the results of the \textit{Planck} 2013 mission~\cite{Planck:2013jfk}, it was shown that both models belong to a broader class of theories known as inflationary, or \textit{cosmological attractors}, more specifically, $\alpha$-attractors~\cite{Kallosh:2013daa, Cecotti:2014ipa, Galante:2014ifa, Roest:2015qya, Carrasco:2015uma, Carrasco:2015rva}.
 
Although the Hubble scale during inflation is estimated as
\[
H \sim \lambda^{1/4} \frac{m_{\rm Pl}}{\xi^{1/2}},
\]
where $\lambda$ is the Higgs self-quartic coupling at the scale of inflation and $m_{\rm Pl}$ is the reduced Planck mass. The scale of inflation $H$ falls well above the tree-level cutoff of the theory in vacuum, $\Lambda \sim m_{\rm Pl}/\xi$~\cite{Burgess:2009ea,Barbon:2009ya,Burgess:2010zq,Hertzberg:2010dc}. Nevertheless, since the perturbative cutoff associated with Higgs fluctuations during inflation can remain well above the inflationary scale, according to Refs.~\cite{Barvinsky:2009ii,Bezrukov:2010jz}, this apparent problem might not be critical. This clearly illustrates several subtleties associated with the effective field theory (EFT) treatment of Higgs inflation.

The non-renormalizable nature of gravity in GR further complicates the picture, introducing ambiguities in low-energy observables~\cite{Bezrukov:2009db,Bezrukov:2014ipa}. These become particularly pivotal during reheating, when transverse electroweak gauge bosons may be produced non-perturbatively in large numbers~\cite{DeCross:2015uza,Ema:2016dny,Sfakianakis:2018lzf}.

Several proposals have been explored to address these issues, including raising the cutoff scale in Higgs inflation scenarios, sometimes up to the Planck scale~\cite{Giudice:2010ka,Barbon:2015fla,Lee:2018esk}. Among these possibilities, a particularly natural and well-motivated approach within the EFT framework is the Higgs, $\mathcal{R}^2$ model~\cite{Wang:2017fuy,Ema:2017rqn,He:2018gyf,Gundhi:2018wyz}. In this framework, inflation is driven jointly by the Higgs field and the \textit{scalaron}, the scalar degree of freedom associated with the geometric $\mathcal{R}^2$ term~\cite{Starobinsky:1980te}, which provides an ultraviolet (UV) completion of Standard Model Higgs inflation, as demonstrated in Refs.~\cite{Salvio:2015kka, Ema:2017rqn,Gorbunov:2018llf}.

Within the context of cosmological attractor phenomena, the Higgs--$\mathcal{R}^2$ framework emerges naturally, yielding remarkably stable predictions for CMB observables that are fully consistent with current \emph{Planck}-BICEP measurements~\cite{Wang:2017fuy,Enckell:2018uic,Rodrigues:2020fle}. In various limits, this framework reduces to either pure Higgs inflation or pure Starobinsky inflation. Moreover, it has been successfully embedded within string theory constructions and reconstructions~\cite{Copeland:2013vva, Ellis:2013nxa,Ellis:2014gxa,Ellis:2015kqa, Salvio:2016vxi, Ellis:2020lnc, Oikonomou:2025htz}.
\begin{figure}[t!]
    \centering
    \includegraphics[width=1\linewidth]{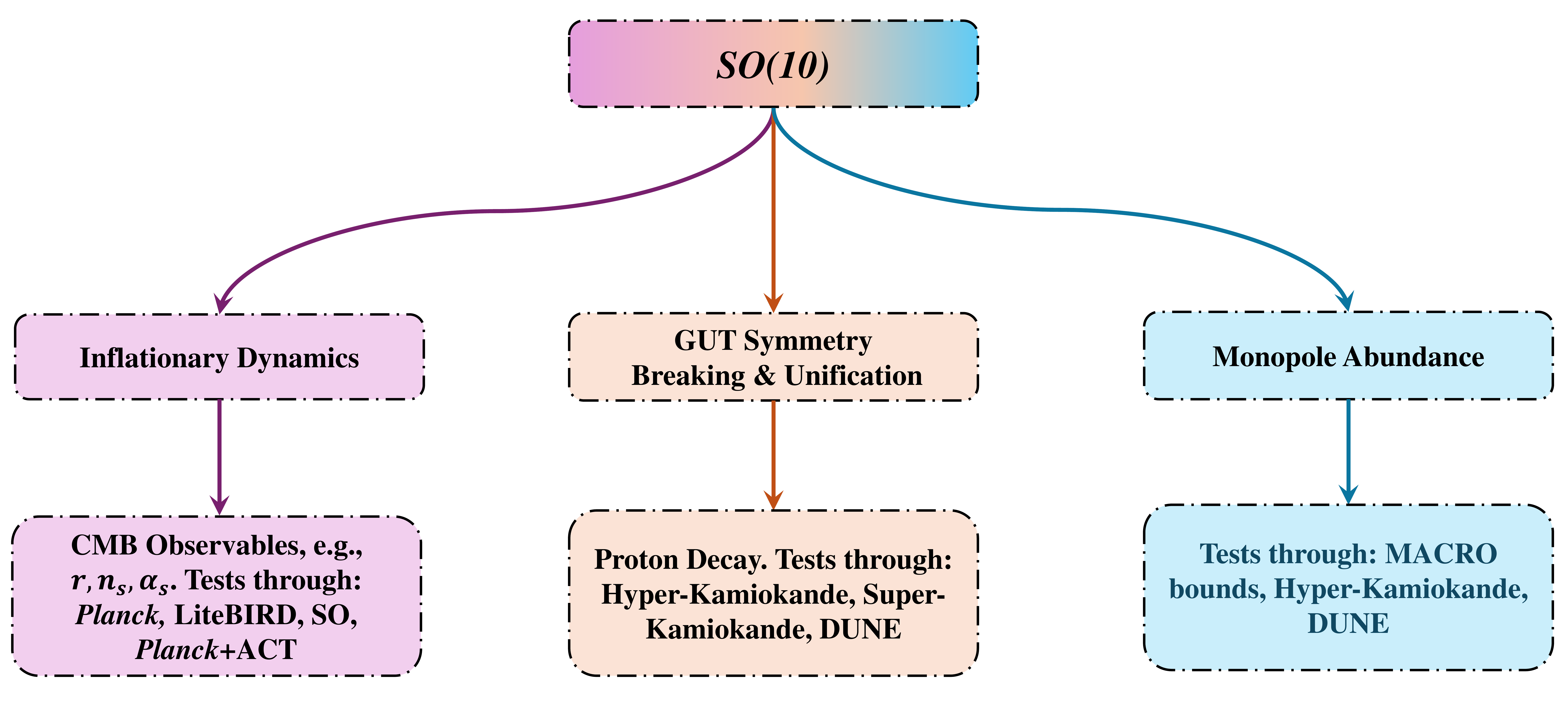}
    \caption{\it Conceptual overview of our unified Palatini $SO(10)$ Higgs--$\mathcal{R}^2$ inflationary framework. The diagram illustrates how a single theoretical framework simultaneously connects and provides complementarity predictions for inflationary dynamics, tested by CMB observations~\cite{BICEP:2021xfz, SPIDER:2017xxz, Planck:2018jri, SimonsObservatory:2018koc, Hazumi_2020, ACT:2025tim}, GUT symmetry breaking, probed by proton decay experiments such as Super-Kamiokande~\cite{Super-Kamiokande:2020wjk} and the upcoming DUNE~\cite{DUNE:2020lwj, DUNE:2020mra, DUNE:2020txw, DUNE:2020ypp} and Hyper-Kamiokande~\cite{Dealtry:2019ldr} experiments; and topological defect formation, constrained by monopole searches, including the MACRO bounds~\cite{MACRO:2002jdv}, as well as future probes such as Hyper-Kamiokande~\cite{Dealtry:2019ldr} and DUNE~\cite{DUNE:2020ypp}.
}
    \label{fig:unified}
\end{figure}

For instance, in the original Higgs inflation model, the non-minimal coupling required to satisfy CMB observables, $\xi \sim \mathcal{O}(10^4)$, can be sufficiently large to induce unitarity-violating interactions~\cite{Bauer:2010jg,McDonald:2020lpz,Mikura:2021clt}. In the metric formulation, the tree-level unitarity cutoff is approximately $\Lambda \sim m_{\rm Pl}/\xi$, which lies below both the inflationary and high-energy reheating scales~\cite{Ito:2021ssc}, making the scenario less appealing. However, within the Palatini formulation of gravity, the cutoff scale is raised to $\Lambda \sim m_{\rm Pl}/\sqrt{\xi}$, ensuring that it remains above the Hubble scale during inflation, $H \sim m_{\rm Pl}/\xi$, in both the vacuum and inflationary regimes. This renders the theory perturbative and reliably predictive up to higher energy scales.

The purpose of this paper is to investigate the impact of inflation on intermediate-scale monopoles in scenarios where the inflaton is a GUT-singlet scalar field ($\phi$) with a Coleman-Weinberg potential~\cite{Coleman:1973jx} and a non-minimal coupling to gravity~\cite{Spokoiny:1984bd,Bezrukov:2007ep,Okada:2010jf,Linde:2011nh,Martin:2013tda,Iso:2014gka,Galante:2014ifa,Campista:2017ovq,Tenkanen:2017jih,Oda:2017zul,Bostan:2018evz,Bostan:2019uvv,Bostan:2019fvk,Bostan:2020pnb,Kubo:2020fdd,Ghoshal:2022qxk,Okada:2022yvq,Shafi:1983bd,Lazarides:1984pq}. This setup is further extended by including the Starobinsky $\mathcal{R}^2$ term, providing a UV-complete framework as discussed in Refs.~\cite{Racioppi:2017spw,Racioppi:2018zoy,Kannike:2018zwn,Racioppi:2019jsp,Gialamas:2020snr,Gialamas:2021rpr,Racioppi:2021ynx,Kannike:2023kzt,Gialamas:2025kef, Gialamas:2025ofz}. In such $SO(10)$ GUT frameworks, the formation of topological defects like monopoles~\cite{Lazarides:1980cc,Rehman:2008qs,Senoguz:2015lba,Lazarides:2019xai,Chakrabortty:2020otp,Lazarides:2021uxv} naturally occurs, and we examine their fate when inflation starts from an unbroken GUT symmetry with sequential intermediate-scale breaking. The intermediate symmetry-breaking scales are determined by the requirement of SM gauge coupling unification at GUT scales, and we analyze how partial inflation affects monopole observability using the MACRO bound~\cite{Ambrosio:2002qq} and related experimental constraints, establishing a natural complementarity between primordial cosmology and particle physics probes.

Our focus is on the \textit{\textbf{interplay between CMB observables and proton decay limits}}, which together impose interlocking constraints on $SO(10)$ models where inflation, driven by a GUT-singlet with a Coleman-Weinberg potential, also determines the survival of intermediate-scale monopoles. \textit{This discussion is summarized schematically in Figure~\ref{fig:unified}.}

\textit{The paper is organized as follows.} In Sec.~\ref{Sec:SO(10)}, we introduce the $SO(10)$ GUT inflation framework, first presenting Coleman-Weinberg inflation with a non-minimal coupling and subsequently extending the analysis to the $\mathcal{R}^2$ model. This section also discusses the Palatini formulation and its implications for scalaron dynamics and inflationary observables. Sec.~\ref{Sec:Monopoles} examines symmetry-breaking patterns and the associated topological defects, focusing on intermediate-scale monopoles that emerge in two representative breaking chains. We determine the intermediate scales from gauge-coupling unification, evaluate the effect of (partial) inflation on monopole abundances, and compare the predictions with current experimental constraints. In Sec.~\ref{Sec:Discussion}, we discuss our results, highlighting the phenomenological implications and the framework's powerful complementarity. We then summarize our main outcomes and outline directions for future work in Sec.~\ref{Sec:Conclusion}. Appendix~\ref{Sec:Slow_roll} collects the generic slow-roll parameters and analytic formulas used in the text. In this paper, we adopt natural units, $c = \hbar = k_B \equiv 1$, and set the reduced Planck mass, $m_{\rm Pl}$, to unity.
\medskip
\section{\(SO(10)\) GUT Inflation}
\label{Sec:SO(10)}
Having established the theoretical advantages of the Palatini formulation, namely its ultraviolet (UV) safety and controlled dynamics in our GUT-extended framework while avoiding the unitarity violations inherent in the metric formulation, we now present our detailed computational framework. In this section, we first outline Coleman–Weinberg inflation with a non-minimal coupling in the Palatini formalism, and then extend the analysis to include the $\mathcal{R}^2$ term, emphasizing its impact on scalaron dynamics and inflationary observables.
\subsection{Coleman-Weinberg inflation in non-minimal coupling}
\label{sec:model}
In the following analysis, we take $\phi$ to be a real scalar field that is a gauge singlet with respect to the SM gauge factors. In explicit $SO(10)$ constructions, such a singlet may either be an elementary $SO(10)$ singlet or correspond to the singlet component of a larger $SO(10)$ multiplet. Phenomenologically, $\phi$ serves two key roles: 
\begin{itemize}
    \item It acts as the inflaton, with its vacuum energy driving accelerated expansion via a Coleman--Weinberg-type potential.
    \item It couples to the GUT-breaking sector, thereby governing the (partial) breaking of the unified symmetry during and after inflation. This coupling directly impacts the formation and dilution of topological defects associated with intermediate symmetry-breaking stages.
\end{itemize}
The action in the Jordan framework\footnote{Throughout the paper, `$J$' refers to the Jordan frame, while `$E$' refers to the Einstein frame.} is given by~\cite{PhysRevD.7.1888,CALLAN197042,PhysRevD.9.1686,FREEDMAN1974354}
\begin{equation}
    S_J =\int\, d^4x\, \sqrt{-g}\,\left( \frac{1}{2}f(\phi) \mathcal{R} - \frac{1}{2}\left(\nabla\phi\right)^2  - V_J(\phi)\right) \ .
\end{equation}
Here, $g$ denotes the determinant of the metric, and $\mathcal{R}$ is the Ricci scalar constructed from $g_{\mu\nu}$. The function $f(\phi)$ encodes a non-minimal coupling between the real GUT-singlet scalar field $\phi$ (the inflaton) and gravity; when $f(\phi)\rightarrow 1$, the scalar field is minimally coupled. The kinetic term is written in compact notation as $(\nabla \phi)^2 \equiv g^{\mu\nu} \partial_\mu \phi\, \partial_\nu \phi$. The Coleman-Weinberg-type radiative potential is then given by
\begin{table}[t!]
\centering
\begin{tabular}{| c | c | c | c | c | c | c | c | c |}
\hline
$\xi$ & $M/m_{\rm Pl}$ & \(A\times10^{-14}\) & $\phi_*/m_{\rm Pl}$ & $\phi_e/m_{\rm Pl}$ & $n_s$ & $r$ & \(-\alpha_s\times10^{-4}\) & $N_*$ \\
\hline
\multirow{3}{*}{\rotatebox[origin=c]{90}{-0.001}}& 50 & \(1.12\) & 37.45 & 48.70 & 0.962 & 0.041 & \( 6.66 \) &50.6\\
& 100 & \(0.41\) & 86.76 & 98.75 & 0.965 & 0.030& \(6.49\)& 50.4 \\
& 500 & \(0.06\) & 486.41 & 499.00 & 0.966 & 0.009 & \(6.10\)& 49.8 \\
\hline
\multirow{3}{*}{\rotatebox[origin=c]{90}{-0.002}}& 50 & \(1.39\) & 37.81 & 48.76 & 0.960 & 0.022 & \(6.30\) &50.3\\
& 100 & \(0.60\) & 87.25 & 98.84 & 0.961 & 0.015 & \(6.17\)&  50.1 \\
& 500 & \(0.11\) & 486.89 & 499.14 & 0.963 & 0.004 & \(6.00\)& 49.5 \\
\hline
\multirow{3}{*}{\rotatebox[origin=c]{90}{-0.003}}& 50 & \(1.53\) & 38.18 & 48.81 & 0.956 & 0.013 & \(6.06\)& 50.1\\
& 100 & \(0.73\) & 87.72 & 98.91 & 0.958 & 0.009 & \(5.99\)& 49.8 \\
& 500 & \(0.13\) & 487.26 & 499.23 & 0.959 & 0.002 & \(5.92\)& 49.4\\
\hline
\multirow{3}{*}{\rotatebox[origin=c]{90}{-0.004}}& 50 & \(1.70\) & 38.66 & 48.85 & 0.951 & 0.009 & \(5.86\)& 50.0\\
& 100 & \(0.80\) & 88.13 & 98.96 & 0.954 & 0.006 & \(5.85\)& 49.8\\
& 500 & \(0.15\) & 487.73 & 499.28 & 0.955 & 0.001 & \(5.84\)& 49.3 \\
\hline
\end{tabular} \\
\caption{\it We compute the inflationary parameters for a real singlet scalar field with a non-minimal coupling to gravity in the Palatini formulation, employing the Coleman-Weinberg potential and assuming a reheating temperature of $T_{\rm reh} = 10^{7}\,\text{GeV}$ and an equation-of-state parameter $\omega_r = 0$.}
\label{tab:infl-para}
\end{table}
\begin{equation}\label{PotentialJORD}
    V_J(\phi) = A \phi^4 \left[\ln \left(\frac{\phi}{M}\right) - \frac{1}{4}\right] + \frac{AM^4}{4} \ .
\end{equation}

Here, $A$ is a (loop-suppressed) coefficient determined by the particle content and couplings of the underlying GUT, while $M$ denotes the renormalization or symmetry-breaking scale--i.e., the scale around which the logarithm is expanded and at which the additive constant $AM^4/4$ sets the vacuum energy.

The non-minimal coupling function is given by
\begin{equation}
    f(\phi) = 1 + \xi (\phi^2 - M^2) .
\end{equation}
The dimensionless non-minimal coupling $\xi$ governs the strength of the interaction between both \(\phi\) and \(\mathcal{R}\), and the shift by $-M^2$ ensures that $f(\phi) \to 1$ in the vacuum at $\phi = M$. To ensure the physical consistency of the inflationary dynamics, we impose the condition
\begin{equation}\label{eq:non_minimal}
f(\phi) = 1 + \xi (\phi^2 - M^2) > 0\,,
\end{equation}
throughout the evolution of the inflaton field \(\phi\). This condition is essential in both the Jordan and Einstein frames, since a vanishing or negative \( f(\phi) \) would render the Weyl transformation ill-defined and lead to pathological gravitational interactions, thereby disrupting the inflationary regime. Consequently, in our analysis, we restrict the non-minimal coupling parameter \(\xi\), whether positive or negative, to values that guarantee \( f(\phi) > 0 \) for all relevant field values during and after inflation, ensuring a stable and physically consistent cosmological evolution.

To express the action in the Einstein framework, we apply a Weyl transformation\footnote{A \textit{tilde } denotes quantities evaluated in the Einstein frame.}\cite{Fujii_Maeda_2003}
\begin{equation}\label{eq:Weyl}
g_{\mu\nu} \to \tilde{g}^{\mu\nu} = f(\phi) g_{\mu\nu}.
\end{equation}
Consequently, the Einstein-frame Lagrangian simplifies to the following form
\begin{equation}
S_E \  = \int\, \sqrt{-\tilde{g}}\,\left(\frac{1}{2} \tilde{\mathcal{R}} - \frac{1}{2} \frac{1}{f(\phi)} \left(\tilde{\nabla}\phi\right)^2 - V_E(\phi)\right)\,.
\end{equation}
Since the kinetic term is non-canonical, a field redefinition is required. The canonical scalar field 
\( \zeta \) is then defined as
\begin{equation}\label{eq:Canonical}
\left(\frac{d\zeta}{d\phi}\right)^2 = \frac{1}{f(\phi)} = \frac{1}{1 + \xi(\phi^2 - M^2)},
\end{equation}
and the field-space metric can be expressed as
\begin{equation}
    Z(\phi) = 1+\xi\left(\phi^2-M^2\right).
\end{equation}
We consider the Palatini  formulation, in which the metric \(g_{\mu\nu}\) and the affine connection \(\Gamma^\rho{}_{\mu\nu}\) are treated as independent variables and varied separately. Varying the action with respect to \(\Gamma^\rho{}_{\mu\nu}\) yields
\begin{figure}[t!]
    \centering
    \includegraphics[width=0.75\linewidth]{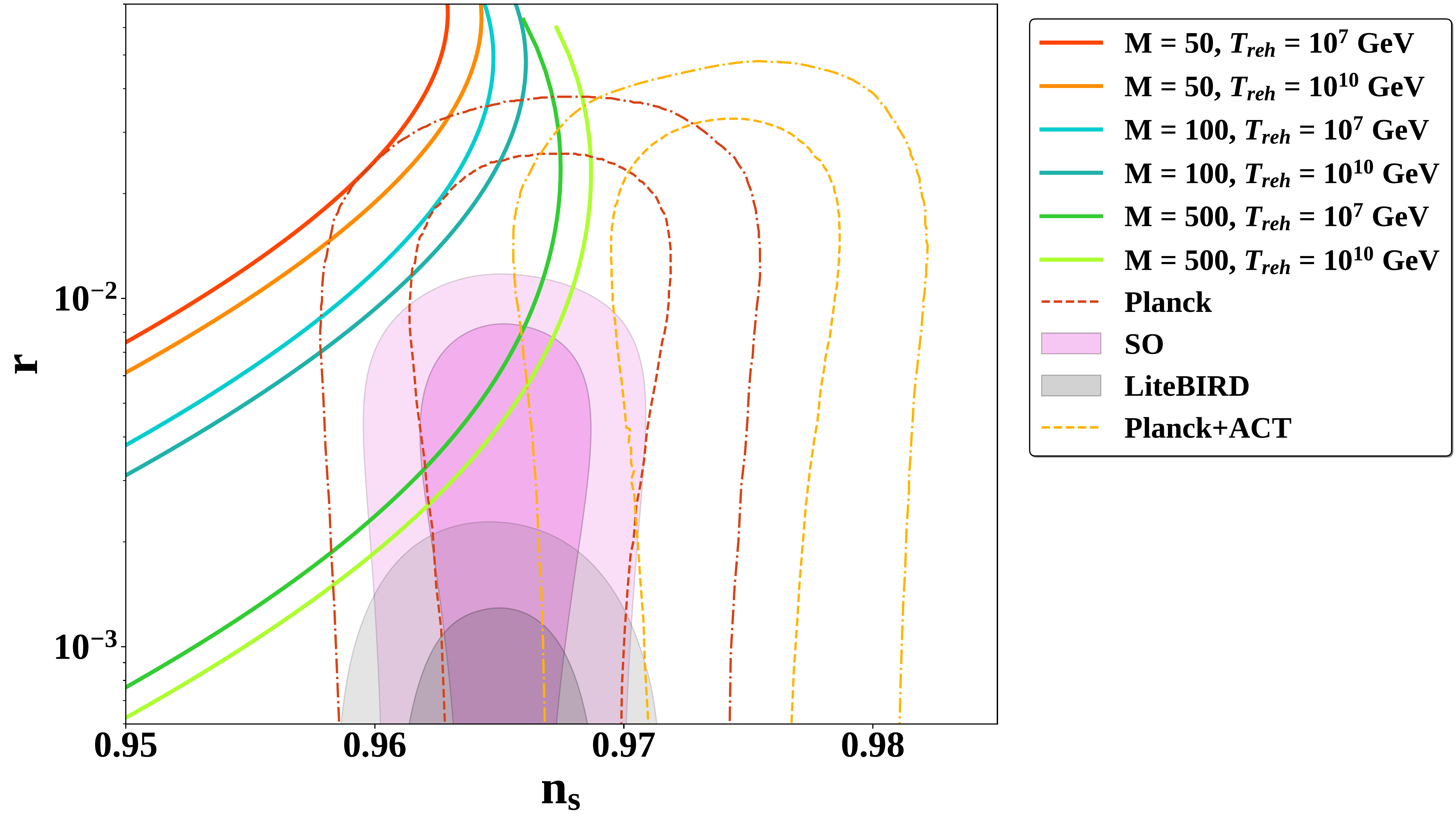}   
     \includegraphics[width=0.75\linewidth]{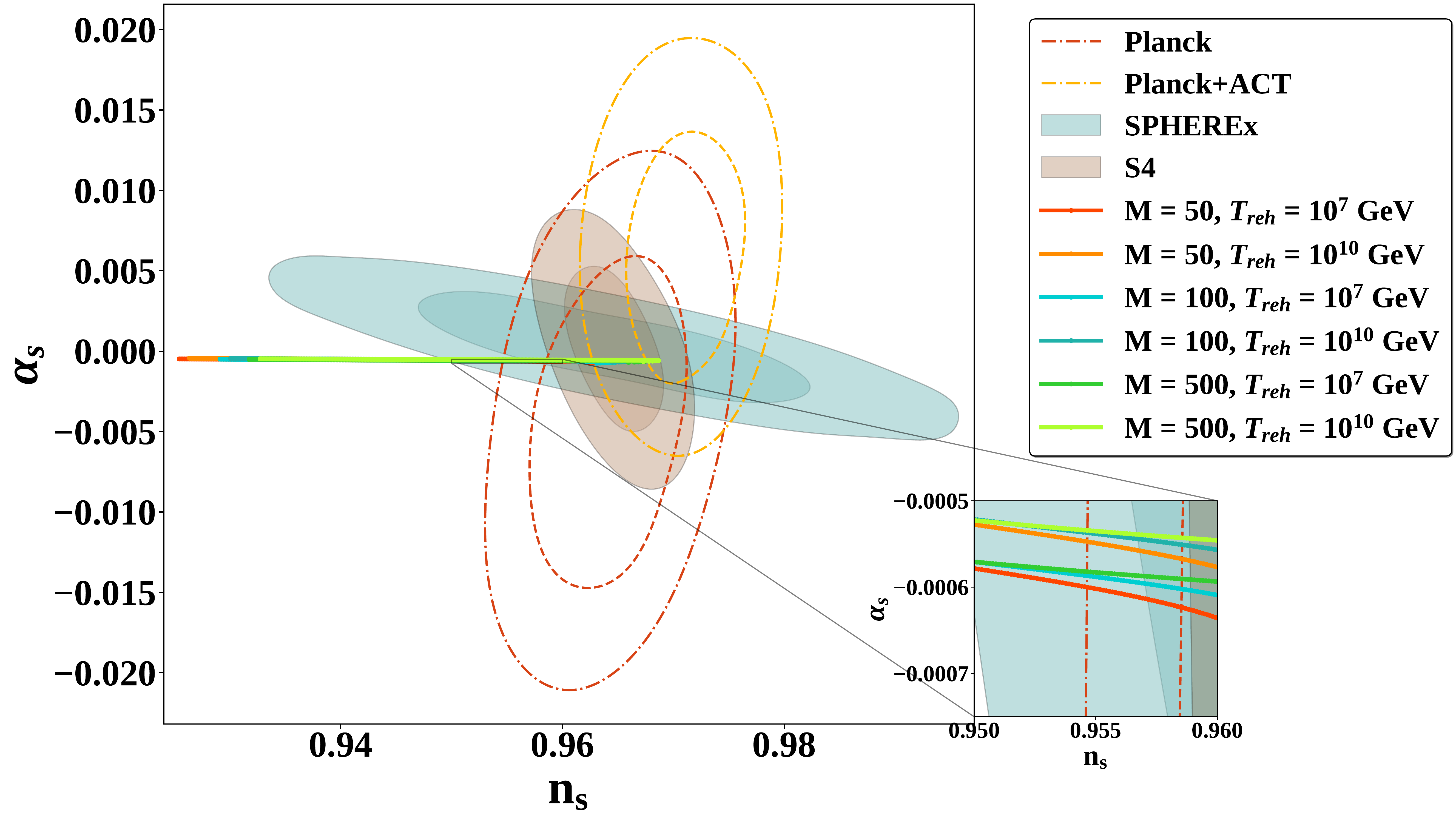}
    \caption{\it Inflationary predictions for the spectral index $n_s$ and the tensor-to-scalar ratio $r$ are shown for different values of $M$ (in $m_{\rm Pl}$ units) and varying reheating temperatures (\textbf{top}), while the running of the spectral index $\alpha_s$ as a function of $n_s$ is shown in the \textbf{bottom} panel.
    \label{fig:rns1}
    }
\end{figure}
\begin{figure}[t!]
    \centering
    \includegraphics[width=1\linewidth]{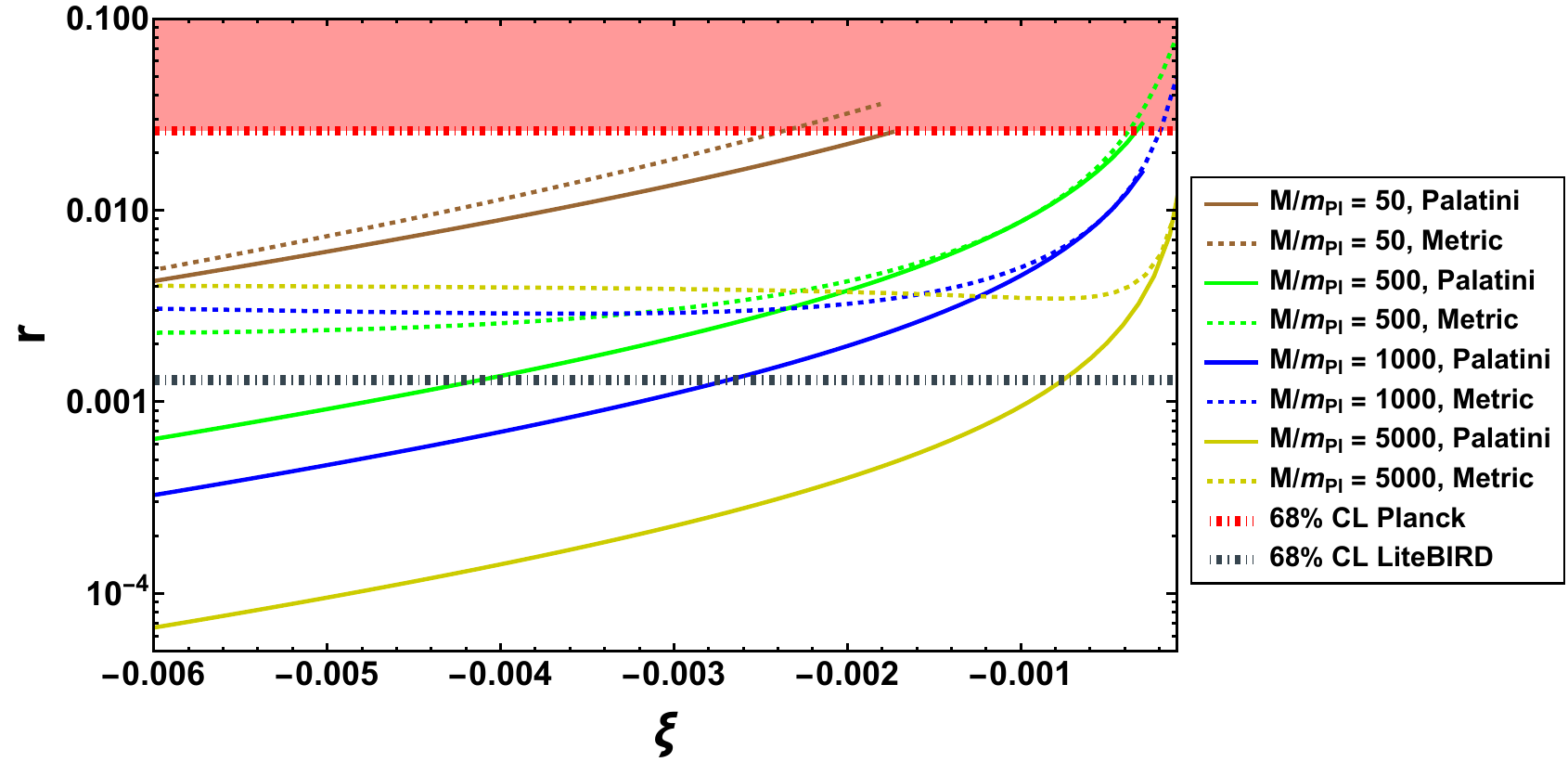}
    \caption{\it The tensor-to-scalar ratio $r$ is shown for different choices of the parameter $M$ as $\xi$ varies. Results are presented for both the metric and Palatini formulations. The red shaded region indicates the parameter space excluded by \emph{Planck} observational constraints~\cite{Planck:2018jri}, while the grey dot-dashed line shows the projected constraints from LiteBIRD~\cite{Hazumi_2020}.
    }
    \label{fig:rxi1}
\end{figure}
\begin{equation}
\nabla^\Gamma_\lambda\!\big(\sqrt{-g}\,f(\phi)\,g^{\mu\nu}\big)=0,
\end{equation}
and the Einstein frame potential is then given by
\begin{equation}\label{eq:potential_palatini_E}
V_E(\phi) = \frac{A \phi^4 \left[\ln \Big(\frac{\phi}{M}\right) - \frac{1}{4}\Big] + \frac{AM^4}{4}}{\Big[ 1 + \xi \big(\phi^2 - M^2\big) \Big]^2} \ .
\end{equation}

We present a concise slow-roll analysis of the Einstein-frame dynamics, with the full definitions of the slow-roll parameters provided in Appendix~\ref{Sec:Slow_roll}. Working consistently within the Palatini formulation and expanding to leading order in \(1/N_*\), where \(N_*\) denotes the number of e-folds between the horizon exit of the CMB pivot scale and the end of inflation, we analytically obtain the following approximate expressions for the inflationary observables
\begin{equation}\label{eq:Palatini_approx}
\begin{split}
r &\simeq \frac{8}{N_*^2},\\
  n_s &\simeq 1 - \frac{2}{N_*}, \\
  N_* = \vartheta
\int_{\phi_e}^{\phi_*} \frac{1-M^2\xi+\xi\phi^2}{\sqrt{\,M^4-\phi^4+4\phi^4\ln(\phi/M)\,}}\,d\phi &\,, \qquad \text{where} \,\quad\vartheta = 4\pi\,\sqrt{\frac{3\times2.1\times10^{-9}}{A}}.
 \end{split}
\end{equation}
We also provide analytic approximations for the inflaton field values at horizon crossing ($\phi_*$) and at the end of inflation ($\phi_e$). Here, \(\phi_*\) denotes the value of the field \(\phi\) at horizon crossing of the CMB pivot scale (i.e., when Eq.~\eqref{perturb1} \(\sim 2.1 \times 10^{-9}\)~\cite{Planck:2018jri}), while \(\phi_e\) corresponds to the field value at which slow-roll ends, defined by \(\epsilon(\phi_e) = 1\), with \(\epsilon\) given in Eq.~\eqref{slowroll1}. Working to leading order in the small displacement of \(\phi\) from the symmetry-breaking scale \(M\), these expressions are obtained by solving the slow-roll relations and imposing the observed amplitude of the curvature perturbations.

The results read
\begin{equation}
\begin{split}
\phi_* &= M - \frac{\varrho}{\sqrt{M\sqrt{A}}}, \qquad {\rm where} \quad \varrho =\left(2.1\times   10^{-9}\times24\pi^2\right)^{1/4}\,, \\
    \phi_e &= M \left(1-\frac{\sqrt{25+2M^2}-5}{M^2}\right)\,.
    \end{split}
\end{equation}
In Fig.~\ref{fig:rns1}, we show the $r$--$n_s$ plane for the potential of Eq.~\eqref{eq:potential_palatini_E} evaluated within the Palatini formulation. This produces a characteristic correlation between the scalar spectral index $n_s$ and the tensor-to-scalar ratio $r$, which depends sensitively on the radiative mass scale\footnote{We identify the renormalization group equation (RGE) scale \(\mu\) with the symmetry-breaking scale \(M\), as it represents the characteristic mass scale at which the radiative corrections that shape the potential are evaluated.} $M$ and the reheating temperature $T_{\rm reh}$. Different colors correspond to distinct mass scales and reheating temperature choices. Increasing $M$ at a fixed non-minimal coupling $\xi$ flattens the Einstein-frame potential, thereby reducing $r$ from $\sim 0.041$ at $M = 50\,m_{\rm Pl}$ down to $\sim 10^{-3}$ for $M = 500\,m_{\rm Pl}$. The black markers on the $M = 50\,m_{\rm Pl}$ trajectory in Fig.~\ref{fig:rns1} denote discrete $\xi$ values yielding $N_* \sim 50$ e-folds of slow-roll inflation. A higher reheating temperature decreases the post-inflationary expansion required to match the CMB pivot scale, shifting each curve slightly toward lower $r$ at fixed $n_s$. Overlaid 68\% (dotted) and 95\% (dashed) confidence contours from \emph{Planck}+BICEP/Keck+BAO~\cite{BICEP:2021xfz}, the Simons Observatory~\cite{SimonsObservatory:2018koc}, and LiteBIRD/\emph{Planck}~\cite{Hazumi_2020} indicate that, for $M \gtrsim 100\,m_{\rm Pl}$ and $T_{\rm reh} \gtrsim 10^{7}$~GeV, the model predictions lie comfortably within the observationally favored region. This confirms the viability of radiatively driven Palatini inflation in reproducing both the observed $n_s$ and the current upper limits on primordial tensor modes. Selected benchmark values for these numerical results are presented in Tab.~\ref{tab:infl-para} and approximated in Eq.~\eqref{eq:Palatini_approx}.

We also compute the running of the scalar spectral index, $\alpha_s \equiv \mathrm{d}n_s/\mathrm{d}\ln k$, and report these values in Tab.~\ref{tab:infl-para}. For the benchmarks shown in the table, we find a small negative running,
\(
\alpha_s \simeq -(5.8\text{--}6.7)\times10^{-4}.
\)
The running exhibits only a weak dependence on the radiative mass scale $M$ and on $\xi$, tending to decrease slightly in magnitude for larger $M$ and for larger $|\xi|$. Overall, $\alpha_s$ remains comfortably within current and projected observational constraints~\cite{SPHEREx:2014bgr, Planck:2018jri, Abazajian:2019eic, ACT:2025tim}.

The family of curves in Fig.~\ref{fig:rns1} illustrates how two physical ingredients, the shape of the Einstein-frame potential (mainly determined by the radiative mass scale $M$) and the post-inflationary expansion history (set by the reheating temperature $T_{\rm reh}$), map onto the observable pair $(r, n_s)$. Increasing $M$ at fixed non-minimal coupling flattens the potential around the inflationary plateau, reducing the slow-roll parameter $\epsilon \propto (V'/V)^2$ (see Eq.~\eqref{slowroll1}) and thereby strongly suppressing $r$. Changes in $T_{\rm reh}$ alter the number of e-folds $N_*$ between horizon crossing and the end of inflation; higher reheating temperatures typically imply larger $N_*$, producing a further reduction of $r$ and a slight shift of $n_s$ toward unity. Consequently, low-$M$ trajectories predict comparatively larger (and in some cases observable) tensor signals, while large-$M$ trajectories reside in a low-$r$ regime comfortably consistent with current bounds. Hence, improved upper limits on $r$ will primarily constrain the lower-$M$ portion of the parameter space.

Figure~\ref{fig:rxi1} shows the tensor-to-scalar ratio $r$ as a function of the non-minimal coupling parameter $\xi$ for four representative mass scales, $M = \{50,\,500,\,1000,\,5000\}\,m_{\rm Pl}$. Solid curves correspond to the Palatini formulation, while dashed curves represent the metric formulation. As $\xi$ approaches zero, all trajectories converge toward the minimally coupled limit; however, the Palatini predictions systematically yield lower values of $r$ than their metric counterparts at small $\xi$. Increasing $M$ further suppresses the tensor modes, reflecting the reduced amplitude of primordial gravitational waves in higher-scale inflationary models. The distinct separation between Palatini and metric predictions underscores the significance of the chosen gravitational framework, with the Palatini formulation providing a natural mechanism to achieve lower $r$, in better agreement with current observational upper limits, as illustrated in Fig.~\ref{fig:rns1} and Tab.~\ref{tab:infl-para}.

These differences arise from the distinct Einstein-frame dynamics in the metric and Palatini formalisms. The effective potential and kinetic normalization that enter the slow-roll parameters differ between the two formulations, making the first slow-roll parameter, $\epsilon$ (and hence $r \simeq 16\epsilon$), generically smaller in the Palatini realization for the same Jordan-frame input. Increasing the symmetry-breaking scale $M$ further suppresses the tensor amplitude because it lowers the overall inflationary energy scale (and the Hubble rate $H$), producing a smaller gravitational-wave signal. These features highlight the phenomenological importance of the gravitational formulation: Palatini implementations more readily yield values of $r$ compatible with current upper limits, while also modifying the mapping between model parameters, $n_s$, reheating history, and monopole dilution, as discussed in Sec. ~\ref{Sec:Monopoles}.
\subsection{Palatini \(\mathcal{R}^2\) case}
\label{sec:R2}
Building on the $SO(10)$ setup described above, we extend the action in Eq.~\eqref{ActionJORD} by including the $\mathcal{R}^2$ term, consistently working in the Palatini formulation (as emphasized in Sec.~\ref{sec:model}). In our construction, the inflaton remains the GUT-singlet scalar $\phi$, whose Coleman-Weinberg potential $V_J(\phi)$ is generated by radiative effects of the GUT field content and whose vacuum expectation value sets the intermediate symmetry-breaking scale; the potential is explicitly given in Eq.~\eqref{PotentialJORD}. The $\mathcal{R}^2$ contribution is most conveniently handled in the Palatini formulation by introducing an auxiliary scalar $\chi$ that parametrizes the scalaron sector. The dimensionless coefficient $\alpha$ controls the strength of this sector, and thus the extent to which the scalaron modifies the effective gravitational coupling during inflation. Working in the Palatini formulation keeps the geometric extension minimal and algebraically tractable. In this approach, the $\chi$ field encodes the $\mathcal{R}^2$ effects, and it provides a transparent description of how the scalaron and the non-minimally coupled inflaton jointly reshape the Einstein-frame potential and the resulting inflationary dynamics. The resulting action is then given by

\begin{equation}
S_J\,=\,\int\,d^4x\,\sqrt{-g}\,\left(\frac{1}{2}\left(\xi\phi^2+\alpha\chi^2\right)\mathcal{R}\,-\frac{1}{2}\left(\nabla\phi\right)^2\,-\frac{\alpha}{4}\chi^4\, - V_J(\phi)\right){\label{ActionJORD}}\ .
\end{equation}
We then perform a Weyl rescaling of the metric to express the action in the Einstein frame
\begin{equation}\label{Transformation}
\tilde{g}_{\mu\nu}\,=\,\Omega^2(\phi)\,g_{\mu\nu}\,,\qquad{\text{and}}\qquad\Omega^2(\phi)\,=\,\xi\phi^2+\alpha\chi^2\,.
\end{equation}
The Weyl rescaling transforms the metric to the Einstein frame, rendering the gravitational term canonical and thereby simplifying the analysis of inflationary dynamics
\begin{equation} 
S_E\,=\,\int\,\mathrm{d}^4x\sqrt{-\tilde{g}}\left( \frac{1}{2}\,\tilde{\mathcal{R}}\,-\frac{1}{2}\frac{\left(\tilde{\nabla}\phi\right)^2}{\Omega^2}\,-V_E(\phi, \chi)\right) ,{\label{ActionEINS}}
\end{equation}
with
\begin{equation}\label{Pot2Parameters}
V_E(\phi,\chi)\,=\,\frac{1}{\Omega^4}\left(V_J(\phi)+\frac{\alpha}{4}\chi^4\right)\,.
\end{equation}
The Einstein-frame action includes a modified kinetic term and potential, with $\Omega^2$ rescaling the fields to yield a canonical Einstein--Hilbert term. Here, $\Omega^2(\phi)$ is defined in Eq.~\eqref{Transformation}, and the auxiliary field is given by the following expression
\begin{equation}
\chi^2\,=\,\frac{4V_J(\phi)\,+\,\xi\phi^2(\nabla\phi)^2}{\left(\xi\phi^2-\alpha(\nabla\phi)^2\right)}\,.{\label{Auxiliary}}
\end{equation}
The auxiliary field $\chi$ is obtained by minimizing the action with respect to $\chi$, thereby eliminating it as an independent degree of freedom. Using the expressions defined in Eqs.~\eqref{Transformation}, \eqref{Pot2Parameters}, and \eqref{Auxiliary}, the action can then be written in the Einstein frame as\footnote{Higher-order derivative terms are neglected here, which is a valid approximation during slow-roll inflation, where the field variations are small.
}
\begin{equation}
S_E\,\approx\,\int\,d^4x\,\sqrt{-\tilde{g}}\left(\,\frac{1}{2}\tilde{R}\,-\frac{1}{2}\frac{(\tilde{\nabla}\phi)^2}{Z_\text{P}(\phi)}\,- \frac{V_J(\phi)}{\xi^2\phi^4+4{\alpha} V_J(\phi)}\,+\,\mathcal{O}((\tilde{\nabla}\phi)^4)\,\right)\ .{\label{ActionEINS-2}}
\end{equation}
The field-space metric \(Z_\text{P}(\phi)\) in the Palatini formalism is defined as
\begin{equation}\label{FieldSPACE}
    Z_\text{P}(\phi) \, = \frac{\xi^2\phi^4\,+\,4\alpha V(\phi)}{\xi\phi^2}\,.
\end{equation}
The field-space metric \(Z_\text{P}(\phi)\) arises from the non-minimal coupling and modifies the kinetic term of the scalar field. We then introduce the canonical scalar field \(\zeta\) as
\begin{equation}\label{eq:Canoical_R2}
d\zeta \, = \, \frac{d\phi}{\sqrt{Z_\text{P}(\phi)}}= \, d\phi\,\sqrt{\frac{\xi\phi^2}{\xi^2\phi^4\, + \, 4\alpha V(\phi)}}\,  \,.
\end{equation}  
Defining the canonical field $\zeta$ normalizes the kinetic term to its standard form, thereby facilitating the application of the slow-roll formalism. From the action in Eq.~\eqref{ActionEINS-2}, it can be concluded that the Einstein-frame potential takes the form
\begin{equation}
  V_E(\phi) = \frac{V_J(\phi)}{\xi^2 \phi^4 + 4 \alpha V_J(\phi)} 
= \frac{A \, \phi^4 \left[\ln \left(\frac{\phi}{M}\right) - \frac{1}{4}\right] + \frac{A M^4}{4}}{\xi^2 \phi^4 + 4 A \alpha \, \phi^4 \left[\ln \left(\frac{\phi}{M}\right) - \frac{1}{4}\right] + A \alpha M^4} \,.
\end{equation}
This effective potential, $V_E(\phi)$, drives inflation in the Einstein frame, incorporating the logarithmic structure of the Coleman-Weinberg potential. Following the slow-roll parameter definitions in Appendix~\ref{Sec:Slow_roll} and working consistently within Palatini gravity, we present the leading-order analytic approximation for the tensor-to-scalar ratio as follows

\begin{equation}\label{eq:tensor_ratio_R2_Palatini}
   r_*  =  \frac{{\rm G}}{\alpha+\,\bar{G}\left(1-\frac{M^4}{\phi_*^4}\right)^{-2/3}}, \quad {\rm where} \quad G = \frac{128}{\Delta^2_{\mathcal{R}}\times(16\sqrt{3}\pi)^2}, \quad {\rm and }\quad \bar{G} = \xi\,\left(\frac{G}{128 A^2}\right)^{1/3}\,.
\end{equation}
\begin{table}[t!]
\centering
\begin{tabular}{| c | c | c | c | c | c | c | c | c | c | c |}
\hline
$\xi$ & $\alpha$ & $M/m_{\rm Pl}$ & $T_{\rm reh}$ (GeV) & $A\times10^{-15}$ & $\phi_*/m_{\rm Pl}$ & $\phi_e/m_{\rm Pl}$ & $n_s$ & $r$ & \(-\alpha_s \times10^{-4}\)& $N$ \\
\hline
\multirow{8}{*}{\rotatebox[origin=c]{90}{0.001}}
  & \multirow{4}{*}{$10^5$}
    & \multirow{2}{*}{50}
      & $10^7$    & 6.76 & 29.82  & 47.75 & 0.955 & 0.260    &  \(7.65\) &51.8 \\
  & & 
      & $10^{10}$ & 6.03 & 29.40  & 47.75 & 0.956 & 0.252    & \(7.36\)& 54.1 \\
\cline{3-11}
  & & \multirow{2}{*}{250}
      & $10^7$    & 6.76 & 149.36 & 238.73 & 0.954 & 0.261    & \(7.64\)& 51.8 \\
  & & 
      & $10^{10}$ & 6.03 & 147.01 & 238.73 & 0.956 & 0.252    & \(7.34\)& 54.1 \\
\cline{2-11}
  & \multirow{4}{*}{$10^{10}$}
    & \multirow{2}{*}{50}
      & $10^7$    & 7.59 & 30.24  & 48.30 & 0.953 & $8.02\times10^{-4}$ & \(7.71\) & 49.3 \\
  & & 
      & $10^{10}$ & 6.92 & 29.93  & 48.28 & 0.954 & $8.02\times10^{-4}$ & \(7.41\)& 51.6 \\
\cline{3-11}
  & & \multirow{2}{*}{250}
      & $10^7$    & 7.59 & 151.18 & 241.50 & 0.953 & $8.02\times10^{-4}$ & \(7.71\)& 49.3 \\
  & & 
      & $10^{10}$ & 6.92 & 149.63 & 241.40 & 0.954 & $8.02\times10^{-4}$ & \(7.41\)& 51.6 \\
\hline
\multirow{8}{*}{\rotatebox[origin=c]{90}{0.0001}}
  & \multirow{4}{*}{$10^5$}
    & \multirow{2}{*}{50}
      & $10^7$    & 1.00 & 43.05  & 49.29 & 0.960 & 0.176    & \(7.58\)& 51.5 \\
  & & 
      & $10^{10}$ & 0.93 & 42.97  & 49.29 & 0.961 & 0.172    & \(7.30\)& 53.8 \\
\cline{3-11}
  & & \multirow{2}{*}{250}
      & $10^7$    & 1.02 & 215.48 & 246.46 & 0.960 & 0.178    & \(7.58\)& 51.5 \\
  & & 
      & $10^{10}$ & 0.92 & 214.65 & 246.46 & 0.962 & 0.170    & \(7.29\)& 53.8 \\
\cline{2-11}
  & \multirow{4}{*}{$10^{10}$}
    & \multirow{2}{*}{50}
      & $10^7$    & 1.10 & 43.22  & 49.49 & 0.958 & $8.01\times10^{-4}$ & \(7.65\) & 49.2 \\
  & & 
      & $10^{10}$ & 1.00 & 43.07  & 49.48 & 0.960 & $8.00\times10^{-4}$ & \(7.36\)& 51.5 \\
\cline{3-11}
  & & \multirow{2}{*}{250}
      & $10^7$    & 1.10 & 216.08 & 247.45 & 0.958 & $8.01\times10^{-4}$ & \(7.65\)& 49.2 \\
  & & 
      & $10^{10}$ & 1.00 & 215.33 & 247.42 & 0.960 & $8.00\times10^{-4}$ & \(7.36\)& 51.5 \\
\hline
\multirow{8}{*}{\rotatebox[origin=c]{90}{$10^{-5}$}}
  & \multirow{4}{*}{$10^5$}
    & \multirow{2}{*}{50}
      & $10^7$    & 0.11 & 47.77  & 49.78 & 0.961 & 0.160    & \(7.55\)& 51.4 \\
  & & 
      & $10^{10}$ & 0.10 & 47.70  & 49.78 & 0.963 & 0.153    & \(7.23\)& 53.8 \\
\cline{3-11}
  & & \multirow{2}{*}{250}
      & $10^7$    & 0.11 & 238.73 & 248.88 & 0.961 & 0.160    & \(7.57\)& 51.4 \\
  & & 
      & $10^{10}$ & 0.10 & 238.55 & 248.88 & 0.962 & 0.155    & \(7.23\)& 53.8 \\
\cline{2-11}
  & \multirow{4}{*}{$10^{10}$}
    & \multirow{2}{*}{50}
      & $10^7$    & 0.12 & 47.81  & 49.84 & 0.959 & $8.00\times10^{-4}$ & \(7.68\)& 49.2 \\
  & & 
      & $10^{10}$ & 0.11 & 47.76  & 49.84 & 0.961 & $8.00\times10^{-4}$ & \(7.37\)& 51.5 \\
\cline{3-11}
  & & \multirow{2}{*}{250}
      & $10^7$    & 0.12 & 239.04 & 249.21 & 0.959 & $8.00\times10^{-4}$ & \(7.68\)& 49.2 \\
  & & 
      & $10^{10}$ & 0.11 & 238.81 & 249.20 & 0.961 & $8.00\times10^{-4}$ & \(7.37\)& 51.5 \\
\hline
\end{tabular}
\caption{\it Inflationary observables for the Coleman-Weinberg potential in the $\mathcal{R}^2$ Palatini formalism, assuming an equation-of-state parameter $\omega_r = 0$.
}
\label{tab:infl-para-all-merged_Palatini_R2_0}
\end{table}
Equation~\eqref{eq:tensor_ratio_R2_Palatini} explicitly demonstrates that the tensor-to-scalar ratio in Palatini \(\mathcal{R}^2\) inflation is suppressed by the factor \(1/\alpha\), while its dependence on the non-minimal coupling \(\xi\) appears only through subleading terms. This analytic structure anticipates the numerical results: for small \(\alpha\), one recovers \(r \sim \mathcal{O}(10^{-1})\), characteristic of minimally coupled quartic models, whereas increasing \(\alpha\) rapidly drives \(r\) toward the universal attractor value \(r \simeq 8 \times 10^{-4}\), independently of \(\xi\) and \(M\). The near-independence of \(n_s\) on \(\alpha\) follows from the same expression, as shown in Eq.~\eqref{eq:spectral_R2_Palatini}, because the slow-roll parameter \(\epsilon\) is suppressed while \(\eta\) remains determined by the shape of the effective potential. In this way, the analytic formula captures the qualitative behavior observed in Tab.~\ref{tab:infl-para-all-merged_Palatini_R2_0}, highlighting that \(\alpha\) is the dominant parameter controlling the tensor amplitude within this framework.

The value of \(\phi_*\) at the time when the pivot scale crosses the horizon is related to the model parameters via the following approximate expression
\begin{equation}
    \phi_* = M\left\{1-\left(\frac{4\,\xi^3}{G\,A}\right)^{1/4}\right\}.
\end{equation}
This equation constrains \(\phi_*\) by requiring that the amplitude of scalar perturbations matches CMB observations, thereby connecting the model parameters to cosmological data~\cite{BICEP:2021xfz, SPIDER:2017xxz, Planck:2018jri, SimonsObservatory:2018koc, Hazumi_2020, ACT:2025tim}. The field value \(\phi_e\) corresponds to the end of inflation, marking the onset of the radiation-dominated era, and is given by the following analytical approximation
 \begin{equation}
    \phi_e  \;\approx\; 
M \,\left\{\,1 \;-\;
\frac{\sqrt{\sqrt{\xi^4 + 64\,A\alpha\,\xi^3}-\xi^2 }}
     {4\,\sqrt{A\alpha}}
\right\}\,.
\end{equation}
Following the discussion in Appendix~\ref{Sec:Slow_roll}, the scalar spectral index can be written as
\begin{equation}\label{eq:spectral_R2_Palatini}
    1\,-n_s = 16\xi\, \frac{\left(1+3x^8-4x^4-8x^4\,{\rm In}(x)\right)}{\left(1-x^4+4x^4{\rm In}(x)\right)^2}, \qquad {\rm where} \qquad x =\frac{\phi_*}{M}.
\end{equation}
The number of $e$-folds can be expressed in the following form
\begin{equation}
\begin{split}
    N_* &= \int_{\phi_e}^{\phi_*}   \frac{\phi^4 -M^4 -   4 \phi^4 \ln\left(\frac{\phi}{M}\right)}{4 M^4 \xi \phi - 4 \xi \phi^5} \, \\&= \,  \left[ 
 \frac{1}{4\xi} \ln\left(\frac{\phi}{M}\right) \ln\left(1 - \frac{\phi^4}{M^4}\right) 
 -\frac{1}{4\xi} \ln\left(\frac{\phi}{M}\right) 
+ \frac{1}{16\xi} \text{Li}_2\left(\frac{\phi^4}{M^4}\right) 
\right]_{\phi_e}^{\phi_*}\,,
\end{split}
\end{equation}
which can be approximated\footnote{here \({\rm Li_2}(z)\) is the dilogarithm function
\[ \text{Li}_2(z) = -\int_0^z \frac{\ln(1 - t)}{t} \, dt,\] and it is given in series form as
\[\text{Li}_2(z) = \sum_{n=1}^{\infty} \frac{z^n}{n^2}, \quad \text{for } |z| \leq 1.\]} as
\begin{equation}
    N_* \simeq  \,  \frac{1}{4\xi}\left[ 
 \ln\left(\frac{\phi}{M}\right) \ln\left(1 - \frac{\phi^4}{M^4}\right) 
 - \ln\left(\frac{\phi}{M}\right) 
+ \frac{1}{4} \left(\frac{\phi}{M}\right)^4
\right]_{\phi_e}^{\phi_*}\,.
\end{equation}
The number of $e$-folds, $N_*$, quantifies the total expansion of the scale factor between the field values $\phi_*$ and $\phi_e$. This quantity is essential for ensuring that inflation persists long enough to resolve the horizon and flatness problems. In standard cosmological scenarios, addressing these issues typically requires $50 \lesssim N_* \lesssim 60$, although the precise value depends on the post-inflationary thermal history.

\begin{figure}[t!]
    \centering
    \includegraphics[width=0.85\linewidth]{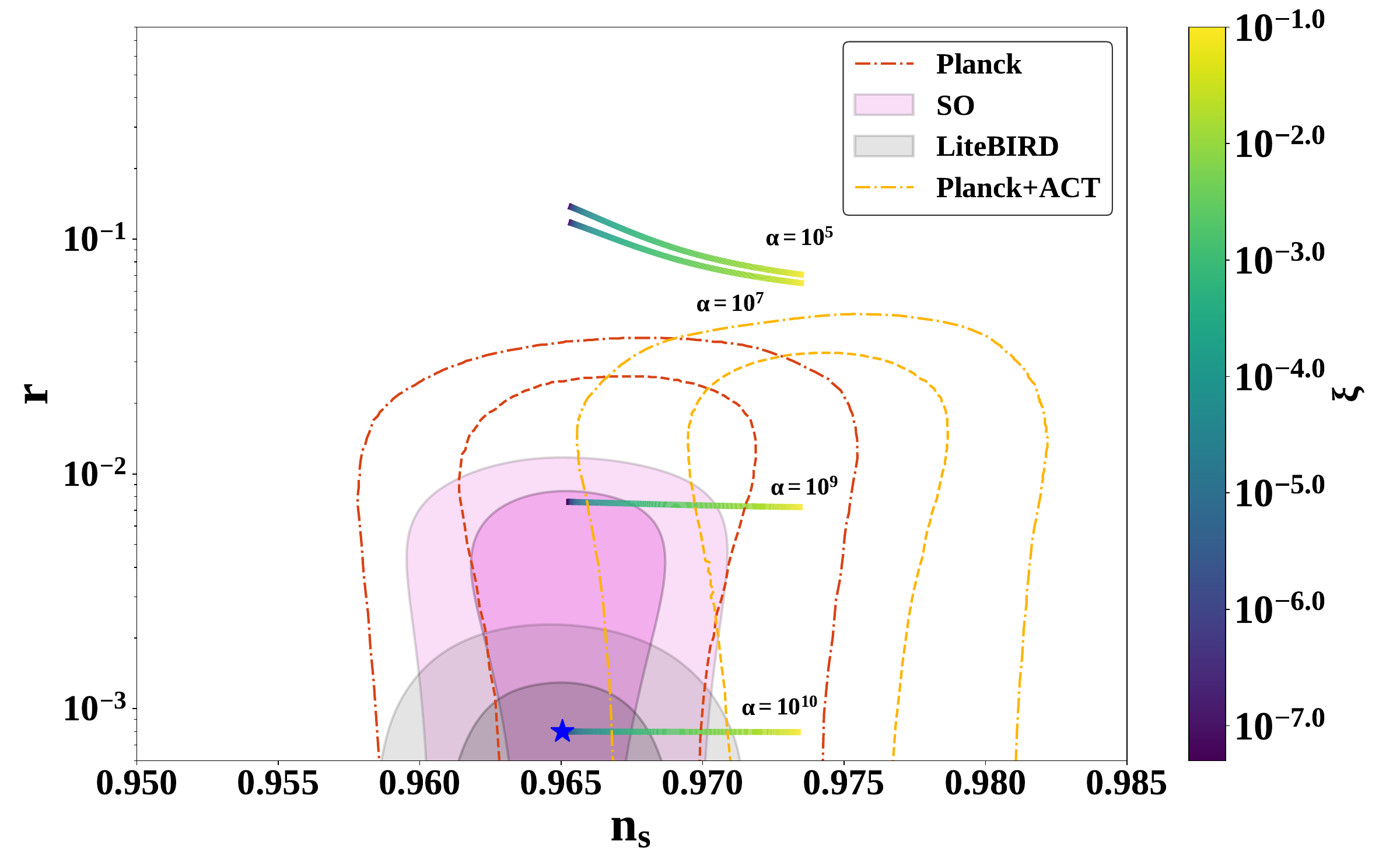}
     \includegraphics[width=0.85\linewidth]{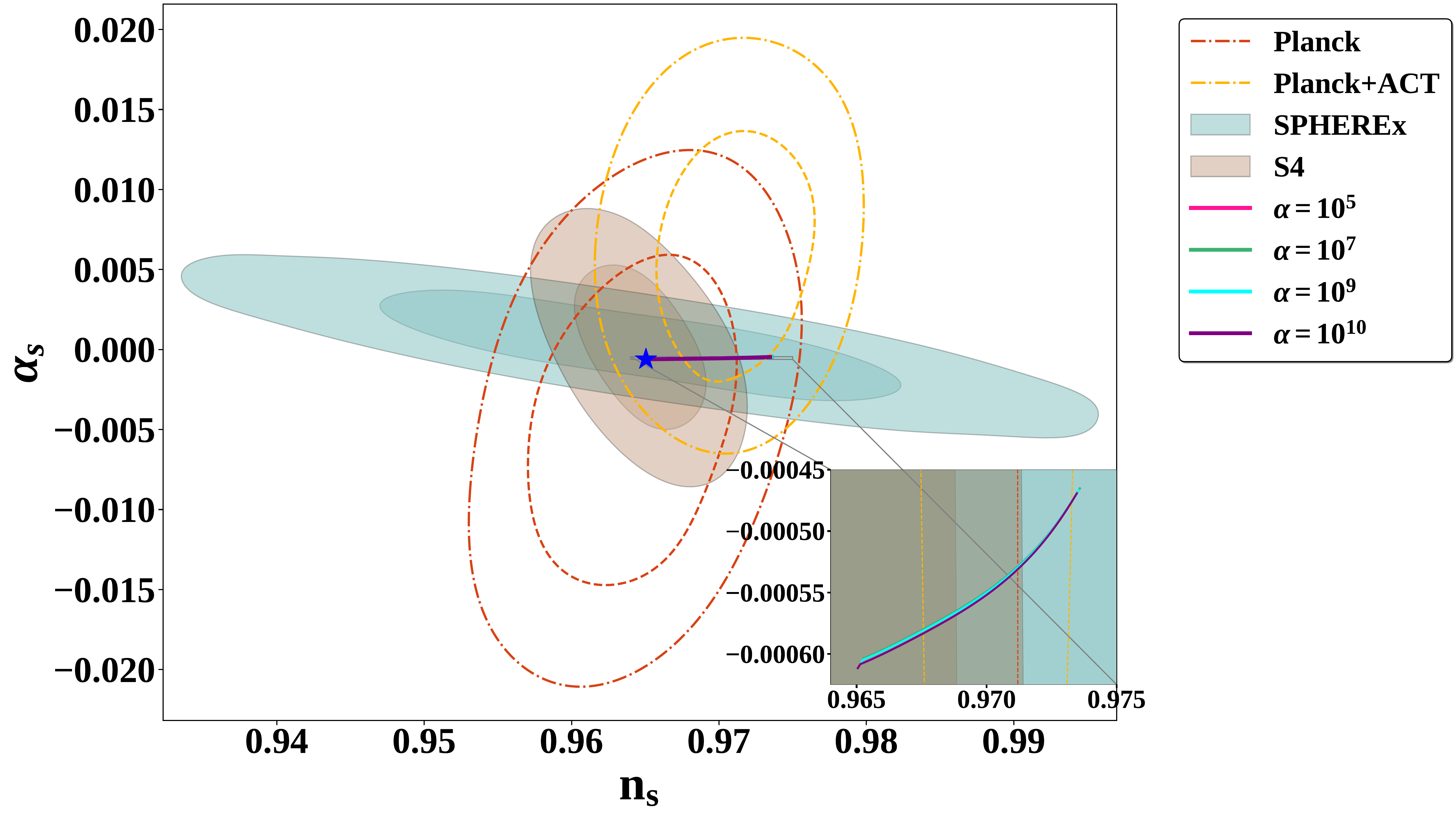}
  \caption{\it Predictions for the inflationary observables of the Coleman-Weinberg potential in the $\mathcal{R}^{2}$ Palatini formalism are shown for the $\phi > M$ case, over a range of values of the non-minimal coupling parameter $\xi$ and for different choices of the $\mathcal{R}^2$ parameter $\alpha$. We fixed $M = 50\,m_{\mathrm{Pl}}$ and $\omega_r = 1/3$. The blue star corresponds to the points shown in Figs.~\ref{fig:mi_1},~\ref{fig:monopoles1},~\ref{fig:monopoles2}, and ~\ref{fig:monopoles3}.
  }
  \label{fig:PalatiniR2_M}
\end{figure}

In the standard Coleman-Weinberg inflation framework considered here, the potential exhibits two distinct inflationary regions determined by the inflaton field value \(\phi\) relative to the symmetry-breaking scale \(M\). For \(\phi < M\), the potential features a hilltop-like structure, characteristic of small-field inflation, where the logarithmic term \(\ln(\phi/M) < 0\) creates a shallow plateau conducive to slow-roll dynamics with relatively low field excursions (see, e.g., \(\phi_*/m_{\rm Pl} \sim 29.4\)--$43.2$ in Tab.~\ref{tab:infl-para-all-merged_Palatini_R2_0}). This region yields spectral indices \(n_s \sim 0.953\)--$0.963$ and moderate tensor-to-scalar ratios \(r \sim \mathcal{O}(10^{-1})\) for smaller \(\alpha\), transitioning to \(r \simeq 8 \times 10^{-4}\) for large \(\alpha = 10^{10}\). Conversely, for \(\phi > M\), the potential adopts a large-field, chaotic-like behavior due to \(\ln(\phi/M) > 0\), leading to significantly larger field values (\(\phi_*/m_{\rm Pl} \sim 77.2\)--$9455.8$, Tab.~\ref{tab:infl-para1-merged_M}) and a flatter potential that drives \(n_s\) closer to unity (\(n_s \sim 0.967\)--$0.974$), aligning with recent \emph{Planck}+ACT data~\cite{ACT:2025tim}. This large-field regime enhances the model's consistency with observations while maintaining the \(\mathcal{R}^2\)-driven suppression of \(r\), as illustrated in Fig.~\ref{fig:PalatiniR2_M}, highlighting the critical role of the field branch in shaping inflationary predictions within our $SO(10)$ GUT framework.

Although the inflaton field $\phi$ can take super-Planckian values in the $\phi > M$ branch, as seen in Tables~\ref{tab:infl-para1-merged_M} and~\ref{tab:mon-infl-alpha1_M}, the model's validity is preserved. The canonical Einstein-frame field $\zeta$ undergoes mildly super-Planckian excursions using Eq.~\eqref{eq:non_minimal} and Eq.~\eqref{eq:Canonical} which is explicitly shown in Eq.~\eqref{eq:Canoical_R2} yielding
\begin{equation}\label{eq:approx}
\Delta \zeta \sim \frac{1}{\sqrt{\xi}}\, \ln\left(\frac{\phi_*}{\phi_e}\right) \approx 10\, m_{\rm Pl}   \,, 
\end{equation}
 comparable to those in large-field models like chaotic inflation, where $\Delta \phi \sim 5$--$15\, m_{\rm Pl}$ is standard. Moreover, the Coleman-Weinberg potential, being a one-loop effective potential, remains perturbative since the loop expansion parameter $A \,\ln(\phi/M) \sim 10^{-12}$--$10^{-14} \ll 1$, with $A \sim 10^{-15}$--$10^{-13}$ and $\ln(\phi/M) \sim 3$--$4$ from the benchmarks. Additionally, the Palatini $\mathcal{R}^2$ formulation elevates the perturbative unitarity cutoff above the inflationary Hubble scale $H \sim 10^{13}$~GeV~\cite{Ema:2017rqn,Gorbunov:2018llf} as shown in Tabs.~\ref{tab:mon-infl-alpha2}, \ref{tab:mon-infl-alpha1}, and~\ref{tab:mon-infl-alpha1_M}, ensuring reliability without undue sensitivity to UV physics.

 In the post-inflationary era, reheating proceeds through the perturbative decay of the inflaton $\phi$, which, as a GUT singlet scalar, couples primarily to the symmetry-breaking multiplet $\varUpsilon_S$ via the portal term $-\frac{1}{2} \varpi_S^2 \phi^2 \varUpsilon_S^2$, as we elaborate with more details in the next Section, specifically around Eq.~\eqref{eq:extra_term}. This interaction enables the dominant two-body decay channel $\phi \to \varUpsilon_S \varUpsilon_S$, provided the kinematic condition $m_\phi > 2 m_{\varUpsilon_S}$ is satisfied, where the inflaton mass near its minimum $\phi = M$ is $m_\phi \approx \sqrt{3 A} M$ from the quadratic approximation of the Coleman-Weinberg potential, and $m_{\varUpsilon_S} \approx \varpi_S M / \sqrt{2 \varkappa_S}$ from the induced mass term, where \(\varkappa_S\) is a dimensionless coupling that control the  self-interaction of the symmetry-breaking
GUT multiplet scalar \(\varUpsilon_S\). The decay width for this channel is~\cite{Bernal:2021qrl, Ghoshal:2023jvf, Ghoshal:2024ycp}
\begin{equation}
\Gamma_\phi \approx \frac{\varpi_S^4 M^2}{32 \pi m_\phi}\,,    
\end{equation}
assuming negligible phase-space suppression. Subsequent rapid decays of $\varUpsilon_S$ to SM particles (via its gauge interactions within the $SO(10)$ multiplet) thermalize the universe. The reheating temperature can then be estimated as~\cite{Bernal:2021qrl, Ghoshal:2023jvf, Ghoshal:2024ycp}
\begin{equation}
T_{\rm reh} \approx \left( \frac{80 \Gamma_\phi^2 m_{\rm Pl}^2}{\pi^2 g_*} \right)^{1/4}\,,    
\end{equation}
 where $g_* \approx 106.75$ accounts for relativistic degrees of freedom. Substituting the model parameters, with $\varpi_S \sim 10^{-4}$ (consistent with $A \sim 10^{-14}$--$10^{-15}$ from the benchmarks and $D=210$) and $M \sim 50$--$250\, m_{\rm Pl}$, yields $T_{\rm reh} \sim 10^{7}$--$10^{10}$~GeV, aligning with the values adopted in our analysis to ensure compatibility with Big Bang nucleosynthesis (BBN) and to avoid overproduction of gravitinos in supersymmetric extensions. This range also minimizes washout effects on any baryon asymmetry generated at intermediate scales, while higher $T_{\rm reh}$ would enhance monopole dilution but potentially conflict with unification constraints.
\begin{table}[t!]
\begin{center}
\begin{tabular}{| c | c | c | c | c | c | c | c | c | c |}
\hline
$\xi$ & $\alpha$ & $\dfrac{M}{m_{\rm Pl}}$ & $A\times10^{-15}$ & $\phi_*/m_{\rm Pl}$ & $\phi_e/m_{\rm Pl}$ & $n_s$ & $r$ & \(-\alpha_s \times 10^{-4}\) &$N$ \\
\hline
\multirow{4}{*}{\rotatebox[origin=c]{90}{0.001}}
  & \multirow{2}{*}{$10^5$}
    & 50  & 4.90 &  28.63 & 47.75 & 0.959 & 0.239    & \(6.36\)& 58.3 \\
  & 
    & 250 & 4.90 & 143.15 & 238.73 & 0.959 & 0.239    & \(6.36\)& 58.3 \\
\cline{2-10}
  & \multirow{2}{*}{$10^{10}$}
    & 50  & 5.75 &  29.21 & 48.23 & 0.957 & 0.000802 & \(6.40\) &55.6 \\
  &
    & 250 & 5.75 & 146.04 & 241.16 & 0.957 & 0.000802 & \(6.40\)& 55.6 \\
\hline
\multirow{4}{*}{\rotatebox[origin=c]{90}{0.0001}}
  & \multirow{2}{*}{$10^5$}
    & 50  & 0.79 &  42.68 & 49.29 & 0.964 & 0.159    & \(6.36\) & 58.0 \\
  &
    & 250 & 0.79 & 213.38 & 246.46 & 0.964 & 0.159   & \(6.36\)& 58.0 \\
\cline{2-10}
  & \multirow{2}{*}{$10^{10}$}
    & 50  & 0.87 &  42.84 & 49.47 & 0.963 & 0.000800 & \(6.41\)& 55.6 \\
  &
    & 250 & 0.87 & 214.19 & 247.36 & 0.963 & 0.000800 & \(6.41\)& 55.6 \\
\hline
\multirow{4}{*}{\rotatebox[origin=c]{90}{$10^{-5}$}}
  & \multirow{2}{*}{$10^5$}
    & 50  & 0.09 &  47.63 & 49.78 & 0.965 & 0.145    & \(6.29\)& 57.9 \\
  &
    & 250 & 0.09 & 238.17 & 248.88 & 0.965 & 0.145    & \(6.29\)& 57.9 \\
\cline{2-10}
  & \multirow{2}{*}{$10^{10}$}
    & 50  & 0.10 &  47.66 & 49.83 & 0.964 & 0.000800 & \(6.34\)& 55.6 \\
  &
    & 250 & 0.10 & 238.38 & 249.18 & 0.964 & 0.000800 & \(6.34\)& 55.6 \\
\hline
\end{tabular}
\caption{\it Inflationary observables for the Coleman-Weinberg potential in the $\mathcal{R}^2$ Palatini formalism. We fixed the equation-of-state parameter at $\omega_r = \tfrac{1}{3}$ and impose the condition $\phi < M$.}
\label{tab:infl-para_R2_negative_instant}
\end{center}
\end{table}

The numerical results in Tab.~\ref{tab:infl-para-all-merged_Palatini_R2_0} confirm that the tensor-to-scalar ratio $r$ is highly sensitive to the Starobinsky parameter $\alpha$, as indicated in Eq.~\eqref{eq:tensor_ratio_R2_Palatini}. For $\alpha = 10^5$, $r$ assumes relatively large values, $0.15 \lesssim r \lesssim 0.26$, depending on the non-minimal coupling $\xi$, whereas for $\alpha = 10^{10}$, it is strongly suppressed to $r \simeq 8 \times 10^{-4}$, essentially independent of other parameters. The non-minimal coupling parameter $\xi$ affects $r$ only moderately and induces minor variations in the spectral index $n_s$, which remains in the range $0.953 \lesssim n_s \lesssim 0.963$. Similarly, the running of the spectral index, $\alpha_s$, exhibits mild dependence on the model parameters, taking values in the range $-7.71 \times 10^{-4} \lesssim \alpha_s \lesssim -7.23 \times 10^{-4}$, with the largest magnitudes occurring for lower reheating temperatures and smaller $\alpha$. The mass scale $M$ primarily rescales the field values $\phi_*$ and $\phi_e$ without significantly affecting the inflationary observables, consistent with the analytic discussion below Eq.~\eqref{eq:tensor_ratio_R2_Palatini}. Reheating temperature $T_{\rm reh}$ introduces only subleading corrections, altering $N_*$ by a few units. Overall, these results demonstrate that the suppression of $r$ is dominated by $\alpha$, while $n_s$ remains consistent with current CMB constraints~\cite{BICEP:2021xfz, Planck:2018jri, SimonsObservatory:2018koc, Hazumi_2020}, and $\alpha_s$ likewise aligns with observational measurements~\cite{SPHEREx:2014bgr, Planck:2018jri, Abazajian:2019eic, ACT:2025tim} across the explored parameter space.

Tab.~\ref{tab:infl-para_R2_negative_instant} reveals a set of clear and consistent behaviors that extend and refine the conclusions drawn for the $\omega_r = 0$ case. Here, $\omega_r$ denotes the equation-of-state (EoS) parameter during the reheating epoch,
\begin{equation}
\omega_r \equiv \frac{p}{\rho},
\end{equation}
which characterizes how the dominant energy component redshifts between the end of inflation and full thermalization. Physically, different values of $\omega_r$ correspond to different reheating dynamics:  
$\omega_r \approx 0$ describes a matter-like epoch (coherent oscillations of the inflaton),  
$\omega_r = 1/3$ corresponds to radiation-dominated reheating (instantaneous or highly efficient thermalization),  
and larger values of $\omega_r$ represent stiffer fluids or nonstandard post-inflationary dynamics.

The equation-of-state parameter $\omega_r$ plays a crucial role in inflationary predictions, as it determines the number of e-folds $N_*$ between the horizon exit of the CMB pivot scale and the end of inflation. In particular, $N_*$ receives a correction that depends on both $\omega_r$ and the ratio of the energy density at the end of inflation to that at the end of reheating. A convenient parametrization is
$$
N_* = N_0 + \frac{1-3\omega_r}{12(1+\omega_r)} \ln\left(\frac{\rho_\mathrm{end}}{\rho_\mathrm{reh}}\right),
$$
where $N_0$ denotes the number of e-folds for instantaneous reheating, $\rho_\mathrm{end}$ is the energy density at the end of inflation, and $\rho_\mathrm{reh}$ is the energy density at the end of reheating. Equivalently, the correction can be expressed as
$$
\Delta N \;\propto\; \frac{1-3\omega_r}{12(1+\omega_r)} \ln\!\left(\frac{\rho_{\rm end}}{\rho_{\rm reh}}\right).
$$

\

This implies that: 
\begin{itemize}
    \item $\omega_r < 1/3$, the correction $\Delta N$ decreases $N_*$ relative to instantaneous reheating; 
    \item $\omega_r = 1/3$, the leading dependence on the reheat temperature cancels;
    \item $\omega_r > 1/3$, the correction increases $N_*$. 
\end{itemize}
 Since the inflationary observables $n_s$ and $r$ are functions of $N_*$, different reheating histories, characterized by $\omega_r$ and $T_{\rm reh}$, shift the predicted values in the $(n_s,r)$ plane.

Numerically, instantaneous reheating increases the required number of e-folds, yielding $N_* \simeq 55.6$--$58.3$ across the scanned parameter space. This shift is accompanied by a small but systematic movement of the spectral index toward unity: $n_s$ increases by a few $10^{-3}$ relative to the $\omega_r = 0$ entries, taking values in the range $0.957$--$0.965$. The running of the spectral index, $\alpha_s$, remains consistently small and negative across all parameter combinations, clustered between approximately $-6.41 \times 10^{-4}$ and $-6.29 \times 10^{-4}$, indicating a mild scale dependence of the tilt. The running shows little sensitivity to $M$, but increases slightly (by about $0.04$--$0.05$) when $\alpha$ is increased from $10^5$ to $10^{10}$, and decreases slightly as $\xi$ is reduced from $10^{-3}$ to $10^{-5}$.

The tensor-to-scalar ratio follows the same hierarchy discussed previously: for the moderate Starobinsky coefficient $\alpha = 10^5$, $r$ lies in the $\mathcal{O}(10^{-1})$ range (e.g., $r \simeq 0.24$ for $\xi = 10^{-3}$), whereas for the large value $\alpha = 10^{10}$, the tensor amplitude essentially collapses to the attractor value $r \simeq 8\times 10^{-4}$, becoming insensitive to both $\xi$ and $M$. The mass scale $M$ primarily rescales the field values $\phi_*$ and $\phi_e$ (see the columns for $M = 50\,m_{\rm Pl}$ and $M = 250\,m_{\rm Pl}$) and has a negligible direct impact on the observables $(n_s, r)$.
\begin{table}[t!]
\begin{center}
\begin{tabular}{| c | c | c | c | c | c | c | c | c | c |}
\hline
$\xi$ & $\alpha$ & $\dfrac{M}{m_{\rm Pl}}$ & $A\times10^{-13}$ & $\phi_*/m_{\rm Pl}$ & $\phi_e/m_{\rm Pl}$ & $n_s$ & $r$ & \(-\alpha_s \times10^{-4}\)& $N$ \\
\hline
\multirow{4}{*}{\rotatebox[origin=c]{90}{0.1}}
  & \multirow{2}{*}{$10^5$}
    & 50  & 64.57 &  1887.84 & 71.85 & 0.974 & 0.070   & \(4.67\)& 57.3 \\
  & 
    & 250 & 64.57 & 9455.75 & 359.24 & 0.974 & 0.070  &  \(4.67\) & 57.3 \\
\cline{2-10}
  & \multirow{2}{*}{$10^{10}$}
    & 50  & 69.18 &   1739.71 & 62.45 & 0.972 & 0.0008 & \(4.71\)& 55.5 \\
  &
    & 250 & 64.57 & 9455.75 & 313.28 & 0.974 & 0.0008 &  \(4.71\)&57.3 \\
\hline
\multirow{4}{*}{\rotatebox[origin=c]{90}{0.01}}
  & \multirow{2}{*}{$10^5$}
    & 50  & 2.45 &  177.01 & 56.95 & 0.972 & 0.076   & \(5.27\)& 57.4 \\
  &
    & 250 & 2.34 &  883.52 & 284.76 & 0.972 & 0.077    & \(5.27\)& 57.4 \\
\cline{2-10}
  & \multirow{2}{*}{$10^{10}$}
    & 50  & 2.51 &  173.04 & 54.55 & 0.971 & 0.0008 & \(5.32\)& 55.5\\
  &
    & 250 & 2.34 & 883.52 & 273.01 & 0.972 & 0.0008 & \(5.32\)& 57.4\\
\hline
\multirow{4}{*}{\rotatebox[origin=c]{90}{$0.001$}}
  & \multirow{2}{*}{$10^5$}
    & 50  & 0.13 &  77.69 & 52.22 & 0.968 & 0.101   & \(5.76\)& 57.6 \\
  &
    & 250 & 0.13 & 389.08 & 261.11 & 0.968 & 0.100    & \(5.75\)&  57.6 \\
\cline{2-10}
  & \multirow{2}{*}{$10^{10}$}
    & 50  & 0.14 &  77.19 & 51.56 & 0.967 & 0.0008 & \(5.83\) & 55.6 \\
  &
    & 250 & 0.13 & 388.75 & 257.89 & 0.968 & 0.0008 & \(5.79\) & 57.6 \\
\hline
\end{tabular}
\caption{\it Inflationary observables for the Coleman-Weinberg potential in the $\mathcal{R}^2$ Palatini formalism. We fix the equation-of-state parameter at $\omega_r = 1/3$ and impose the condition $\phi > M$.}
\label{tab:infl-para1-merged_M}
\end{center}
\end{table}

Instantaneous reheating ($\omega_r = 1/3$) reduces the post-inflationary redshift required to connect horizon exit to the present, thereby increasing $N_*$ for fixed model parameters. In slow-roll single-field inflation, a larger $N_*$ typically shifts the pivot exit point $\phi_*$ slightly up the plateau of the effective potential (or equivalently to a region that yields a slightly smaller tilt), producing the observed increase in $n_s$ and the mild reduction of $r$ for moderate-$\alpha$ models. Working in the $\phi < M$ branch changes the sign and magnitude of the logarithmic term in the Coleman–Weinberg potential, slightly modifying the mapping between $(\xi, A)$ and $\phi_*$. However, because the Einstein-frame potential retains its plateau-like shape over the considered parameter range, the slow-roll parameters remain in the same qualitative regime, and the dominant driver of $r$ continues to be $\alpha$. In short, reheating shifts the location of horizon exit (and thus $n_s$ and $N_*$), the $\phi < M$ branch modifies the numerical value of $\phi_*$ (and the small multiplicative correction in the analytic formula), but none of these effects overcome the strong $\alpha$-dependence of the tensor amplitude.

Tab.~\ref{tab:infl-para1-merged_M} and the corresponding Fig.~\ref{fig:PalatiniR2_M} show that, for instantaneous reheating and inflaton evolution along the $\phi > M$ branch, the model predicts a larger spectral index than in the $\phi < M$ or $\omega_r = 0$ cases, with $n_s$ shifted into the region favored by recent \emph{Planck}+ACT data~\cite{ACT:2025tim}. The required number of $e$-folds remains $N_* \simeq 55.5$--$57.6$ across the explored parameter range, but the pivot exit point $\phi_*$ occurs at substantially larger values, often by orders of magnitude, than on the $\phi < M$ branch. The tensor-to-scalar ratio retains the hierarchical behavior discussed previously. The running of the spectral index, $\alpha_s$, remains consistently negative, ranging approximately from $-4.7 \times 10^{-4}$ to $-5.8 \times 10^{-4}$, with a mild increase in magnitude as $\xi$ decreases, reflecting the steeper curvature of the potential for smaller non-minimal couplings. The dependence on $\alpha$ is weak, as the running is primarily governed by the slow-roll parameters evaluated at horizon exit.

\begin{figure}[t!]
    \centering
    \includegraphics[width=0.65\linewidth]{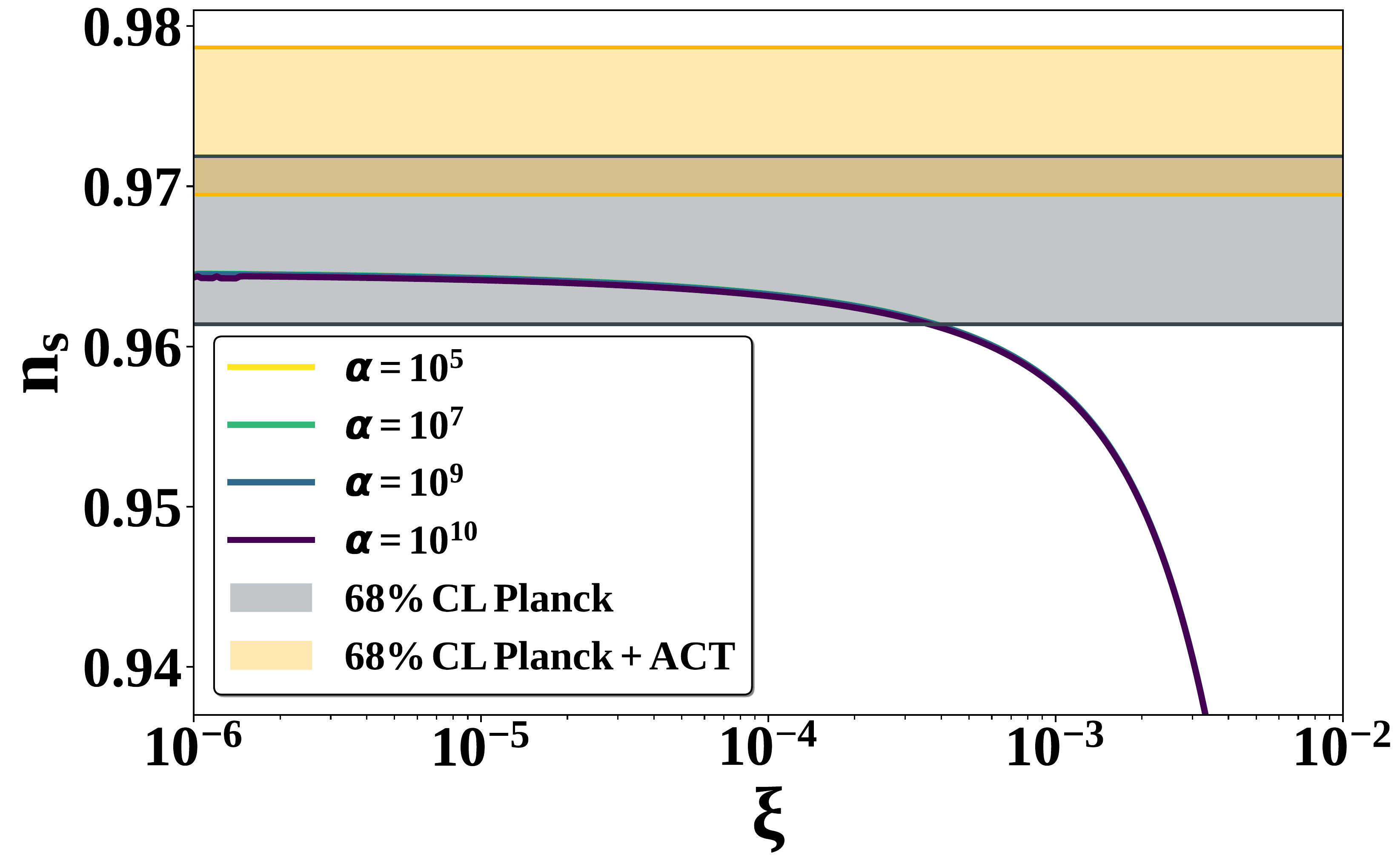}
        \includegraphics[width=0.65\linewidth]{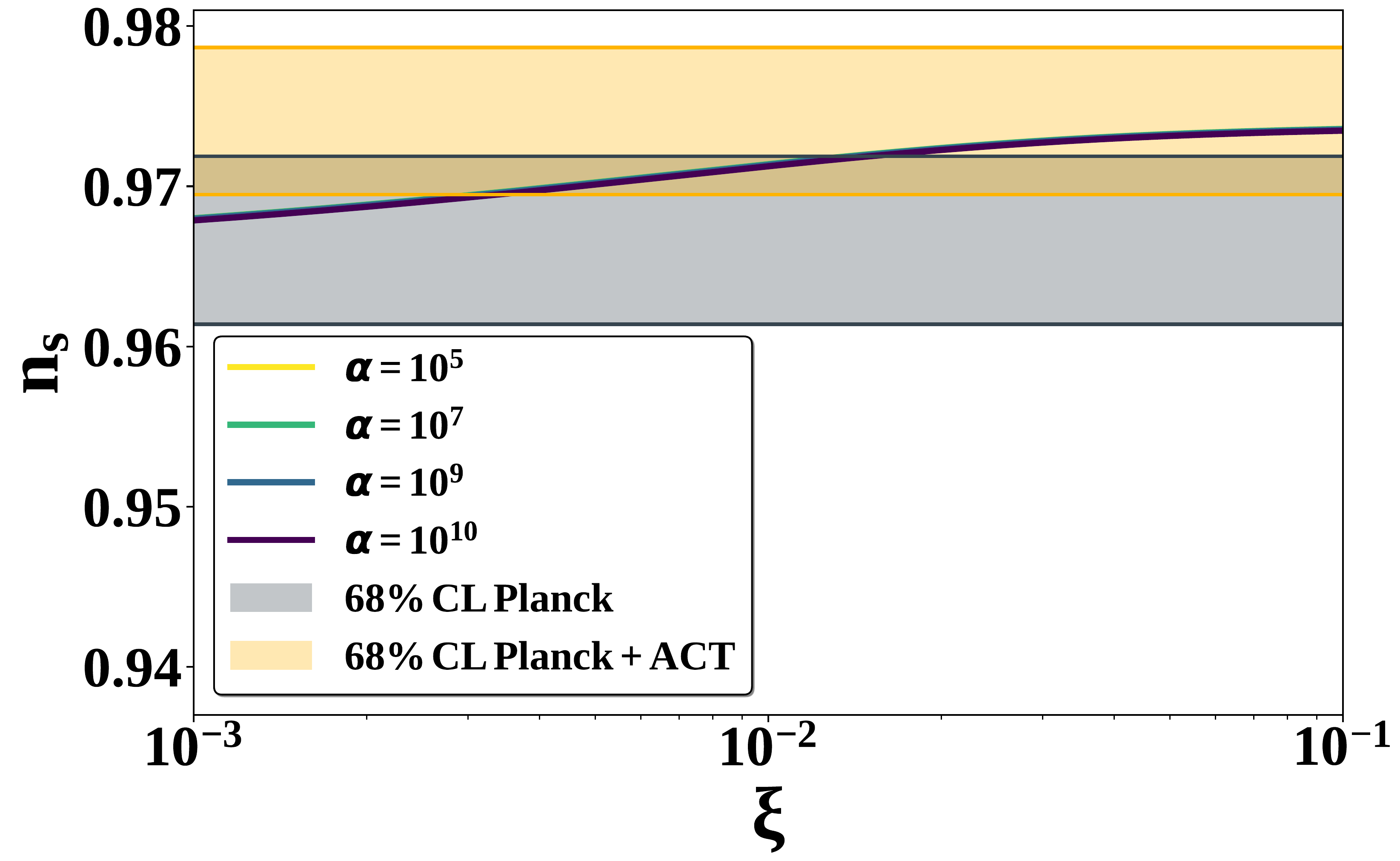}
  \caption{\it $n_s$ as a function of $\xi$ is shown for the two considered cases in the $\mathcal{R}^2$ framework: $\phi < M$ (\textbf{top}) and $\phi > M$ (\textbf{bottom}). The shaded regions correspond to the viable observational constraints from \emph{Planck}~\cite{Planck:2018jri} and \emph{Planck}+ACT~\cite{ACT:2025tim}.
}
  \label{fig:Palatini_ns}
\end{figure}

Two effects combine to shift $n_s$ toward the higher values preferred by \emph{Planck}+ACT observations~\cite{ACT:2025tim} in this branch. First, instantaneous reheating ($\omega_r = 1/3$) reduces the post-inflationary redshifting needed to match the pivot scale, effectively increasing $N_*$ relative to delayed reheating; in slow-roll, plateau-like potentials, a larger $N_*$ generally drives the spectral index closer to unity. Second, the $\phi > M$ branch alters the sign and magnitude of the Coleman–Weinberg logarithm (so that $\ln(\phi/M) > 0$), flattening the Einstein-frame potential near the pivot for the parameter range we explore. Together, these effects shift the horizon-exit point to a flatter region of the potential, producing the observed increase in $n_s$ while leaving the leading behavior of $r$ dominated by $\alpha$.

This same flattening also affects the running, $\alpha_s$, making it slightly less negative than in steeper regions of the potential. Nevertheless, $\alpha_s$ remains within the observational bounds from \emph{Planck}~\cite{Planck:2018jri}, \emph{Planck}+ACT~\cite{ACT:2025tim}, and future experiments such as CMB-S4~\cite{Abazajian:2019eic} and SPHEREx~\cite{SPHEREx:2014bgr}, as illustrated in the lower panel of Fig.~\ref{fig:PalatiniR2_M}. For a comparison between the two cases, \(\phi < M\) and \(\phi > M\), we present the predicted scalar spectral index \(n_s\) as a function of the non-minimal coupling \(\xi\) in Fig.~\ref{fig:Palatini_ns}. Notably, the \(\phi > M\) scenario remains consistent with the latest \emph{Planck}+ACT constraints~\cite{ACT:2025tim} over a broader range of \(\xi\), yielding spectral values that lie within the observationally allowed region. In contrast, the \(\phi < M\) case falls below the \emph{Planck}+ACT bounds and aligns only with the \emph{Planck}-allowed region~\cite{Planck:2018jri}.

\section{Symmetry Breaking and Topological Defects}
\label{Sec:Monopoles}
\begin{figure}[t!]
    \centering
    \includegraphics[width=1\linewidth]{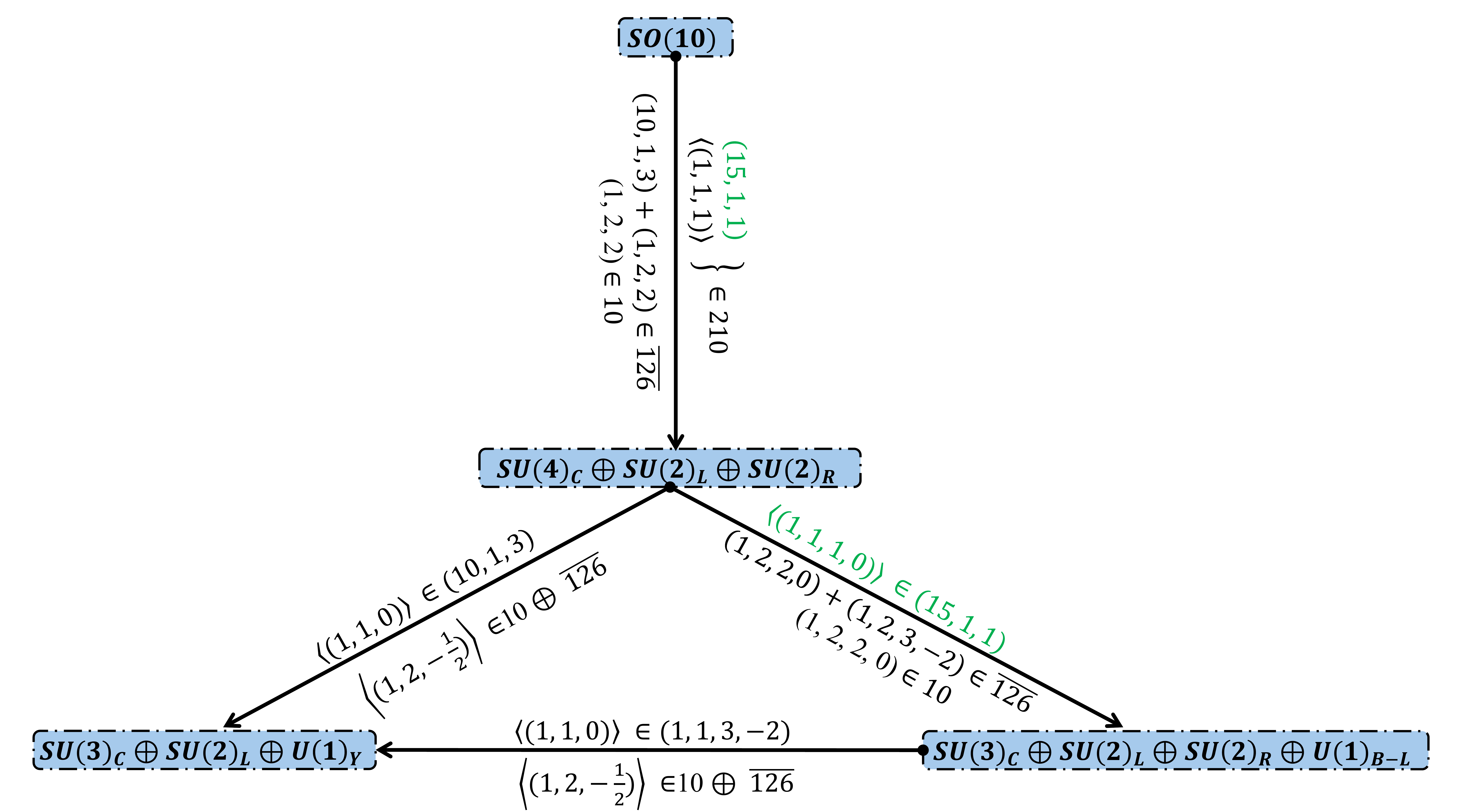}
    \caption{\it Breaking patterns of $SO(10)$ and their associated scalar multiplets.}
    \label{fig:Symmetry_breaking}
\end{figure}

We now turn to the symmetry-breaking structure of our \( SO(10) \) construction and the topological defects that can arise as the unified symmetry is reduced to the Standard Model. Building on the \( SO(10) \) framework with non-minimal Coleman-Weinberg inflation introduced in Sec.~\ref{Sec:SO(10)}, which extends the analysis in~\cite{Maji:2022jzu} that such models can produce observable intermediate-mass monopoles (\(\sim 10^{14} \, \text{GeV}\)) consistent with CMB constraints, this section examines two representative breaking chains (summarized in Fig.~\ref{fig:Symmetry_breaking}). We identify the scalar multiplets responsible for each step and analyze when and how monopoles are produced and subsequently diluted by inflation. Starting from the unified group \( SO(10) \), a typical first breaking proceeds to the Pati–Salam subgroup \( SU(4)_C \times SU(2)_L \times SU(2)_R \); different choices of GUT Higgs representations then reduce the symmetry further either to the SM gauge group or to an intermediate left–right symmetric group \( SU(3)_C \times SU(2)_L \times SU(2)_R \times U(1)_{B-L} \).

In our inflationary scenario, the timing of each symmetry-breaking step is controlled by the inflaton–Higgs portal and by the inflaton trajectory described in Sec.~\ref{Sec:SO(10)}. Consequently, the multiplet indicated on a given arrow both specifies the topological charge of the defect that would be produced at that step and identifies the field whose dynamics, through the couplings \(\varpi_S, \varkappa_S\) and the radiative scale \(M\) introduced below, determine the degree to which any produced defect population is diluted by subsequent inflation. We now consider the following breaking patterns of \(SO(10)\) and the corresponding scalar multiplets.

While alternative breaking paths, such as via the \(SU(5)\) subgroup using the adjoint 45 or the 210 representation, are possible within \(SO(10)\), we focus on the Pati--Salam intermediate group \(SU(4)_C \times SU(2)_L \times SU(2)_R\) for several compelling reasons. First, proton decay in Pati--Salam models is typically mediated by Higgs scalars rather than gauge bosons, leading to suppressed rates compared with the gauge-mediated decays in \(SU(5)\), which often conflict with experimental limits from Super-Kamiokande. Second, the Pati--Salam pathway preserves the rank-5 structure of \(SO(10)\) until the B-L breaking stage, enabling staged symmetry breaking with tunable intermediate scales that align well with gauge coupling unification and allow the formation of topological monopoles at energies accessible to partial inflationary dilution, as explored in this work. \(SU(5)\) breaking, a single-step rank reduction to 4, tends to produce GUT-scale monopoles that require full inflationary dilution to avoid overabundance, limiting phenomenological richness for relic observables. These features make the Pati--Salam scenario particularly advantageous for embedding inflation within \(SO(10)\), as it enhances compatibility with neutrino data, proton lifetime bounds, and cosmological constraints while offering greater flexibility in addressing the hierarchy problem and matter-antimatter asymmetry~\cite{Lazarides:2019xai, Senoguz:2015lba}.

The term in the potential that governs the interaction between the inflaton \(\phi\) and the scalar \(\varUpsilon_S\), which is responsible for gauge symmetry breaking\footnote{Concretely, \(\varUpsilon_S\) denotes the component(s) of that multiplet that acquire a vacuum expectation value (VEV) and thereby reduce \(SO(10)\) to the intermediate subgroups shown in Fig.~\ref{fig:Symmetry_breaking}. Different choices of \(S\) determine the residual symmetry, and consequently the topological charge and mass spectrum of the associated defects.} is
\begin{equation}\label{eq:extra_term}
    V_J(\phi,\varUpsilon_S) = - \frac{1}{2}\varpi_S^2\phi^2\varUpsilon_S^2 + \frac{\varkappa_S}{4}\varUpsilon_S^4\,.
\end{equation}
Here, the parameters \(\varpi_S\) and \(\varkappa_S\) are dimensionless couplings that control, respectively, the GUT singlet Higgs–inflaton (\(\phi\)) ``portal'' and the self-interaction of the symmetry-breaking GUT multiplet scalar \(\varUpsilon_S\). In particular, \(\varpi_S\) governs the strength of the quadratic mixing between \(\phi\) and \(\varUpsilon_S\), rendering the effective mass of the \(\varUpsilon_S\) field dependent on \(\phi\). Through this portal, the evolution of \(\phi\) during inflation triggers the spontaneous breaking of the GUT symmetry. Phenomenologically, larger \(\varpi_S\) (or smaller \(\varkappa_S\)) leads to earlier or stronger symmetry breaking as \(\phi\) evolves, thereby affecting the formation, typical mass scale, and subsequent dilution of monopoles during inflation.

For simplicity, we take \(\varUpsilon_S\) to be a canonically normalized real scalar field transforming in a \(D\)-dimensional representation of the gauge group, which we choose as \(D=210\) for the \(SO(10)\) case. We forbid bare mass terms in our framework by imposing scale invariance on both the gauge-singlet and gauge-charged fields. Excluding such terms allows us to control the number of $e$-folds occurring during a given symmetry-breaking event. Consequently, the potential in Eq.~\eqref{eq:extra_term} induces a vacuum expectation value (VEV) for \(\varUpsilon_S\) given by
 \begin{equation}\label{eq:VEV}
    \left<\varUpsilon_S\right> = \frac{\varpi_S}{\sqrt{\varkappa_S}} M\,.
\end{equation}
This implies that the timing of symmetry breaking is controlled by the evolution of \(\phi\), as well as by any additional mass or thermal contributions omitted above. In Eq.~\eqref{eq:VEV}, we take \(\varkappa_S = 0.25\) to ensure that the \(\varUpsilon_S\) potential remains stabilized at large field values. The magnitude of \(\varkappa_S\) determines the steepness of the \(\varUpsilon_S\) potential and, consequently, the VEV that develops once symmetry breaking occurs. When the inflaton settles at its minimum, $M$, the coefficient
\begin{equation}  
A = \frac{\varpi_S^4\,D}{16\pi^2}\,.
\end{equation}
In Eq.~\eqref{PotentialJORD}~\cite{Lazarides_2019} is dominated by the coupling of the scalar field responsible for GUT symmetry breaking. Accordingly, the unification scale \(M_U\) is determined by the resulting VEV of \(\varUpsilon_S\), leading to the relation~\cite{Senoguz:2015lba}
\begin{equation}\label{eq:MU}
M_U \;=\; \sqrt{\frac{8\pi}{\varkappa_S}}\left(\frac{A M^4} {D}\right)^{1/4}\sim \sqrt{4\pi}\left(\frac{AM^4}{4}\right)^{1/4}\,.
\end{equation}
This succinctly explains how the GUT-breaking scale is determined by the interplay of the symmetry-breaking potential with the overall mass and normalization factors. Physically, this sets the masses of the heavy GUT gauge bosons and directly influences the production of monopoles, their typical mass, and whether cosmological processes, such as inflationary dilution, make them observable today.

During inflation, the scalar field \(\varUpsilon_S\) acquires an effective mass given by
\begin{equation}\label{m_eff}
    m_{\rm eff}^2 = 2\left(\varpi_S^2\phi^2- \sigma_{\varUpsilon_S}T_H^2\right)\,,
\end{equation}
\begin{figure}[t!]
   \centering
\includegraphics[width=0.85\linewidth]{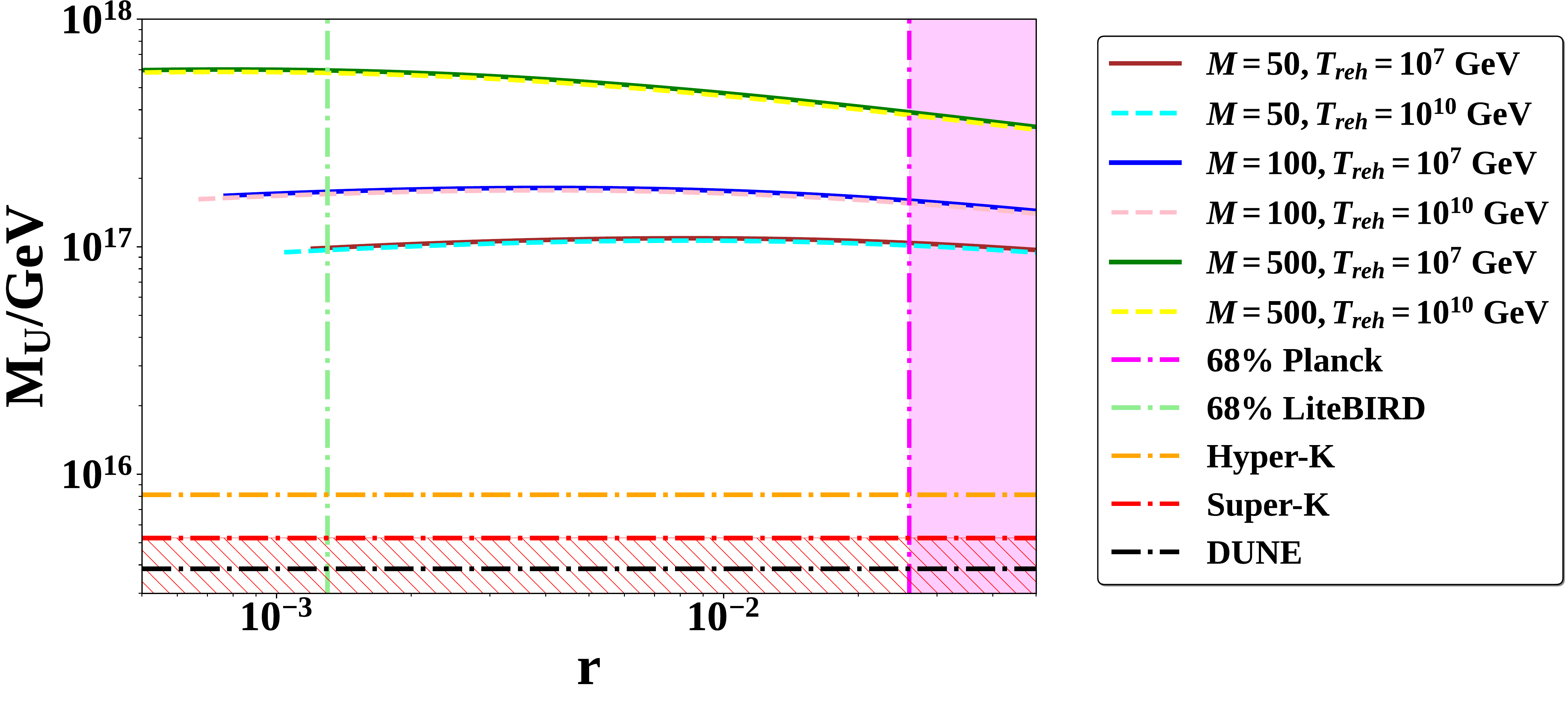}
\caption{\it $M_U$ vs. $r$ for the Palatini case with different choices of $M$ (in $m_{\rm Pl}$ units).}
 \label{fig:mi_0}
\end{figure}
and it encodes the two competing effects that control the immediate stability of the GUT-breaking multiplet \(\varUpsilon_S\) during inflation. The term proportional to \(\varpi_S^2 \phi^2\) arises from the inflaton–portal coupling and therefore tracks the instantaneous value of the inflaton \(\phi\). The sign and magnitude of \(m_{\rm eff}^2\) determine the local behavior of \(\varUpsilon_S\): in the usual convention, a negative \(m_{\rm eff}^2\) signals a tachyonic instability that drives \(\varUpsilon_S\) away from the symmetric origin, triggering spontaneous breaking of the gauge symmetry, whereas a positive \(m_{\rm eff}^2\) keeps \(\varUpsilon_S\) stabilized near zero.

Two dynamical regimes are especially significant for cosmology. If \(|m_{\rm eff}| \gg H\), then \(\varUpsilon_S\) is heavy and adiabatically follows its instantaneous minimum, suppressing defect formation; whereas if \(|m_{\rm eff}| \lesssim H\), the field is light and accumulates large superhorizon fluctuations, increasing the likelihood of domain formation and topological defects via the Kibble mechanism. Consequently, the competition encoded in Eq.~\eqref{m_eff} determines whether monopoles associated with the chosen \(SO(10)\) breaking chain are produced and whether they are subsequently diluted by inflationary expansion.

In Eq.~\eqref{m_eff}, \(T_H\) is the Hawking temperature, \(T_H = H/2\pi\), and we take \(\sigma_{\varUpsilon_S} \sim 1\). The Hubble parameter can then be computed through the following expression
\begin{equation}\label{eq:Hubble}
    H(\phi) = \sqrt{\frac{V_E(\phi)}{3}}\,.
\end{equation}
In the slow–roll approximation (kinetic energy \(\ll V_E\)), this relation follows from the Friedmann equation and provides the instantaneous expansion rate, which is directly controlled by the potential energy. Physically, \(H\) sets the Hubble radius \(H^{-1}\) (the causal horizon and the freeze-out condition \(k = aH\)), determines the Hawking temperature defined above, and establishes the time scale for classical evolution and the dilution of relics.

The Ginzburg criterion is given by~\cite{Ginz}
\begin{figure}[t!]
   \centering
\includegraphics[width=0.85\linewidth]{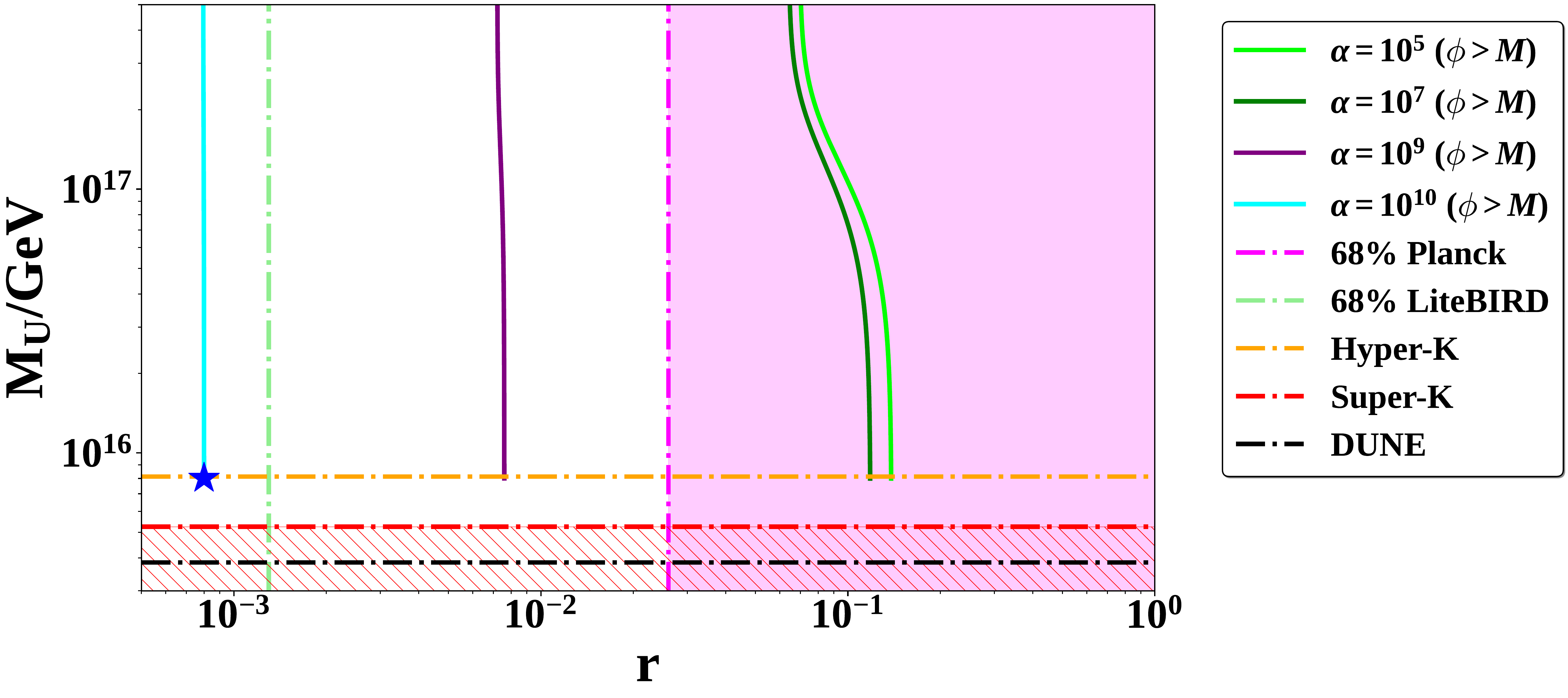}
  \caption{\it $M_U$ as a function of $r$ in the Palatini $\mathcal{R}^2$ case. We fix $M = 50\,m_{\rm Pl}$ and $\omega_r = 1/3$. The blue star corresponds to the points shown in Figs.~\ref{fig:PalatiniR2_M},~\ref{fig:monopoles1},~\ref{fig:monopoles2}, and~\ref{fig:monopoles3}.      }
   \label{fig:mi_1}
\end{figure}
\begin{equation}\label{ginz}
   \Theta^3\Delta V > T_H\,,
\end{equation}
where the correlation length can be given by
\begin{equation}\label{minima}
    \Theta = {\rm min}\left(H^{-1}, m_{\rm eff}^{-1}\right)\,.
\end{equation}
The above criterion implies that the energy contained within a single correlation volume, \(\Theta^3 \Delta V\), must exceed the typical Hawking temperature \(T_H\) for classical ordering to proceed. If this condition is satisfied, the GUT Higgs \(\varUpsilon_S\) can settle into its broken vacuum and produce monopoles. During inflation, one can take \(\Theta \sim H^{-1}\)~\cite{SENOGUZ2016169}. Moreover, \(\Delta V\) is defined as the difference between the potential evaluated at \(\varUpsilon_S = 0\) and at the vacuum expectation value \(\langle \varUpsilon_S \rangle\), which is given by
\begin{equation}\label{dektav}
    \Delta V = \frac{m_{\rm eff}^4}{16 \varkappa_S}\,.
\end{equation}
\begin{table}[t!]
\begin{center}
\begin{tabular}{| c | c | c | c | c | c | c | c | c | c | }
\hline
$\xi$ & $\frac{M}{m_{\rm Pl}}$ & $\phi_+/m_{\rm Pl}$ & $\phi_-/m_{\rm Pl}$ & \thead{$H_+$\\($10^{13}$ GeV)} & \thead{$H_-$\\($10^{13}$ GeV)} & \thead{$M_{I+}$\\($10^{13}$ GeV)} & \thead{$M_{I-}$\\($10^{13}$ GeV)} & $N_+$ & $N_-$ \\
\hline
\multirow{3}{*}{\rotatebox[origin=c]{90}{-0.001}}& 50 & 43.59&  42.13 & 3.77 & 4.18 & 5.64 & 6.47 & 10.9& 17.1\\
& 100 & 93.47 & 91.94 & 3.48 & 3.76 & 4.84& 5.33 & 10.8 & 17.1 \\
& 500 & 493.46 &  491.86 & 2.17 & 2.23 & 2.87 & 2.95 & 10.7 & 16.9 \\
\hline
\multirow{3}{*}{\rotatebox[origin=c]{90}{-0.002}}& 50 &  43.66 &  42.25 & 3.04 & 3.27 & 4.54 & 5.04 & 10.8 & 17.0 \\
& 100 & 93.56 & 92.07 & 2.69 & 2.82 & 3.74 & 3.99 & 10.8 & 16.9 \\
& 500 & 493.56 &  491.99 & 1.49 & 1.51 & 1.97 & 2.00 & 10.6 & 16.7 \\
\hline
\multirow{3}{*}{\rotatebox[origin=c]{90}{-0.003}}& 50 & 43.73 &42.36 & 2.50 & 2.64 & 3.73 & 4.06 & 10.7 & 16.9 \\
& 100 & 93.64 &  92.18 & 2.18 & 2.26 & 3.03 & 3.19 & 10.7 & 16.9 \\
& 500 &  493.64 & 492.11 & 1.13 & 1.14 & 1.49 & 1.51 & 10.5 & 16.6 \\
\hline
\multirow{3}{*}{\rotatebox[origin=c]{90}{-0.004}}& 50 & 43.80 & 42.46 & 2.17 & 2.26 & 3.23 & 3.47 & 10.7 & 16.9 \\
& 100 & 93.71 & 92.29 & 1.81 & 1.85 & 2.51 & 2.62 & 10.6 & 16.8 \\
& 500 &  493.71 & 492.22 & 0.92 & 0.93 & 1.22 & 1.23 & 10.5 & 16.6 \\
\hline
\end{tabular}
\caption{\it Intermediate symmetry breaking occurs at the scale $M_I$, producing monopoles with a yield 
\(
Y_M = n_M/s
\) 
that can saturate the MACRO bound, $Y_M^+ \simeq 10^{-27}$ (parameters with subscript `$+$'), and an adopted threshold for observability, $Y_M^- = 10^{-35}$ (parameters with subscript `$-$'). Inflation is driven by a real GUT-singlet scalar with a Coleman–Weinberg potential and a non-minimal coupling to gravity, for different values of $\xi$ and $M$. The corresponding values of the inflaton field $\phi$, the Hubble parameter $H$, and the number of $e$-folds experienced by the monopoles are indicated. The reheating temperature is set to $T_{\rm reh} = 10^7$~GeV.
}
\label{tab:mon-infl}
\end{center}
\end{table}
Using Eqs.~\eqref{m_eff}, \eqref{ginz}, \eqref{minima}, and \eqref{dektav}, the symmetry-breaking scale can be expressed as
\begin{align}\label{eq:breaking-scale}
M_I\sim \left<\varUpsilon_S\right> =  \sqrt{\left(4\pi\sqrt{2\pi \varkappa_S} + \sigma_{\varUpsilon_S}\right)} \frac{M}{\sqrt{\varkappa_S}\phi_I} \frac{H_I}{2\pi} \,.
\end{align}
Here, the subscript $I$ denotes the corresponding intermediate-scale values. The comoving number density, or the so-called monopole yield, after reheating, is given by
\begin{equation}\label{eq:YM}
    Y_M \simeq \frac{H_I^3\,{\rm exp}\left(-3N_I\right)\left(\frac{\tau}{t_r}\right)^2}{s}\,,
\end{equation}
where \(s = \frac{2\pi^2}{45} g_*\, T_{\rm reh}^3\), and we proceed with the choice \(T_{\rm reh} = 10^7~\mathrm{GeV}\), as variations in the reheating temperature affect the results only mildly. Furthermore, \(N_I\) is obtained from Eq.~\eqref{perturb2} by replacing \(\phi_*\) with \(\phi_I\). The reheating time is \(t_r \simeq \Gamma_\phi^{-1}\), where \(\Gamma_\phi\) is the decay width of \(\phi\), and it is given by~\cite{lazarides1999inflation, Lazarides2002}
\begin{equation}
    t_r =\sqrt{\frac{45}{2\pi^2g_*}} \frac{m_{\rm Pl}}{T_{\rm reh}^2}\,,
\end{equation}
and the cosmic time at the end of inflation can be estimated as
\begin{equation}
    \tau \simeq \int_{\phi_e}^{\phi_*} \frac{3H(\phi)}{V'} \left(\frac{d\zeta}{d\phi}\right)^2 d\phi\,.
\end{equation}
\begin{figure}[t!]
    \centering
    \includegraphics[width=0.88\linewidth]{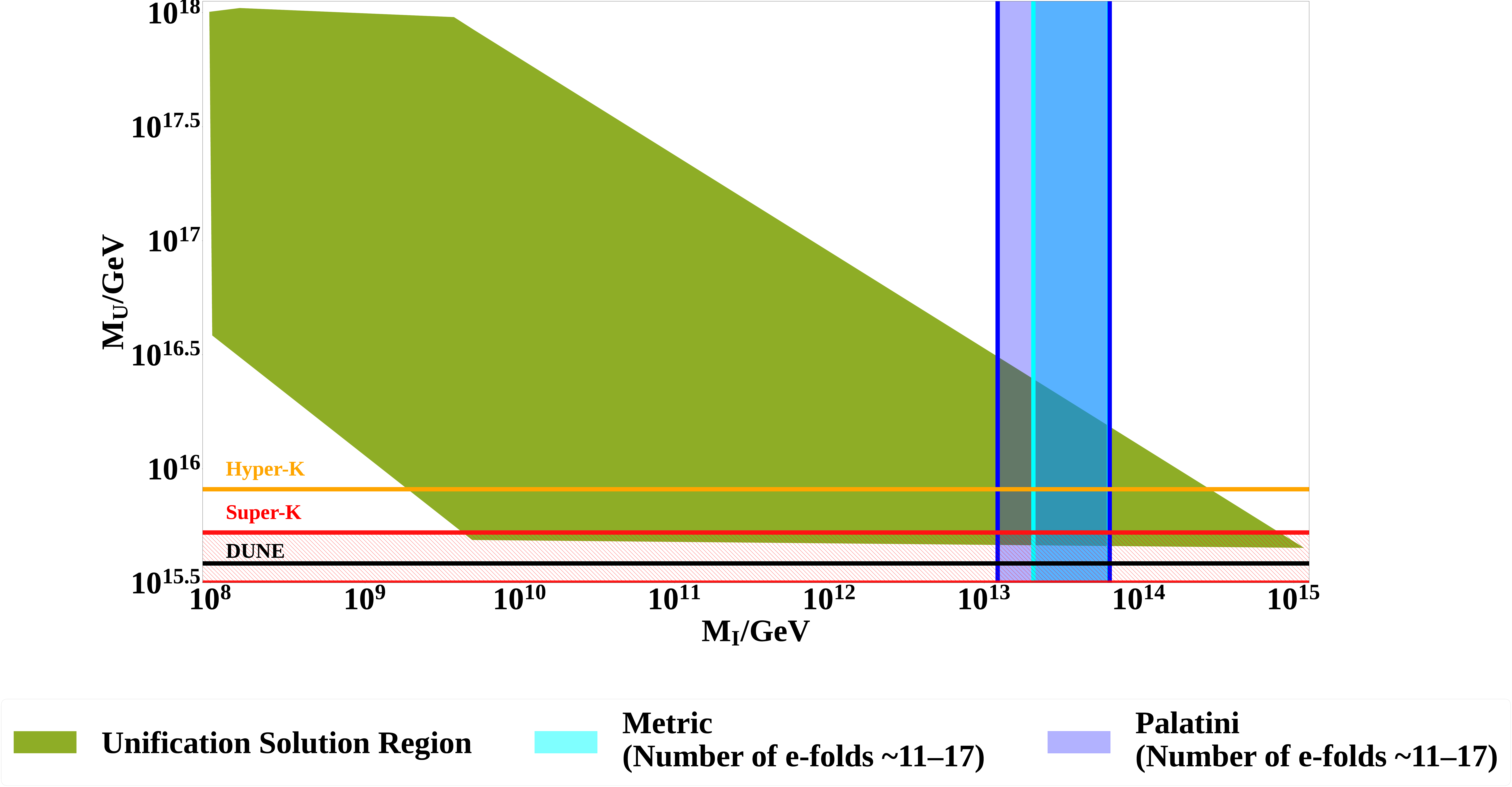}
    \includegraphics[width=0.88\linewidth]{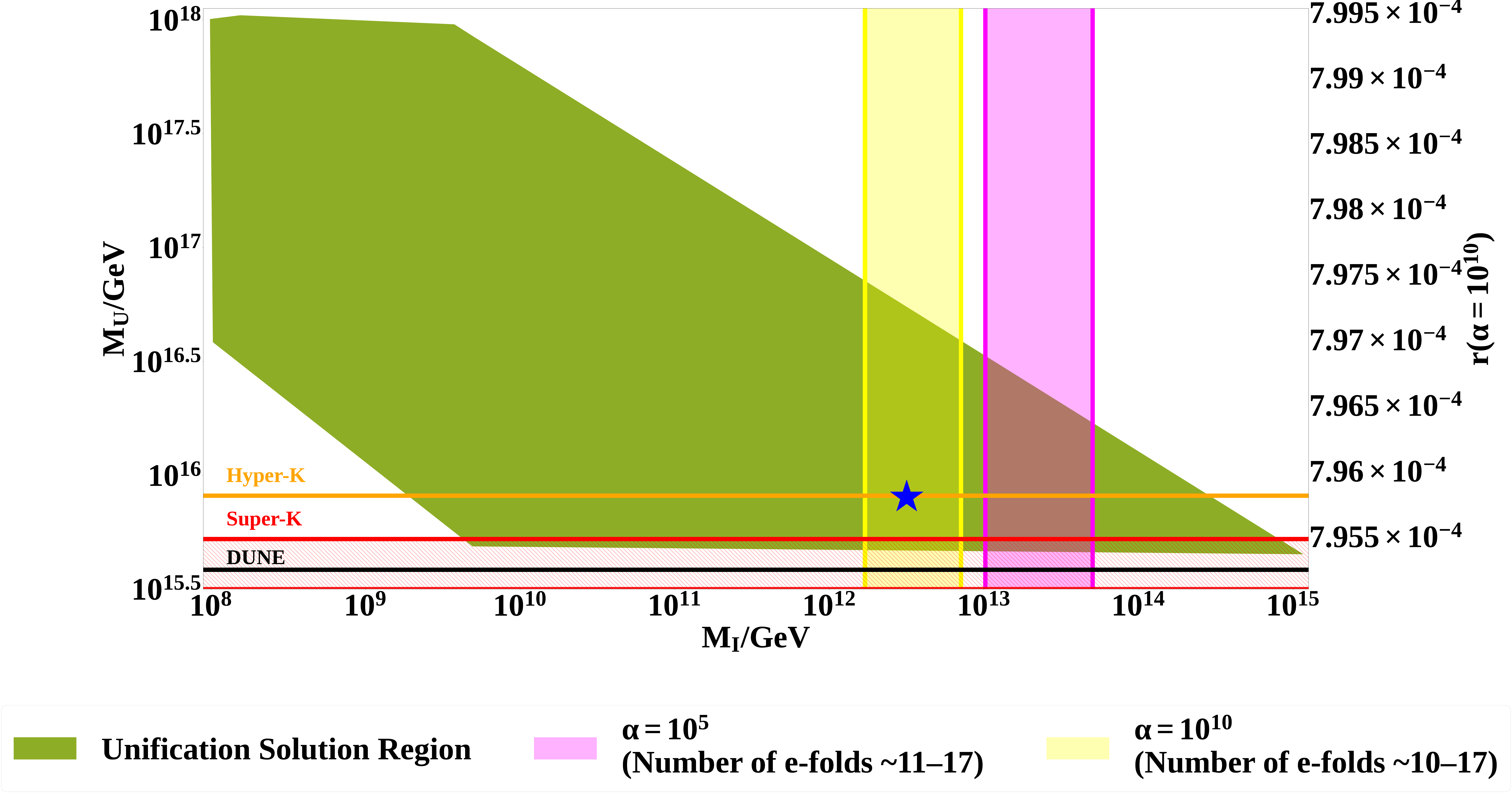}
    \caption{\it Unification solutions for the breaking chain 
    $SO(10) \xrightarrow{M_U} SU(4)_C \times SU(2)_L \times SU(2)_R 
    \xrightarrow{M_I} SU(3)_C \times SU(2)_L \times U(1)_Y.$  The vertical shaded regions indicate the range of the intermediate scale $M_{I}$ associated with monopoles for the different cases shown. The \emph{bottom} panel corresponds to the \(\phi > M\) branch. The blue star corresponds to \((M_U/\text{GeV}, \, r) \sim (8.7 \times10^{15},\, 8\times10^{-4})\), and consistent with the ones shown in Figs.~\ref{fig:PalatiniR2_M},~\ref{fig:mi_1},~\ref{fig:monopoles2}, and~\ref{fig:monopoles3}. The horizontal shaded region, e.g., for DUNE, refers to the ruled-out values.}    
    \label{fig:monopoles1}
\end{figure}
To connect the intermediate symmetry-breaking scale \(M_I\) with observable inflationary parameters, such as the tensor-to-scalar ratio \(r\), we leverage the dynamics of the inflaton field and the Hubble parameter during inflation.

The intermediate symmetry-breaking scale \(M_I\), at which topological defects like monopoles may form, is approximately given by~\cite{SENOGUZ2016169}
\begin{equation}\label{eq:simplified-MI}
M_I \approx \frac{H_*}{2\pi} \frac{M}{\phi_*}\,,
\end{equation}
where the subscript ``\(_*\)'' denotes quantities evaluated when the pivot scale exits the horizon. Here, \(H(\phi)\) is the Hubble parameter given by Eq.~\eqref{eq:Hubble}, and \(\phi_*\) is the inflaton field value determined by imposing \(\Delta_\mathcal{R}^2 \approx 2.1\times10^{-9}\)~\cite{Planck:2018vyg}, as discussed in detail in App.~\ref{Sec:Slow_roll}. This expression arises as an approximation of the more complete relation in Eq.~\eqref{eq:breaking-scale}.

Grand unified theories based on \(SU(5)\)~\cite{PhysRevLett.32.438}, \(SO(10)\)~\cite{FRITZSCH1975193}, or \(E(6)\)~\cite{SHAFI1978301} predict the existence of topologically stable magnetic monopoles whose masses are typically an order of magnitude above the unification scale~\cite{HOOFT1974276,Polyakov:1974ek,LAZARIDES198087,Shafi1984}. The inflationary scenario considered here completely dilutes these GUT monopoles. In \(SO(10)\) and \(E(6)\) models, topologically stable monopoles arise during the spontaneous breaking of an intermediate symmetry whenever the second homotopy group \(\pi_2(\mathcal{M})\) of the vacuum manifold \(\mathcal{M}\) is nontrivial.

The MACRO experiment~\cite{MACRO:2002jdv} places an upper limit on the monopole flux of \(2.8\times 10^{-16}~\mathrm{cm}^{-2}\,\mathrm{s}^{-1}\,\mathrm{sr}^{-1}\) for monopole masses \(m_M \sim 10^{14}~\mathrm{GeV}\). This flux limit corresponds to an upper bound on the comoving monopole number density (the yield) of \(Y_M^+ = n_M/s \simeq 10^{-27}\)~\cite{Kolb:1990vq}, where \(n_M\) denotes the monopole number density and \(s\) the entropy density. For observability, we adopt a conservative lower threshold \(Y_M^- \simeq 10^{-35}\), corresponding to a monopole flux of \(10^{-24}~\mathrm{cm}^{-2}\,\mathrm{s}^{-1}\,\mathrm{sr}^{-1}\), and the results shown in this section is computed accordingly. Future experiments, notably Hyper-Kamiokande~\cite{Dealtry:2019ldr} and DUNE~\cite{DUNE:2020ypp}, will significantly extend the search for non-relativistic magnetic monopoles. By recasting their sensitivity to monopole-catalyzed nucleon decay, these detectors can probe monopole fluxes down to \( 2.3 \times 10^{-23}~\mathrm{cm}^{-2}\,\mathrm{s}^{-1}\,\mathrm{sr}^{-1}\) and \(1.1 \times 10^{-22}~\mathrm{cm}^{-2}\,\mathrm{s}^{-1}\,\mathrm{sr}^{-1}\), respectively, thereby exploring a substantial portion of the theoretically interesting parameter space below the MACRO limit.

\begin{figure}[t!]
    \centering
     \includegraphics[width=0.88\linewidth]{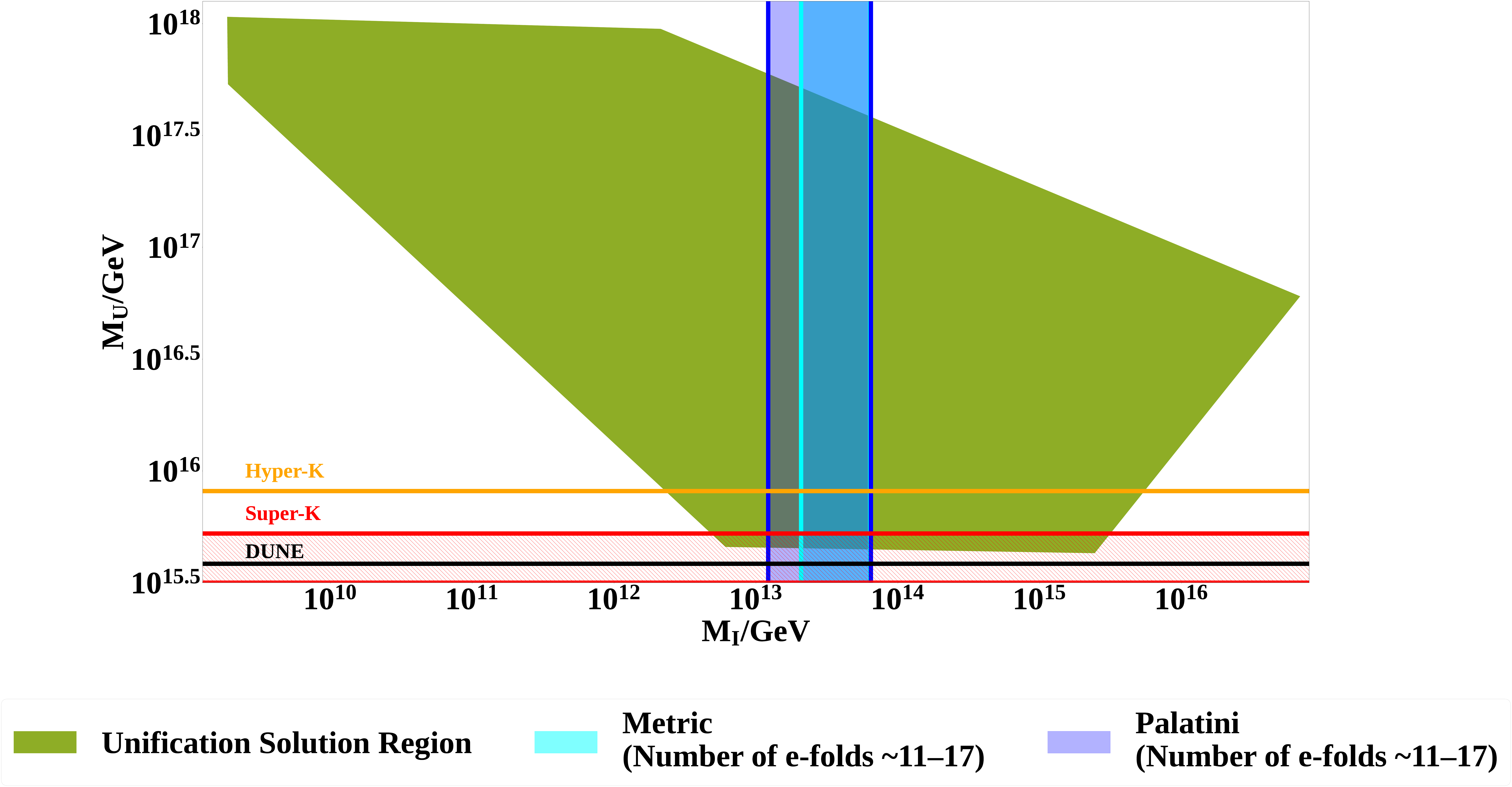}
    \includegraphics[width=0.88\linewidth]{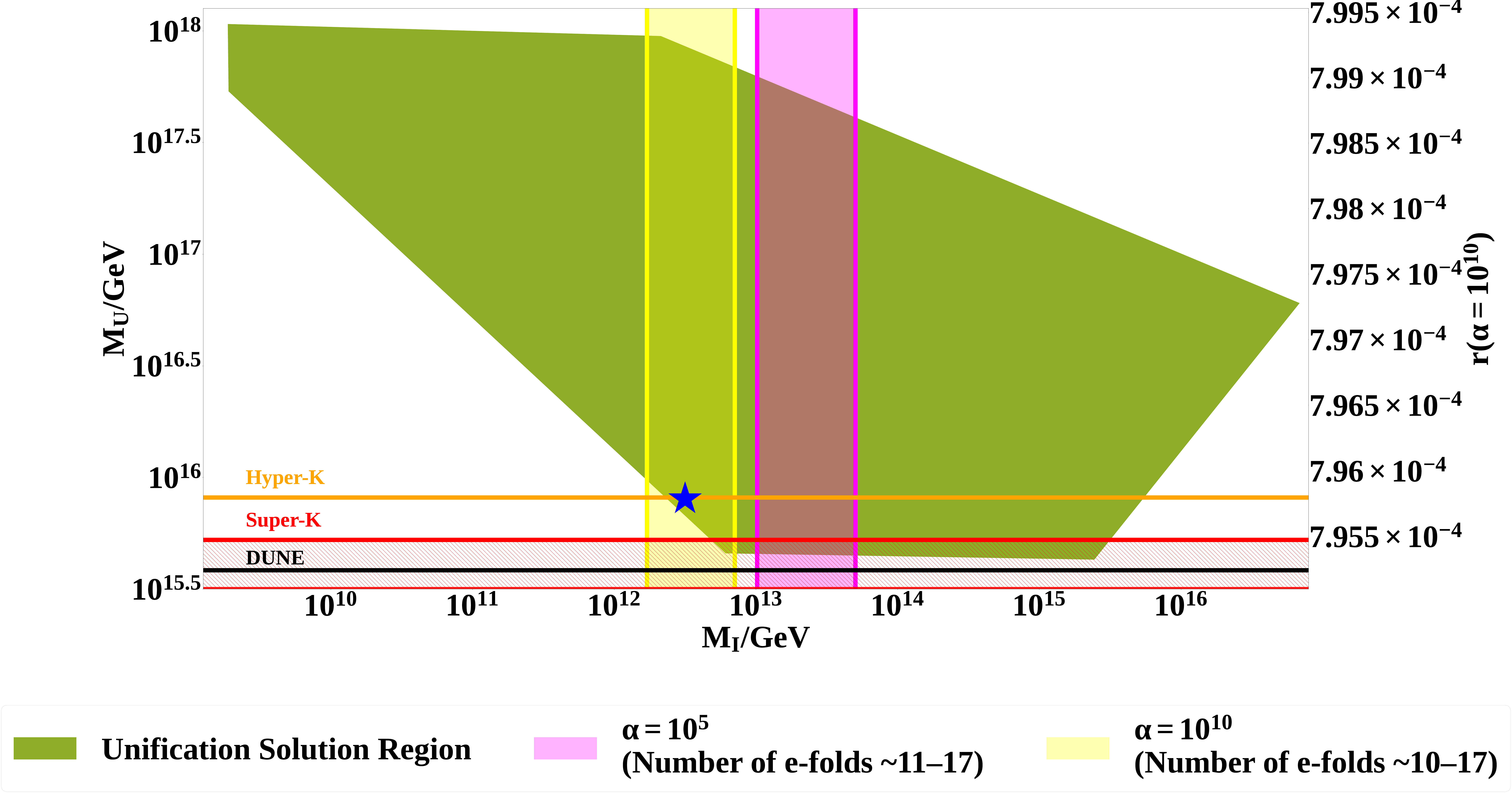}
\caption{\it Unification solutions (plot of $M_U$ vs.\ $M_I$) for the breaking chain
$SO(10)\xrightarrow{M_U} SU(4)_C\times SU(2)_L\times SU(2)_R \xrightarrow{M_I}
SU(3)_C\times SU(2)_L\times SU(2)_R\times U(1)_{B-L}\xrightarrow{M_{II}}
SU(3)_C\times SU(2)_L\times U(1)_Y$. The vertical shaded regions indicate the range of the intermediate scale $M_{I}$ associated with monopoles for the different cases shown. The \emph{bottom} panel corresponds to the \(\phi > M\) branch. The blue star corresponds to \((M_U/\text{GeV}, \, r) \sim (8.7 \times10^{15},\, 8\times10^{-4})\), and consistent with the ones shown in Figs.~\ref{fig:PalatiniR2_M},~\ref{fig:mi_1},~\ref{fig:monopoles1}, and~\ref{fig:monopoles3}. The horizontal shaded region for DUNE refers to the ruled-out values.} \label{fig:monopoles2}
\end{figure}

Additionally, Figs.~\ref{fig:monopoles1},~\ref{fig:monopoles2}, and~\ref{fig:monopoles3} present unification solution\footnote{Our analysis employs the following methodologies: two-loop $\beta$-coefficients are computed following Refs.~\cite{Maji:2022jzu, PhysRevD.25.581}; matching conditions are obtained from Refs.~\cite{Weinberg:1980wa, Hall:1980kf}; and the derived unification solutions are constrained by electroweak observables, guided by Refs.~\cite{PhysRevD.99.095008, Chakrabortty:2017mgi, Gialamas:2021enw}.} in the form of plots showing the Grand Unification scale as a function of the intermediate symmetry-breaking scale for two different breaking chains, which are depicted in Fig.~\ref{fig:Symmetry_breaking}, with the parameter \(R\) varying in the range \([1/2, 2]\). This parameter represents the ratio $R \equiv g_R / g_L$ of the \(SU(2)_R\) to \(SU(2)_L\) gauge couplings at the intermediate symmetry-breaking scale, accounting for threshold corrections and possible D-parity breaking effects~\cite{Chakrabortty:2017mgi}. The figure consists of two main panels. The upper panel displays results for both the metric and Palatini formulations of gravity, assuming a reheating equation-of-state parameter \(\omega_r = 0\). The lower panel focuses on the Palatini formalism augmented with an \(\mathcal{R}^2\) term, corresponding to the instant reheating branch where the inflaton field \(\phi\) evolves beyond its minimum value \(M\) (\(\phi > M\)). Additionally, in the bottom panel, we incorporate Eq.~\eqref{eq:MU} to illustrate the corresponding tensor-to-scalar ratio\footnote{In Sec.~\ref{Sec:SO(10)}, we present the complete inflationary computations for two choices of \(\omega_r\) and various reheat temperature scenarios. In the following analysis, we proceed with the case of \(\phi > M\), as this regime lies within the sensitivity of current cosmological observations---including the recent \emph{Planck}+ACT constraints~\cite{ACT:2025tim}---and is accessible to forthcoming experiments~\cite{SimonsObservatory:2018koc, Hazumi_2020}. } \(r\) values as a function of \(M_U\), providing a direct link to inflationary observables.

This integration shows that \(r\) exhibits a mild, nearly monotonic decrease with decreasing \(M_U\), ranging from approximately \(8 \times 10^{-4}\) at higher \(M_U \sim 10^{17.5}~\mathrm{GeV}\) to slightly lower values around \(7.95 \times 10^{-4}\) at \(M_U \sim 10^{16}~\mathrm{GeV}\). For instance, at the highlighted point marked by a blue star in the \emph{bottom} panels in Figs.~\ref{fig:monopoles1},~\ref{fig:monopoles2}, and~\ref{fig:monopoles3}, corresponding to \(M_U \approx 8.7 \times 10^{15}~\mathrm{GeV}\), \(r \approx 8 \times 10^{-4}\). This value of \(r\) is well below the current upper limits from CMB observations, and aligns well with the further tightened future observations, such as LiteBIRD~\cite{Hazumi_2020} and SO~\cite{SimonsObservatory:2018koc}.  This behavior reflects the underlying dynamics, where a lower unification scale leads to a compressed inflationary energy density and adjusted inflaton field values to maintain CMB normalization, thereby suppressing tensor modes through the slow-roll parameter \(\epsilon \approx r/16\).  Concurrently, the associated \(M_U \sim 8.7 \times 10^{15}~\mathrm{GeV}\) implies proton lifetimes in \(SO(10)\) models on the order of \(10^{34}\)--\(10^{35}\) years, based on the approximate relation in Eq.~\eqref{eq:MU}, which falls within the projected sensitivity of next-generation proton decay searches like Hyper-Kamiokande~\cite{Dealtry:2019ldr}.

This framework thus establishes a compelling complementarity between CMB observables and proton decay searches: a detection or a stringent limit on \(r\) at the \(10^{-3}\) level from future CMB experiments would constrain inflationary parameters tied to \(M_U\), while simultaneous advances in proton decay experiments could independently probe the same GUT scale, offering a multi-messenger test of the underlying \(SO(10)\) unification and inflationary paradigm presented in this work.

Notably, for \(\alpha = 10^{10}\), we present results anchored to \(M = 50\, m_{\rm Pl}\) as the benchmark, since variations up to \(M = 250\, m_{\rm Pl}\) induce only mild shifts in the curves due to the dominant \(\mathcal{R}^2\) curvature term, which stabilizes the potential against changes in the mass parameter, as seen in Tab.~\ref{tab:mon-infl-alpha1_M}.

Furthermore, the plots in this section incorporate key experimental constraints to delineate viable parameter space. Horizontal lines indicate the current exclusion limit on \(M_U\) from Super-Kamiokande~\cite{Super-Kamiokande:2020wjk} (\emph{red}) proton decay searches, which provides the constraint \(\tau (p \to e^+ \pi^0) \geq 2.4 \times10^{34}\) years, and hence implies that the unification scale \(M_U \gtrsim10^{15.7} \, \text{GeV}\). However, the future Hyper-Kamiokande experiment~\cite{Dealtry:2019ldr} (\emph{orange}) is projected to be sensitive to unification scales up to $
M_U \gtrsim 10^{15.9}~\text{GeV}$, probing proton lifetimes in the \(\tau(p \to e^+ \pi^0)\) channel up to \(10^{35}\) years at the \(3\sigma\) confidence level. We additionally show  constraints by DUNE (\emph{black})~\cite{DUNE:2020lwj, DUNE:2020mra, DUNE:2020txw, DUNE:2020ypp}  which probe  \(\tau(p\to K^+\, \bar{\nu}) \geq 10^{34} \) years.

The overlaid shaded bands for \(\alpha = 10^5\) (magenta, \(N \sim 11\text{--}17\) e-folds) and \(\alpha = 10^{10}\) (yellow, \(N \sim 10\text{--}17\) e-folds) delineate the viable intermediate scales in which the monopole flux after production satisfies the MACRO~\cite{MACRO:2002jdv} bound and the observability threshold, via partial dilution during the inflationary e-folds following symmetry breaking at \(M_I\). Physically, larger \(\alpha\) enhances the quadratic curvature contribution, flattening the potential and yielding lower, more stable \(M_U\) across \(r\), with extended e-fold dilution windows that favor monopole detectability in future experiments. In contrast, smaller \(\alpha\) allows higher \(M_U\) and produces peaked behavior due to Coleman–Weinberg dominance before transitioning to \(\mathcal{R}^2\) flattening. Altogether, this analysis connects GUT phenomenology with CMB anisotropies and potentially testable bounds on the tensor-to-scalar ratio \(r\)~\cite{Enckell:2018hmo,Tenkanen:2019jiq,Lykkas:2021vax,Eadkhong:2023ozb}, thereby linking high-scale symmetry breaking to early-Universe reheating and monopole cosmology.

Tab.~\ref{tab:mon-infl} and the subsequent tables list the number of $e$-folds required to achieve the monopole yields \(Y_M^+\) and \(Y_M^-\), together with the corresponding values of the inflaton field, the Hubble parameter, and the symmetry-breaking scale for representative choices of \(\xi\) and \(M\) that yield successful inflation.

The results in Tab.~\ref{tab:mon-infl} and in the unification plots (Figs.~\ref{fig:monopoles1} and \ref{fig:monopoles2}, \textbf{top panels}) indicate that, for a reheating equation-of-state \(\omega_r = 0\), there exists a narrow but physically interesting window of intermediate scales \(M_I\), where monopoles are only \emph{partially} diluted by inflation. Within this window, they can simultaneously satisfy gauge coupling unification and proton-decay constraints. The vertical shaded bands in the unification plots denote the ranges of \(M_I\) corresponding to the targeted monopole yields (cyan for the metric choice, blue for the Palatini choice); these bands overlap nontrivially with the green unification region and, importantly, lie above the Super–Kamiokande~\cite{Super-Kamiokande:2020wjk} exclusion limit and, in some cases, within the projected Hyper–Kamiokande~\cite{Dealtry:2019ldr} sensitivity. This overlap demonstrates that, for physically motivated choices of the non-minimal coupling \(\xi\) and the model mass parameter \(M\), the scenario naturally produces intermediate-scale monopoles that are neither completely diluted by inflation nor in conflict with current proton-decay limits.

Upon closer inspection, Tab.~\ref{tab:mon-infl} reveals two systematic trends that determine the fate of monopoles. First, at fixed \(\xi\) and increasing \(M\) (rows with \(M/m_{\rm Pl}=50\to100\to500\)), the Hubble scale during the relevant stage of inflation \(H_I\), the inferred symmetry-breaking scale \(M_I\), and the number of $e$-foldings experienced by the monopoles \(N\) all \emph{decrease}. This behavior aligns with the scaling implicit in the breaking-scale estimate, Eq.~\eqref{eq:breaking-scale}. Larger $M$ shifts the inflaton field values \(\phi_I\) to larger absolute values while lowering the inflationary energy density (and hence $H_I$) for the same CMB normalization; the net effect is a reduction in both $M_I$ and $N$. Physically, this means that larger model mass parameters delay the intermediate transition (leading to a lower $M_I$) and thus decrease the inflationary dilution of monopoles.

Second, at fixed $M$ and increasing \(|\xi|\) (columns from \(\xi=-0.001\) to \(\xi=-0.004\)), the Hubble parameter, \(M_I\), and \(N\) likewise decrease. A stronger non-minimal coupling flattens the effective Einstein-frame potential, lowering the inflationary energy scale, which reduces the correlation length and the effective symmetry-breaking scale computed via Eq.~\eqref{eq:breaking-scale}. Consequently, monopoles formed at the intermediate transition experience fewer $e$-foldings. This explains why larger \(|\xi|\) entries in Tab.~\ref{tab:mon-infl} approach the lower boundary of the partial-inflation window; for sufficiently large \(|\xi|\) and $M$, the scenario may exit the partially-inflated regime entirely, potentially overproducing monopoles or requiring additional adjustments to the model.

The existence of a narrow ``partial-inflation'' window has significant physical implications. It indicates that early-Universe dynamics can leave a remnant population of intermediate–scale monopoles that encode information about both the GUT symmetry-breaking sector and the inflationary history. An observable monopole flux would thus serve as a twofold probe: it would pinpoint a specific range of intermediate breaking scales \(M_I\) (and hence particular $SO(10)$ representations and VEVs responsible for the breaking) while simultaneously constraining the effective energy scale during the relevant epoch and the amount of post-formation inflationary dilution. Conversely, current and projected proton-decay bounds, combined with unification requirements, carve out a limited viable parameter space; non-observation in forthcoming experiments would meaningfully restrict allowed combinations of the non-minimal coupling, the model mass scale, and the Higgs–inflaton portal parameters that control the instability of the GUT multiplet \(\varUpsilon_S\).

Figure~\ref{fig:mi_0} illustrates the unification scale $M_U$ as a function of the tensor-to-scalar ratio $r$ in the Palatini formulation of gravity, computed for various choices of the model mass parameter $M = \{50, 100, 500\} \, m_{\rm Pl}$ and reheat temperatures $T_{\rm reh} = \{10^7, 10^{10}\}$~GeV, assuming a Coleman-Weinberg inflaton potential with non-minimal coupling to gravity.  For fixed $T_{\rm reh}$, increasing $M$ shifts the curves upward, yielding higher $M_U$ values across the $r$ range, as larger $M$ enhances the overall mass normalization in the potential and VEV, thereby elevating the GUT-breaking scale while maintaining consistency with slow-roll parameters. Conversely, at fixed $M$, higher $T_{\rm reh}$ lowers $M_U$ very modestly, originating from the altered post-inflationary expansion history that affects the e-fold number $N_*$ between horizon exit and reheating, requiring a compensatory adjustment in $\phi_*$ that diminishes the portal-induced VEV. The horizontal dashed lines mark experimental constraints: both the Super-Kamiokande~\cite{Super-Kamiokande:2020wjk} and Hyper-Kamiokande~\cite{Dealtry:2019ldr} exclusion limits from proton decay searches. Additionally, the vertical dot-dashed lines represent the cosmological constraints on the tensor-to-scalar ratio \( r \), translated into an effective $M_U$ bound via inflationary correlations, indicating that the viable parameter space lies within these constraints. Overall, the figure shows a broad compatibility range for \( r \), where $M_U$ stays within the observational limits for all considered $M$ and $T_{\rm reh}$, with larger $M$ favoring higher unification scales that could be tested by future experiments, thereby connecting GUT physics to observable CMB anisotropies.

Tab.~\ref{tab:mon-infl-alpha2} summarizes the numerical outcomes when a Starobinsky-type \(\mathcal{R}^2\) term is included in the \emph{Palatini} formulation (parameterized by \(\alpha\)). As in the Coleman–Weinberg-only case, the Palatini \(\mathcal{R}^2\) model admits a narrow but phenomenologically interesting window of $e$-foldings for which monopoles are only partially inflated. Across the entries in Tab.~\ref{tab:mon-infl-alpha2}, the number of $e$-foldings experienced by monopoles that reproduce the MACRO-consistent benchmark (\(Y_M^+\)) lies close to \(N_+\simeq 10\)–\(11\), while the adopted observability threshold (\(Y_M^-\)) corresponds to \(N_-\simeq 16\)–\(17\).  

The new control parameter \(\alpha\) has a clear quantitative effect on both the inflationary scale and the inferred intermediate symmetry-breaking scale \(M_I\). For the relatively small choice \(\alpha=10^5\), the table shows Hubble rates in the range \(H\sim\mathcal{O}(5)\times10^{13}\,\mathrm{GeV}\) and inferred \(M_I\) up to \(\sim 10^{14}\,\mathrm{GeV}\) (e.g., \(M_{I+}\simeq 7.05\times10^{13}\,\mathrm{GeV}\) for \(\xi=0.001, M=250\)). In contrast, for the much larger value \(\alpha=10^{10}\), the Hubble rate is reduced to a nearly constant value \(H\sim 0.69\times10^{13}\,\mathrm{GeV}\) across the entries, and the corresponding \(M_I\) values collapse to the \(\sim 10^{13}\,\mathrm{GeV}\) scale (e.g., \(M_{I+}\simeq 1.12\times10^{13}\,\mathrm{GeV}\) for \(\xi=0.001, M=250\)). Physically, increasing \(\alpha\) in the Palatini setup suppresses the effective inflationary energy scale in the Einstein-frame dynamics, and this reduction translates into smaller symmetry-breaking scales, as computed via Eq.~\eqref{eq:breaking-scale}.

The dependence on the non-minimal coupling \(\xi\) and on the mass parameter \(M\) is qualitatively similar to the Coleman–Weinberg-only case, but exhibits Palatini-specific modulations. At fixed \(\alpha\), increasing \(M\) tends to shift the inflaton field values \(\phi_I\) and can slightly raise both \(H_I\) and \(M_I\) (for example, for \(\alpha=10^5, \xi=0.001\), the entries with \(M=50\to250\) show \(H_+\) increasing modestly from \(\sim 4.31\) to \(\sim 4.32 \times 10^{13}\,\mathrm{GeV}\) and \(M_{I+}\) from \(\sim 7.03\) to \(\sim 7.05\times 10^{13}\,\mathrm{GeV}\)). However, for very large \(\alpha\) (e.g., \(10^{10}\)), this sensitivity to \(M\) is almost entirely erased: both \(H_I\) and \(M_I\) become nearly independent of \(M\), remaining essentially constant within numerical precision. The non-minimal coupling \(\xi\) produces only modest shifts in \(H_I\) and \(M_I\). Two numerical features stand out in Tab.~\ref{tab:mon-infl-alpha2}:
\begin{figure}[t!]
    \centering
     \includegraphics[width=0.88\linewidth]{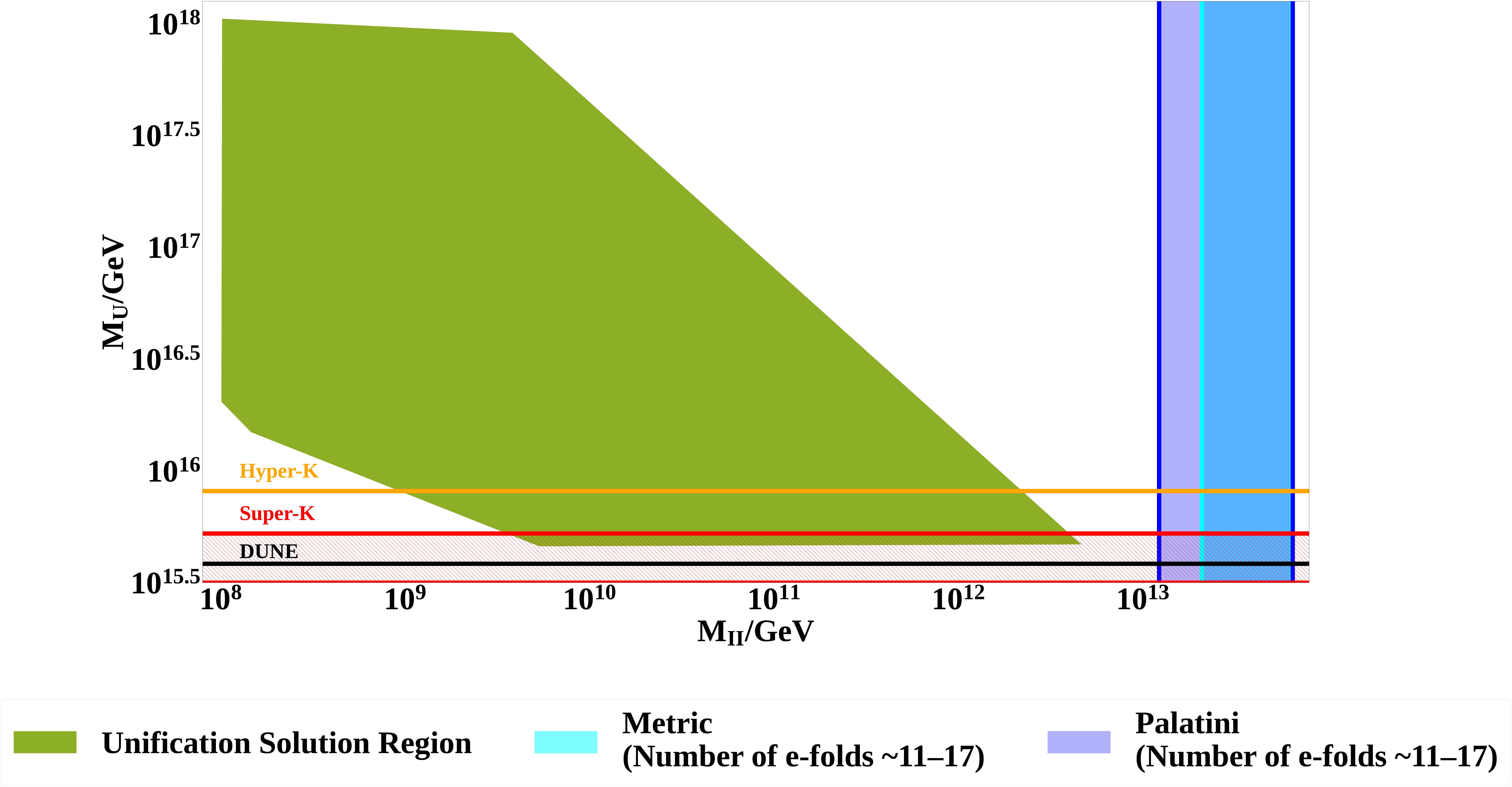}
          \includegraphics[width=0.88\linewidth]{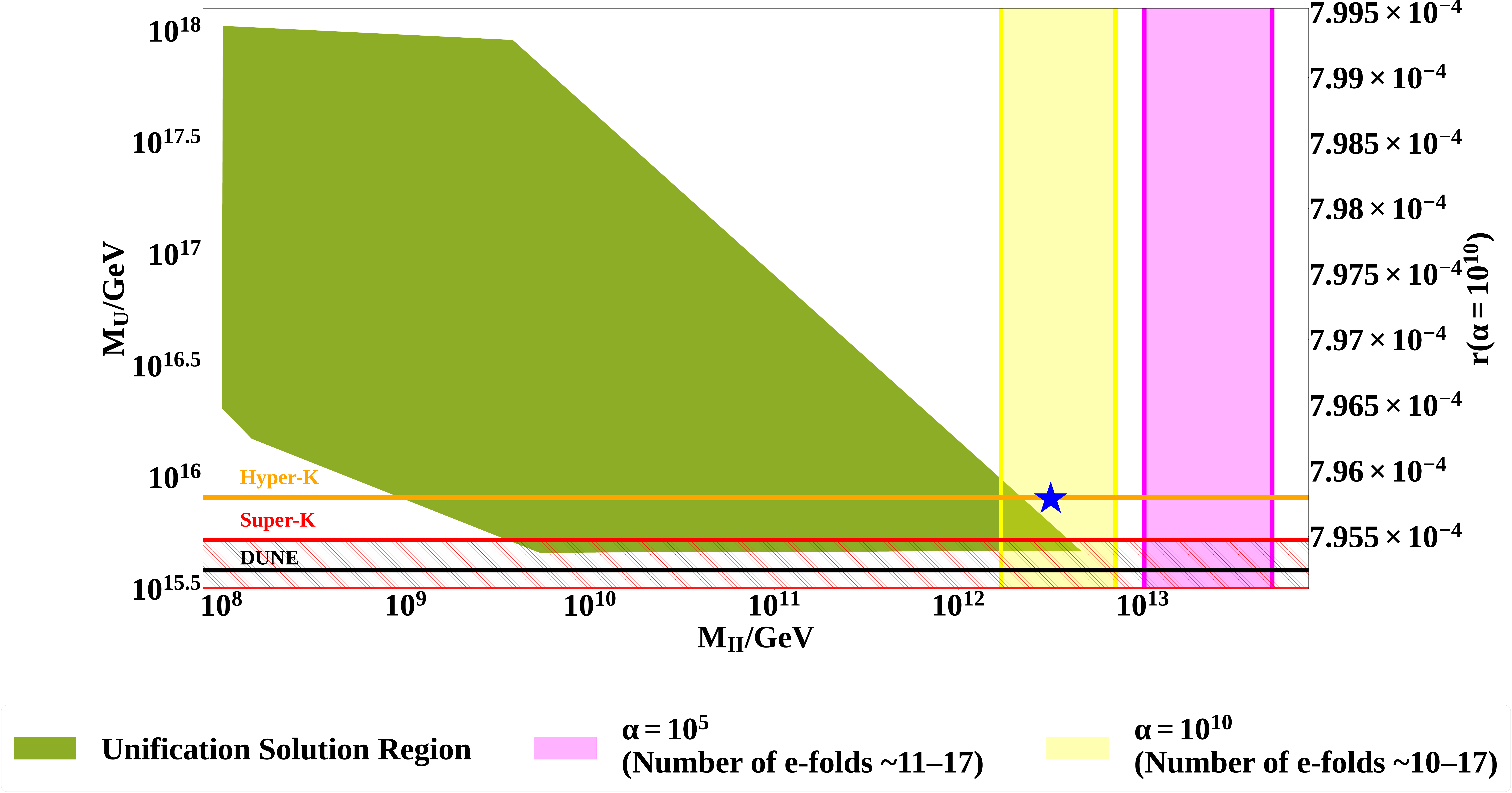}
\caption{ \it Unification solutions (plot of $M_U$ vs.\ $M_I$) for the breaking chain $SO(10)\xrightarrow{M_U} SU(4)_C\times SU(2)_L\times SU(2)_R \xrightarrow{M_I} SU(3)_C\times SU(2)_L\times SU(2)_R\times U(1)_{B-L}\xrightarrow{M_{II}} SU(3)_C\times SU(2)_L\times U(1)_Y$.  The vertical shaded regions indicate the range of the intermediate scale $M_{II}$ associated with monopoles for the different cases shown. The \emph{bottom} panel corresponds to the \(\phi > M\) branch. The blue star corresponds to \((M_U/\text{GeV}, \, r) \sim (8.7 \times10^{15},\, 8\times10^{-4})\), and consistent with the ones shown in Figs.~\ref{fig:PalatiniR2_M},~\ref{fig:mi_1},~\ref{fig:monopoles1}, and~\ref{fig:monopoles2}. The horizontal shaded region for DUNE indicates the ruled-out values.} \label{fig:monopoles3}
\end{figure}
\begin{enumerate}
  \item For \(\alpha = 10^5\), the largest variations in \(H\) and \(M_I\) occur with changes in \(M\) and \(\xi\); typical \(M_{I+}\) values range from a few \(\times 10^{13}\,\mathrm{GeV}\) up to approximately \(10^{14}\,\mathrm{GeV}\).
 
  \item For \(\alpha=10^{10}\), the inflationary scale is reduced to \(H\sim 0.69\times10^{13}\,\mathrm{GeV}\), and the inferred \(M_I\) values cluster tightly around \(\sim 0.7\text{--}1.12\times10^{13}\,\mathrm{GeV}\). In this case, the monopole-producing transition occurs at slightly lower intermediate scales, yielding \(N_+\simeq 10.3\) and \(N_-\simeq 16.5\), still within the partial-inflation window but with a systematically smaller spread in \(M_I\).
\end{enumerate}

Taken together, these patterns demonstrate that \(\alpha\) serves as an effective control parameter: small-to-moderate values of \(\alpha\) sustain a high inflationary energy scale, allowing larger intermediate symmetry-breaking scales \(M_I\) and greater sensitivity to the model parameters, whereas very large \(\alpha\) suppresses the inflationary energy and compresses the viable range of \(M_I\).

Tab.~\ref{tab:mon-infl-alpha1} presents the same diagnostics as Tab.~\ref{tab:mon-infl-alpha2}, but for a radiation-like reheating equation of state, \(\omega_r = 1/3\), with the inflaton evolving along the \(\phi < M\) branch. The results are illustrated in Fig.~\ref{fig:monopoles1} and Fig.~\ref{fig:monopoles2} (\textbf{bottom panels}). The key observations for this scenario can be summarized as follows:
\begin{itemize}
\item Consistency of the partial-inflation window: The characteristic $e$-fold ranges remain consistently $N_+\simeq 10$--$11$ and $N_-\simeq 16$--$17$ across various parameter choices in our model, i.e., the MACRO-consistent and adopted-observability benchmarks are essentially unchanged compared to the $\omega_r=0$ results in Tab.~\ref{tab:mon-infl-alpha2}.
  
  \item Small quantitative shifts. For moderate $\alpha$ (e.g., $\alpha=10^5$), the Hubble rates $H$ and the inferred intermediate scales $M_I$ are slightly reduced relative to the $\omega_r=0$ case, producing negligible changes in $N$. The exponential sensitivity of the monopole yield remains largely controlled by $N$, which continues to lie within the same $\mathcal{O}(10)$ window.
  
  \item Large-$\alpha$ regime insensitive. For very large $\alpha$ (e.g., $\alpha=10^{10}$), both $H$ and $M_I$ remain tightly clustered and essentially independent of $M$ and $\xi$, so the change in the reheating equation of state has minimal numerical impact.
  
  \item Physical origin of the shifts. The modest reductions in $H$ and $M_I$ for $\omega_r=1/3$ arise from the altered relation between the reheating time, the inflationary dilution factor, and the mapping between CMB normalization and the inflaton value $\phi_I$ at the monopole-forming epoch. Since $N$ is only slightly affected, the exponential suppression of the monopole yield remains governed by the same approximate $N$-window.
\end{itemize}
In short, adopting $\omega_r=1/3$ in the $\phi<M$ branch does not alter the existence of the partial-inflation window. It induces only small, parameter-dependent shifts in $H$ and $M_I$, most noticeable for moderate $\alpha$, while leaving the main phenomenological conclusion unchanged: a non-empty region of parameter space persists where partially-inflated monopoles remain compatible with both gauge coupling unification and proton-decay constraints.
\begin{table}[t!]
\begin{center}
\begin{tabular}{| c | c | c | c | c | c | c | c | c | c | c |}
\hline
$\xi$ & $\alpha$ & $\frac{M}{m_{\rm Pl}}$ & $\phi_+/m_{\rm Pl}$ & $\phi_-/m_{\rm Pl}$ & \thead{$H_+$\\($10^{13}$ GeV)} & \thead{$H_-$\\($10^{13}$ GeV)} & \thead{$M_{I+}$\\($10^{13}$ GeV)} & \thead{$M_{I-}$\\($10^{13}$ GeV)} & $N_+$ & $N_-$ \\
\hline
\multirow{4}{*}{\rotatebox[origin=c]{90}{0.001}} & \multirow{2}{*}{\(10^5\)} & 50 & 39.95 & 37.60 & 4.31 & 5.74 & 7.03 & 9.95  & 10.8 & 17.2 \\
& & 250 & 199.75 & 188.00 & 4.32 & 5.76 & 7.05 & 9.97 & 10.8 & 17.2 \\
\cline{2-11}
& \multirow{2}{*}{\(10^{10}\)} & 50 & 40.26 &  37.93 & 0.69 & 0.70 & 1.12 & 1.20 & 10.3 & 16.5 \\
& & 250 & 201.31 & 189.70 & 0.69 & 0.70 & 1.12 & 1.20 & 10.3 & 16.4 \\
\hline
\multirow{4}{*}{\rotatebox[origin=c]{90}{0.0001}} & \multirow{2}{*}{\(10^5\)} & 50 & 46.69 &  45.88 & 4.51 & 5.73 & 6.29 & 8.13 & 10.9 & 17.2 \\
& & 250 & 233.46 & 229.43 & 4.62 & 5.87 & 6.45 & 8.33 & 10.9 & 17.2 \\
\cline{2-11}
& \multirow{2}{*}{\(10^{10}\)} & 50 & 46.81 & 46.00 & 0.69 & 0.70 & 0.70 & 0.99 & 10.3 & 16.5 \\
& & 250 & 234.02 & 230.02 & 0.69 & 0.70 & 0.97 & 0.99 & 10.3 & 16.4 \\
\hline
\multirow{4}{*}{\rotatebox[origin=c]{90}{\(10^{-5}\)}} & \multirow{2}{*}{\(10^5\)} & 50 & 48.94 & 48.68 & 4.53 & 5.68 & 5.55 & 7.61 & 10.9 & 17.2 \\
& & 250 & 244.70 & 243.39 & 4.53 & 5.69 & 6.03 & 7.61 & 10.9 & 17.2 \\
\cline{2-11}
& \multirow{2}{*}{\(10^{10}\)} & 50 & 48.98 & 48.72 & 0.69 & 0.70 & 0.92 & 0.93 & 10.3 & 16.5 \\
& & 250 &  244.88 & 243.57 & 0.69 & 0.70 & 0.92 & 0.93 & 10.3 & 16.5 \\
\hline
\end{tabular}
\caption{\it Same as Tab.~\ref{tab:mon-infl}, but now for the Palatini  \(\mathcal{R}^2\) case. \(T_{\rm reh} = 10^7\) GeV.}
\label{tab:mon-infl-alpha2}
\end{center}
\end{table}

Turning to the unification plots, the green polygon representing the allowed \((M_I, M_U)\) solutions intersects the vertical shaded bands (which denote the monopole-producing intermediate scales) in regions where the unification scale \(M_U\) safely exceeds the limits set by Super-Kamiokande~\cite{Super-Kamiokande:2020wjk} and Hyper-Kamiokande~\cite{Dealtry:2019ldr}. This is an important phenomenological result: the same intermediate-scale physics that produces partially inflated monopoles in our Coleman–Weinberg plus non-minimal framework can also be linked to unification scenarios that are testable in next-generation proton decay experiments. In particular, the plots show that, for the adopted ranges of \(R\) and particle content, the viable \(M_I\) values for observable monopoles cluster around \(10^{13}\)–\(10^{14}\,\mathrm{GeV}\) (corresponding to the cyan and blue bands in Figs.~\ref{fig:monopoles1} and~\ref{fig:monopoles2}). This region coincides with the range, where the monopole \(e\)-fold number lies within the \(\mathcal{O}(10)\) window required for partial inflation (cf.\ Tab.~\ref{tab:mon-infl}).

We now consider the regime $\phi > M$. Tab.~\ref{tab:mon-infl-alpha1_M} and the corresponding unification plot for the $M_{II}$ breaking (Fig.~\ref{fig:monopoles3}) summarize the key results in this domain and illustrate how allowing the inflaton to traverse beyond $M$ shifts the monopole-producing scale.
\begin{table}[t!]
\begin{center}
\begin{tabular}{| c | c | c | c | c | c | c | c | c | c | c |}
\hline
$\xi$ & $\alpha$ & $\frac{M}{m_{\rm Pl}}$ & $\phi_+/m_{\rm Pl}$ & $\phi_-/m_{\rm Pl}$ & \thead{$H_+$\\($10^{13}$ GeV)} & \thead{$H_-$\\($10^{13}$ GeV)} & \thead{$M_{I+}$\\($10^{13}$ GeV)} & \thead{$M_{I-}$\\($10^{13}$ GeV)} & $N_+$ & $N_-$ \\
\hline
\multirow{4}{*}{\rotatebox[origin=c]{90}{0.001}} & \multirow{2}{*}{\(10^5\)} & 50 & 39.96  & 37.61 & 3.66 & 4.88 & 5.97 & 8.46 & 10.8 & 17.2 \\
& & 250 & 199.78 & 188.03 & 3.67 & 4.88 & 5.98 & 8.46 & 10.8 & 17.2 \\
\cline{2-11}
& \multirow{2}{*}{\(10^{10}\)} & 50 & 40.22 & 37.90 & 0.69 & 0.70 & 1.12 & 1.20 & 10.4 & 16.5 \\
& & 250 & 201.10 & 189.49 & 0.69 & 0.69 & 1.12 & 1.20 & 10.4 & 16.5 \\
\hline
\multirow{4}{*}{\rotatebox[origin=c]{90}{0.0001}} & \multirow{2}{*}{\(10^5\)} & 50 & 46.69 & 45.89 & 4.00 & 5.09 & 5.59 & 7.23 & 10.8 & 17.2 \\
& & 250 & 233.46 & 229.43 & 4.11 & 5.22 & 5.73 & 7.40 & 10.9 & 17.2 \\
\cline{2-11}
& \multirow{2}{*}{\(10^{10}\)} & 50 & 46.79 & 45.99 & 0.69 & 0.70 & 0.96 & 0.99 & 10.4 & 16.5 \\
& & 250 & 233.97 & 220.95 & 0.69 & 0.70 & 0.96 & 0.99 & 10.4 & 16.5 \\
\hline
\multirow{4}{*}{\rotatebox[origin=c]{90}{\(10^{-5}\)}} & \multirow{2}{*}{\(10^5\)} & 50 & 48.94 & 48.68 & 4.10 & 5.15 & 5.46 & 6.89 & 10.9 & 17.2 \\
& & 250 & 244.70 &  243.39 & 4.10 & 5.15    & 5.46 & 6.89 & 10.9 & 17.2 \\
\cline{2-11}
& \multirow{2}{*}{\(10^{10}\)} & 50 & 48.97 & 48.71 & 0.69 & 0.70 & 0.92 & 0.93 & 10.4 & 16.6 \\
& & 250 & 244.86 & 243.56 & 0.69 & 0.70 & 0.92 & 0.93 & 10.4 & 16.5 \\
\hline
\end{tabular}
\caption{\it Same as Tab.~\ref{tab:mon-infl}, but now for the Palatini  \(\mathcal{R}^2\) case. We fixed \(\omega_r = 1/3\) and \(\phi <M\).}
\label{tab:mon-infl-alpha1}
\end{center}
\end{table}

The main observation is that the inferred intermediate scale $M_I$ is driven significantly lower than in the $\phi < M$ case, sometimes down to $\mathcal{O}(10^{12}\,\mathrm{GeV})$ for certain parameter choices. This follows directly from the breaking-scale estimate in Eq.~\eqref{eq:breaking-scale}: in the $\phi > M$ regime, the larger $\phi_I$ suppresses $M_I$ for otherwise comparable $H_I$ and $M$, and very large $\phi$ entries in Tab.~\ref{tab:mon-infl-alpha1_M} make this behavior explicit.

Despite the lower $M_I$, the characteristic partial-inflation $e$-fold numbers remain essentially unchanged: $N_+\simeq 10$--$11$ (MACRO-consistent) and $N_-\simeq 16$--$17$ (adopted observability threshold). Thus, the mapping from monopole yield to the required $N$-window is robust against moving into the $\phi > M$ regime.

The dependence on $\alpha$, $\xi$, and $M$ is qualitatively similar to the $\phi < M$ case. Moderate or small $\alpha$ (e.g., $10^5$) yields a larger spread in $H_I$ and $M_I$ with parameter variation, whereas very large $\alpha$ (e.g., $10^{10}$) compresses $H_I$ and $M_I$ to nearly constant values. Varying $\xi$ or $M$ produces only modest adjustments in $N$ in this regime.

From a phenomenological perspective, the lower $M_I$ bands still intersect the green unification region in parts of parameter space. Consequently, the $\phi > M$ case opens a window for monopole production at lower intermediate scales that remain compatible with gauge coupling unification and proton-decay bounds. Practically, the $\phi > M$ regime provides a simple mechanism to shift monopole-producing transitions to lower energies (via the $1/\phi_I$ factor) without substantially altering the required $e$-fold numbers for partial inflation.

Importantly, Fig.~\ref{fig:monopoles3} displays the $M_{II}$ unification solutions: in the $\phi > M$, $\omega_r=1/3$ Palatini runs, the downward shift of the monopole-producing scale not only moves the $M_I$ band into overlap with the unification polygon (as seen in earlier $M_I$ plots, e.g., Figs.~\ref{fig:monopoles1} and~\ref{fig:monopoles2}, bottom panels), but also positions the $M_{II}$ band itself within the observationally relevant region, as shown in Fig.~\ref{fig:monopoles3} (bottom panel).
\begin{table}[t!]
\begin{center}
\begin{tabular}{| c | c | c | c | c | c | c | c | c | c | c |}
\hline
$\xi$ & $\alpha$ & $\frac{M}{m_{\rm Pl}}$ & $\phi_+/m_{\rm Pl}$ & $\phi_-/m_{\rm Pl}$ & \thead{$H_+$\\($10^{13}$ GeV)} & \thead{$H_-$\\($10^{13}$ GeV)} & \thead{$M_{I+}$\\($10^{13}$ GeV)} & \thead{$M_{I-}$\\($10^{13}$ GeV)} & $N_+$ & $N_-$ \\
\hline
\multirow{4}{*}{\rotatebox[origin=c]{90}{0.1}} & \multirow{2}{*}{\(10^5\)} & 50 & 275.70 & 402.82 & 4.31 & 4.83 & 1.02 & 0.78 & 10.9 & 17.2 \\
& & 250 & 1377.75 & 2013.42 & 4.30 & 4.83 & 1.02 & 0.78 & 10.9 & 17.2 \\
\cline{2-11}
& \multirow{2}{*}{\(10^{10}\)} & 50 & 260.69 & 382.96 & 0.69 & 0.70 & 0.17 & 0.12 & 10.4 & 16.5 \\
& & 250 & 1305.62 &  1919.19 & 0.69 & 0.70 & 0.17 & 0.12 & 10.4 & 16.6 \\
\hline
\multirow{4}{*}{\rotatebox[origin=c]{90}{0.01}} & \multirow{2}{*}{\(10^5\)} & 50 & 91.29 & 104.38 & 4.16 & 4.81 & 2.97 & 3.00 & 10.9 & 17.2 \\
& & 250 & 456.64 & 521.95 & 4.18 & 4.82 & 2.98 & 3.01 & 10.9 & 17.2 \\
\cline{2-11}
& \multirow{2}{*}{\(10^{10}\)} & 50 & 89.59 & 102.60 & 0.69 & 0.70 & 0.50 & 0.44 & 10.4 & 16.5 \\
& & 250 & 448.28 & 513.07 & 0.69 & 0.70 & 0.50 & 0.44 & 10.4 & 16.6 \\
\hline
\multirow{4}{*}{\rotatebox[origin=c]{90}{0.001}} & \multirow{2}{*}{\(10^5\)} & 50 & 61.33 & 64.36 & 4.14 & 4.98 & 4.40 & 5.04 & 10.9 & 17.2 \\
& & 250 & 306.70 & 321.80 & 4.11 & 4.94 & 4.37 & 5.00 & 10.9 & 17.2 \\
\cline{2-11}
& \multirow{2}{*}{\(10^{10}\)} & 50 & 60.94 & 63.95 & 0.69 & 0.70 & 0.74 & 0.71 & 10.4 & 16.5 \\
& & 250 & 304.74 & 319.82 & 0.69 & 0.70 & 0.74 & 0.71 & 10.4 & 16.6 \\
\hline
\end{tabular}
\caption{\it Same as Tab.~\ref{tab:mon-infl}, but now for the Palatini  \(\mathcal{R}^2\) case. We fixed \(\omega_r = 1/3\) and \(\phi > M\).}
\label{tab:mon-infl-alpha1_M}
\end{center}
\end{table}

Physically, this implies that the same mechanism which lowers $M_I$ in the $\phi > M$ regime also shifts the subsequent breaking scale $M_{II}$ into a parameter-space window that is simultaneously compatible with gauge coupling unification and close to the Hyper–Kamiokande sensitivity. In short, for $\phi > M$ we obtain a novel twofold phenomenological opportunity: observable monopole-producing transitions for the $M_I$ case \emph{and} for the $M_{II}$ case, as shown in Fig.~\ref{fig:monopoles3}, a reach that the other scenarios did not achieve.

Figure~\ref{fig:mi_1} depicts the unification scale $M_U$ versus the tensor-to-scalar ratio $r$ in the Palatini $\mathcal{R}^2$ framework, with a fixed model mass parameter $M = 50 \, m_{\rm Pl}$ and reheating equation-of-state parameter $\omega_r = 1/3$, showcasing curves for various Starobinsky coupling strengths $\alpha = 10^5, 10^7, 10^9, 10^{10}$ for  the  $\phi > M$ inflaton field branch.  Experimental bounds are overlaid as horizontal lines for  the Super-Kamiokande~\cite{Super-Kamiokande:2020wjk}, future Hyper-Kamiokande observations~\cite{Dealtry:2019ldr},   probes by DUNE~\cite{DUNE:2020lwj, DUNE:2020mra, DUNE:2020txw, DUNE:2020ypp}. However, the vertical lines provide the cosmological constraints by LiteBIRD and \textit{Planck}~\cite{Planck:2018jri, Hazumi_2020}, highlighting that small-$\alpha$ scenarios access high $M_U$ values compatible with current limits but potentially ruled out by tighter CMB constraints on $r$, whereas large-$\alpha$ cases are within the CMB region that can be even testable by future experiments~\cite{SimonsObservatory:2018koc, Hazumi_2020}. This plot underscores the $\alpha$-dependent tuning of the inflationary scale that shapes GUT phenomenology, enabling partial monopole dilution in the $N \sim 10-17$ e-fold window for intermediate $M_I$ scales linked to $M_U$, with the $\phi > M$ branch enabling access to lower unification scales that enhance overlap with unification polygons and next-generation observability (cf. Figs.~\ref{fig:monopoles1},~\ref{fig:monopoles2}, and~\ref{fig:monopoles3}), thus providing a mechanism to probe intertwined high-scale symmetry breaking and early-Universe dynamics. Notably, reaching low $M_U$ with viable $r$ values is only possible when incorporating the $\mathcal{R}^2$ term; in both the metric and Palatini formulations, without it, as shown in Fig.~\ref{fig:mi_0}, this could not be achieved.

An additional remark on the role of $M_{II}$ is useful because this second intermediate scale governs the subsequent breaking
\[
SU(3)_C\times SU(2)_L\times SU(2)_R\times U(1)_{B-L}\xrightarrow{M_{II}} SU(3)_C\times SU(2)_L\times U(1)_Y,
\]
and therefore sets the mass scale and formation epoch of any defects produced at that stage. $M_{II}$ is important for two reasons relevant to monopole phenomenology. First, if the $M_{II}$ transition produces one-dimensional defects (cosmic strings), monopoles formed earlier at $M_I$ can become attached to those strings; monopole–antimonopole pairs connected by strings experience a strong string tension that drives rapid annihilation or confinement, greatly reducing the surviving monopole flux compared to the case with no subsequent string-forming transition. Second, even in the absence of strings, the value of $M_{II}$ affects the renormalization-group running and thus the allowed unification polygon in the $(M_I, M_U)$ plane; shifting $M_{II}$ indirectly changes which $M_I$ bands overlap with the unification region and the proton-decay sensitivity lines. In short, $M_{II}$ serves both as a direct dynamical handle on post-monopole evolution (annihilation or confinement) and as an indirect organizer of the parameter-space intersections that determine whether partially inflated monopoles remain phenomenologically viable and potentially observable.
\medskip
\section{Discussion}
\label{Sec:Discussion}
In this work, we have presented a comprehensive, UV-complete framework for cosmological inflation within an $SO(10)$ Grand Unified Theory. By unifying cosmic inflation with the GUT Higgs sector and augmenting the gravitational action with a $\mathcal{R}^2$ term, we have constructed a viable cosmological scenario in which the dynamics of the early universe, particle physics grand unification, and the production of topological defects are intrinsically correlated. This addresses several key challenges in modern cosmology. The inclusion of the $\mathcal{R}^2$ term provides a natural UV completion, raising the effective cutoff scale, while the inflaton is given a well-defined microphysical origin as a gauge singlet within the $SO(10)$ Higgs sector. This construction directly links the inflationary energy scale to the grand-unification scale.

Regarding the naturalness of the parameters, the large values of $\alpha \sim 10^{5}-10^{10}$ are consistent with the standard Starobinsky inflation framework, where $\alpha \approx m_{\rm Pl}^2 / (3 M_s^2)$ and the scalaron mass $M_s \sim 10^{13}$~GeV is set by the requirement to match the observed CMB scalar perturbation amplitude $\Delta_\mathcal{R}^2 \approx 2.1 \times 10^{-9}$~\cite{Planck:2018jri}. This large $\alpha$ arises naturally from the hierarchy between the Planck scale and the inflationary Hubble scale, ensuring a light scalaron that dominates the dynamics and suppresses the tensor-to-scalar ratio $r$ to observable levels without fine-tuning. For the parameter $M \gtrsim 50\, m_{\rm Pl}$, although super-Planckian, it represents the renormalization scale in the Coleman-Weinberg potential, where the physical GUT scale remains sub-Planckian, $M_U \sim (A M^4)^{1/4} \sim 10^{16}$~GeV, with the loop-suppressed coefficient $A \sim 10^{-14}-10^{-15}$ determined by the small Higgs--inflaton portal coupling $\varpi_S \sim 10^{-4}$ and the dimensionality $D=210$ of the $SO(10)$ representation. In the $SO(10)$ GUT context, $M$ is the vacuum expectation value of a gauge singlet scalar and such super-Planckian VEVs are accommodated in effective theories with UV completions such as the Palatini $\mathcal{R}^2$ term, which raises the perturbative cutoff above the inflationary scale and stabilizes the dynamics against quantum gravity effects, as discussed in Refs.~\cite{Ema:2017rqn,Gorbunov:2018llf}.

A major outcome of this work is establishing a quantitative connection between cosmological observables ($n_s$, $r$) and the parameters of the GUT symmetry-breaking chain ($M_U$, $M_I$). The slow-rolling inflaton $\phi$ triggers spontaneous symmetry breaking, dictating both the formation epoch and the subsequent inflationary dilution of topological monopoles. We identify a precise \textit{partial-inflation} window in which monopoles are sufficiently diluted to avoid overclosure yet remain potentially observable. The Palatini formulation has important phenomenological consequences, altering the mapping from the Jordan to the Einstein frame. This leads to systematically lower predictions for the tensor-to-scalar ratio $r$ and, critically, produces a direct, predictive correlation between the grand unification scale $M_U$ and $r$ (shown in the \emph{bottom} panels of Figs.~\ref{fig:monopoles1}, \ref{fig:monopoles2}, and~\ref{fig:monopoles3}). Consequently, observational constraints on $r$ from the CMB can directly inform the allowed parameter space for GUT-scale physics.

Our work extends previous studies on Coleman-Weinberg inflation in the Palatini formulation, such as Refs.~\cite{Racioppi:2017spw} and \cite{Gialamas:2020snr}, by embedding the mechanism within an $SO(10)$ GUT, thereby linking primordial cosmology to GUT-scale particle physics. In \cite{Racioppi:2017spw}, linear inflation emerges naturally from a non-minimally coupled Coleman-Weinberg potential with a dynamically generated Planck scale, and the Palatini formulation yields distinct e-fold predictions compared to the metric case, potentially discriminable for non-minimal couplings $\xi > 1$ through future CMB sensitivity on the tensor-to-scalar ratio $r$. Building on this, \cite{Gialamas:2020snr} incorporates an $\mathcal{R}^2$ term in Palatini gravity to suppress $r$ and restore compatibility with \emph{Planck} data, predicting an attractor behavior for large $\alpha \sim 10^{10}$ where $r$ approaches $\mathcal{O}(10^{-3})$ while maintaining viable $n_s$. Our framework improves upon these by identifying the inflaton as a GUT-singlet scalar that not only drives inflation but also induces intermediate-scale symmetry breaking, leading to topological monopole formation and partial dilution through $N_I \sim 10$--$17$ e-folds. This differs from prior works, which primarily focused on inflationary observables and reheating without GUT embedding. We establish a novel correlation between $r$ and the unification scale $M_U$, enabling complementarity between CMB probes and proton decay experiments. Unlike similar analyses emphasizing standalone cosmological predictions, our unified setup enhances testability by allowing cross-validation across cosmology and particle physics, while preserving the $\mathcal{R}^2$-driven attractor for $r$ and incorporating monopole abundances as additional observables.

This establishes a profound complementarity between CMB observables and proton decay searches. At the benchmark $M_U = 8.7 \times 10^{15}$ GeV, the model yields $r \approx 8 \times 10^{-4}$. This $M_U$ corresponds to proton lifetimes accessible to future probes such as Hyper-Kamiokande. Thus, independent cosmological and particle physics probes can cross-validate the model: a CMB detection of $r$ would pinpoint $M_U$ for targeted proton decay searches, while proton decay observations would predict specific $r$ values testable against CMB data, thereby enhancing the framework's falsifiability. Furthermore, this benchmark is shown in both Figs.~\ref{fig:PalatiniR2_M} and~\ref{fig:mi_1}. It not only yields the characteristic tensor ratio $r \approx 8 \times 10^{-4}$ and corresponds to testable proton lifetimes, but also lies within the region where intermediate-scale monopoles are sufficiently diluted to potentially observable abundances, as can be seen from Figs.~\ref{fig:monopoles1},~\ref{fig:monopoles2}, and~\ref{fig:monopoles3}. This triple consistency, across CMB observables, proton decay limits, \emph{and} monopole phenomenology, demonstrates the robust predictive power of our unified framework.

Future work will extend this framework to explore other cosmological relics, such as cosmic strings, and to compute the characteristic gravitational wave spectra generated during associated phase transitions.
\section{Conclusion}
\label{Sec:Conclusion}
In this paper, we present a UV-complete framework for cosmological inflation within an $SO(10)$ GUT, using Palatini gravity with an $\mathcal{R}^2$ term. The inflaton originates as a gauge singlet from the GUT Higgs sector, linking inflationary energy to the grand unification scale.

A key result is the correlation between cosmological observables ($n_s, r$) and GUT symmetry-breaking ($M_U, M_I$). The slow-rolling inflaton enables a partial-inflation window ($N_I \sim 10$--$17$ e-folds), diluting topological monopoles to avoid overclosure while keeping them potentially observable.
Main results:
\begin{itemize}
\item \textbf{Dominant $\mathcal{R}^2$ Control:} For large $\alpha$ ($\sim 10^{10}$), $r$ reaches a universal attractor $r \simeq 8 \times 10^{-4}$, detectable by upcoming CMB experiments~\cite{BICEP:2021xfz, SimonsObservatory:2018koc, Hazumi_2020}.
\item \textbf{Predictive CMB Observables:} Scalar spectral index $0.955 \lesssim n_s \lesssim 0.974$ (Fig.~\ref{fig:Palatini_ns}), aligning with current data~\cite{Planck:2018jri, ACT:2025tim}.
\item \textbf{Observable Monopole Window:} Relic monopole abundance $Y_M^+ \sim 10^{-27}$ (MACRO bound) to $Y_M^- \sim 10^{-35}$ (potentially observable) for $N_I \sim 10$--$17$ e-folds post-formation (Eq.~\eqref{eq:YM}, Tabs.~\ref{tab:mon-infl}--\ref{tab:mon-infl-alpha1_M}).
\item \textbf{Complementarity via Proton Decay:} Intermediate scales $M_I \sim 10^{13}$--$10^{14}$ GeV (Tabs.~\ref{tab:mon-infl}--\ref{tab:mon-infl-alpha1_M}), from unification and partial inflation, yield $M_U$ above Super-Kamiokande limits~\cite{Super-Kamiokande:2020wjk} and testable by Hyper-Kamiokande~\cite{Dealtry:2019ldr}.
\end{itemize}
Our model addresses the tension between \emph{Planck}-BICEP ($n_s \sim 0.955$--$0.965$)~\cite{Planck:2018jri, BICEP:2021xfz} and \emph{Planck}+ACT ($n_s \sim 0.967$--$0.974$)~\cite{ACT:2025tim} via $\phi < M$ and $\phi > M$ branches of the Coleman-Weinberg potential, maintaining $r \simeq 8\times10^{-4}$ within bounds.

The framework establishes a falsifiable complementarity among CMB
observations, monopole searches, and proton decay experiments
(Fig.~\ref{fig:unified}). For the benchmark value
$M_U = 8.7 \times 10^{15}\,\mathrm{GeV}$, the predicted tensor-to-scalar
ratio $r \simeq 8 \times 10^{-4}$ lies within the reach of forthcoming
CMB experiments such as the Simons Observatory and LiteBIRD, while the
corresponding proton lifetime for $p \to e^+ \pi^0$ falls within the
sensitivity of Hyper-Kamiokande, up to $10^{35}$ years~\cite{Dealtry:2019ldr}.
To clearly illustrate this complementarity, we highlight this
representative benchmark point, marked by a \emph{blue star}, in the
inflationary predictions shown in Fig.~\ref{fig:PalatiniR2_M},
correlate it with the unification scale in Fig.~\ref{fig:mi_1}, and
consistently emphasize this benchmark across the unification-related
plots in Figs.~\ref{fig:monopoles1}-\ref{fig:monopoles3}.

This complementarity enables cross-validation: CMB detection of $r$ pinpoints $M_U$ for proton decay searches; non-observation pushes $M_U$ thresholds, potentially excluding model regions; positive detection fixes $M_U$, predicts $r$, and allows consistency tests against CMB data.
\medskip
\acknowledgments
The authors thank Qaisar Shafi for collaboration during the early stages of this study, Antonio Racioppi for valuable comments and suggestions, and Rinku Maji for helpful feedback.
\clearpage
\appendix
\section{Slow-roll parameters}
\label{Sec:Slow_roll}
In this appendix, we provide a general discussion of the slow-roll approximation parameters used in the various scenarios examined throughout this manuscript. Inflationary predictions can be derived using these slow-roll parameters as follows~\cite{Lyth:2009zz}
\begin{equation}\label{slowroll1}
\epsilon =\frac{m_{\rm Pl}^2}{2}\left( \frac{V_{\zeta} }{V}\right) ^{2}\,, \quad
\eta = m_{\rm Pl}^2 \frac{V_{\zeta \zeta} }{V}  \,, \quad
\kappa ^{2} = m_{\rm Pl}^4 \frac{V_{\zeta} V_{\zeta\zeta\zeta} }{V^{2}}\,.
\end{equation}
Here, the subscript $\zeta$ denotes derivatives with respect to the canonical scalar field $\zeta$. In the slow-roll approximation, the inflationary observables, the scalar spectral index $n_s$, the tensor-to-scalar ratio $r$, and the running of the spectral index $\alpha_s$, are given by the following expressions
\begin{align}
    n_s &= 1 - 6 \epsilon_\zeta + 2 \eta_\zeta, \\
    r &= 16 \epsilon_\zeta, \\
    \alpha_s &\equiv \frac{\mathrm{d} n_s}{\mathrm{d} \ln k} = 16 \epsilon_\zeta \eta_\zeta - 24 \epsilon_\zeta^2 - 2 \kappa_\zeta^2,
\end{align}
where the slow-roll parameters are defined as
\begin{align}
    \epsilon_\zeta &\equiv \frac{1}{2} \left( \frac{V_\zeta'}{V} \right)^2, &
    \eta_\zeta &\equiv \frac{V_\zeta''}{V}, &
    \kappa_\zeta^2 &\equiv \frac{V_\zeta' V_\zeta'''}{V^2}.
\end{align}
\begin{eqnarray}\label{nsralpha1}
n_s = 1 - 6 \epsilon + 2 \eta \,,\qquad
r = 16 \epsilon, \nonumber\qquad
\alpha_s = 16 \epsilon \eta - 24 \epsilon^2 - 2 \kappa^2\,.
\end{eqnarray}
The number of e-folds $N_*$ in the slow-roll approximation is given by
\begin{equation} \label{efold1}
N_*=\int^{\zeta_*}_{\zeta_e}\frac{\rm{d}\zeta}{\sqrt{2\epsilon(\zeta)}}\,, \end{equation}
where the subscript ``$_*$'' denotes the quantities when the pivot scale exits the horizon, and $\zeta_e$ is the value of the inflaton at which inflation ends. The end of inflation can be computed from the condition
\begin{equation}
    \epsilon(\zeta_e) = 1.
\end{equation}
Furthermore, the amplitude of the curvature perturbations can be calculated using the relation
\begin{equation}
    \Delta_\mathcal{R}^2 = \frac{1}{24 \pi^2} \frac{V_*}{\epsilon_\zeta^*},
\end{equation}
where $V_*$ and $\epsilon_\zeta^*$ are evaluated at horizon exit.
\begin{equation} \label{perturb1}
\Delta_\mathcal{R}^2= \frac{V_E(\zeta)}{24\pi^2\epsilon(\zeta)}\,.
\end{equation}
The best-fit value for the pivot scale $k_* = 0.05~\mathrm{Mpc}^{-1}$ is 
\begin{equation}
    \Delta_\mathcal{R}^2 \approx 2.1\times 10^{-9},
\end{equation}
as obtained from the \emph{Planck} measurements~\cite{Planck:2018vyg}.

However, obtaining an analytical expression for an inflationary potential originally defined as $V_J(\phi)$ when expressed in terms of the canonical field $V_E(\zeta)$ is not always straightforward, or even possible, depending on the specific form of the potential. In such cases, one can instead perform a numerical analysis of the model’s predictions directly in terms of the original scalar field $\phi$, rather than transforming to the canonical field $\zeta$. Consequently, the slow-roll parameters must be expressed in terms of $\phi$; these parameters take the form~\cite{Linde_2011}
\begin{equation}
    \epsilon_\phi = \frac{1}{2} \left( \frac{V_\phi}{V} \right)^2, \qquad
    \eta_\phi = \frac{V_{\phi\phi}}{V}, \qquad
    \kappa_\phi^2 = \frac{V_\phi V_{\phi\phi\phi}}{V^2},
\end{equation}
where the derivatives are taken with respect to \(\phi\). The slow-roll parameters can then be expressed in terms of \(Z(\phi)\) as follows
\begin{eqnarray}
\epsilon=Z\epsilon_{\phi}\,,\quad
\eta=Z\eta_{\phi}+{\rm sgn}(V')Z'\sqrt{\frac{\epsilon_{\phi}}{2}}, \nonumber \\
\kappa^2=Z\left(Z\zeta^2_{\phi}+3{\rm sgn}(V')Z'\eta_{\phi}\sqrt{\frac{\epsilon_{\phi}}{2}}+Z''\epsilon_{\phi}\right).
\end{eqnarray}
Furthermore, the number of $e$-folds $N_*$ and the amplitude of the curvature perturbations $\Delta_\mathcal{R}^2$ can be given in terms of $\phi$ in the following forms
\begin{align}
    N_* &= \int_{\phi_e}^{\phi_*} \frac{Z(\phi)\, V(\phi)}{V^\prime(\phi)}\, d\phi, \\
    \Delta_\mathcal{R}^2 &= \frac{1}{12\pi^2} \frac{V^3(\phi_*)}{\left[ V^\prime(\phi_*) \right]^2 Z(\phi_*)}.
\end{align}
\begin{eqnarray}\label{perturb2}
N_*&=&\rm{sgn}(V')\int^{\phi_*}_{\phi_e}\frac{\mathrm{d}\phi}{\sqrt{2\,Z(\phi)\,\epsilon_{\phi}}}\,,\\
\label{efold2} \Delta_\mathcal{R}&=&\frac{1}{2\sqrt{3}\pi}\frac{V^{3/2}}{\sqrt{Z}|V^{\prime}|}\,.
\end{eqnarray}
To calculate the numerical values of inflationary observables, such as the spectral index $n_s$ and the tensor-to-scalar ratio $r$, the number of $e$-folds $N_*$ corresponding to the pivot scale is also required. Assuming a standard thermal history after inflation, $N_*$ for the pivot scale $k_* = 0.05~\mathrm{Mpc}^{-1}$ can be expressed as~\cite{Liddle_2003}
\begin{eqnarray} \label{efolds}
N_*\approx61.5+\frac12\ln\frac{\rho_*}{m_{\rm Pl}^4}-\frac{1}{3(1+\omega_r)}\ln\frac{\rho_e}{m_{\rm Pl}^4} +\left(\frac{1}{3(1+\omega_r)}-\frac14\right)\ln\frac{\rho_r}{m_{\rm Pl}^4}.
\end{eqnarray}
Here, $\rho_e = (3/2) V(\phi_e)$ is the energy density at the end of inflation, $\rho_r$ is the energy density at the end of reheating, and $\rho_* \approx V(\phi_*)$ is the energy density when the scale corresponding to $k_*$ exits the horizon. The parameter $\omega_r$ denotes the equation-of-state during reheating. 

The energy densities $\rho_r$ and $\rho_*$ can be expressed as
\begin{align}
    \rho_r &= \frac{\pi^2}{30}\, g_*\, T_\mathrm{reh}^4, \\
    \rho_* &\simeq V(\phi_*)= \frac{3 \pi^2 \, \Delta_R^2 \, r
}{2},
\end{align}
where $g_* = 106.75$ is the effective number of relativistic degrees of freedom, and $T_\mathrm{reh}$ denotes the reheat temperature.
\bibliography{main}
\bibliographystyle{JHEP}
\end{document}